\documentclass[aps,superscriptaddress,pre,twocolumn,nofootinbib,floatfix]
{revtex4}

\usepackage{amsmath}
\usepackage{graphicx}
\usepackage{color}
\usepackage{hyperref}
\newcommand{\be}{\begin{equation}}
\newcommand{\ee}{\end{equation}}
\newcommand{\bea}{\begin{eqnarray}}
\newcommand{\eea}{\end{eqnarray}}

\begin{document}
\sloppy


\title{Phase transitions in self-gravitating systems and bacterial populations
\\
surrounding a central body}

\author{Pierre-Henri Chavanis}
\author{Julien Sopik}
\author{Cl\'ement Sire}
\affiliation{Laboratoire de Physique Th\'eorique, CNRS \&
Universit\'e de Toulouse -- Paul Sabatier, France}

\begin{abstract}
We study the nature of phase transitions in a self-gravitating
classical gas in the presence of a central body.  The central body
can mimic  a black hole at the center of a galaxy or a rocky core
(protoplanet) in the context of planetary formation. In the chemotaxis of bacterial
populations, sharing formal analogies with self-gravitating systems, the central
body can be
a supply of ``food'' (chemoattractant). We consider both microcanonical (fixed
energy) and 
canonical (fixed temperature)  descriptions and study the inequivalence
of statistical ensembles. At high energies (resp. high temperatures), the system
is in a ``gaseous'' phase and at low energies (resp.  low temperatures) it
is in a condensed phase with a ``cusp-halo'' structure, where the cusp
corresponds to the rapid increase of the density of the gas at the
contact with the central body. For a fixed density $\rho_{*}$ of the central body, we show the existence of two critical points in the phase
diagram, one in each ensemble, depending on the core
radius $R_{*}$: for small radii $R_{*}<R_{*}^{\rm MCP}$, there exist both
microcanonical and canonical phase transitions (that are zeroth and first order); for intermediate radii
$R_{*}^{\rm MCP}<R_{*}<R_{*}^{\rm CCP}$, only canonical phase transitions are
present; and for large radii $R_{*}>R_{*}^{\rm CCP}$, there is no phase
transition at all.  We study how the nature of these phase transitions changes as a function
of the dimension of space. We also discuss the analogies and the differences with
phase transitions in the self-gravitating Fermi gas [P.\,H. Chavanis, Phys. Rev. E
{\bf 65}, 056123 (2002)].
\end{abstract}

\maketitle

\section{Introduction}

Self-gravitating systems have a very peculiar thermodynamics due to the
unshielded long-range attractive nature of the gravitational interaction
\cite{paddy,dvs1,dvs2,katzrevue,ijmpb}. Their study is
interesting both from the viewpoint of astrophysics and statistical
mechanics  since a collection of $N$ stars (composing  globular
clusters, galaxies...) constitutes a fundamental physical system with
long-range
interactions~\cite{houches,cdr,campabook}.
The subject started with the seminal paper of Antonov~\cite{antonov} who
considered the thermodynamics of a self-gravitating classical gas
confined within a spherical box of radius $R$. Since the system is isolated, it
must be
studied in the microcanonical ensemble. The box is necessary to prevent
the evaporation of the gas and make the problem well-posed
mathematically.\footnote{There is no statistical equilibrium state for a
self-gravitating system in an infinite domain because the gas has the
tendency to evaporate \cite{amba,spitzer1940,chandra} (this is already the case
for an ordinary gas if it is not
confined within a container). In this sense, the strict statistical equilibrium
state of a stellar system is made of two stars in Keplerian motion with all the
other stars dispersed at infinity. 
Antonov~\cite{antonov} argued that the
evaporation of stellar systems is a slow process so
that, on intermediate timescales, everything happens as if the system
were confined within a bounded container. In another interpretation, 
the box can mimic the effect of a tidal radius beyond which the particles are
captured by a neighboring object (e.g., a galaxy in the vicinity of a globular
cluster).} Following Ogorodnikov~\cite{ogo1,ogo2}, Antonov studied the problem
of
maximizing the Boltzmann entropy at fixed mass and energy, and
discovered that there is no global entropy maximum. There is not even an
extremum of entropy (canceling the first order variations of entropy at
fixed mass and energy) if the energy is below a critical value
$E_{c}=-0.335GM^2/R$~\cite{antonov}. Lynden-Bell and Wood~\cite{lbw}
(see also  Lynden-Bell~\cite{lb}) interpreted
this mathematical result in terms of a  gravitational
collapse that they called {\it gravothermal catastrophe}. Below the Antonov
threshold $E<E_c$, the system takes a ``core-halo'' structure. The core, which
has a negative specific heat, becomes hotter as it loses energy to the
profit of the halo. Therefore, it continues to lose energy and evolves
away from equilibrium. By this process, the system becomes hotter and
hotter and more and more centrally condensed (as a consequence of the virial
theorem).
From thermodynamical arguments, the end-product of the gravothermal catastrophe is expected to be 
a {\it binary
star} having a small mass ($2m\ll M$) but a huge binding energy
\cite{henonbinary}. The potential energy released by the binary star is
redistributed
in the halo in the form of kinetic energy (the system heats up) leading to an
infinite  (very large) entropy.
Cohn~\cite{cohn} studied the dynamical evolution of the gravothermal catastrophe in the
microcanonical ensemble (fixed energy) by numerically solving the
orbit-averaged-Fokker-Planck equation. He found that the
collapse is
self-similar and that the density profile develops a finite time
singularity (core collapse). The central
density becomes infinite in a
finite time, scaling as $\rho\sim r^{-2.23}$, but the  core mass $M_0(t)\sim
(t_{\rm coll}-t)^{0.41}$ tends to zero at the collapse
time.\footnote{Previously,
Larson~\cite{larson}, Hachisu {\it et al.}~\cite{hachisu} and Lynden-Bell and
Eggleton~\cite{lbe} studied the same problem by using fluid
equations and obtained similar results.} Therefore, the
divergence of the central density is simply due to a {\it few} stars approaching
each other.  In fact, a binary star is formed in the post-collapse regime and
releases so much energy that the halo re-expands in a
self-similar manner~\cite{inagakilb}.\footnote{Heggie
and Stevenson~\cite{hs} confirmed these results by constructing self-similar
solutions of the orbit-averaged-Fokker-Planck equation in the pre-collapse and
post-collapse regimes.} Finally, a
series of gravothermal oscillations follows~\cite{sugimoto}. On the other hand,
at sufficiently high energies $E>E_c$, there exist statistical equilibrium
states in the form of local entropy maxima at fixed mass and energy (there may
also be saddle points of
entropy but they are unstable)~\cite{antonov}. These states are metastable but
they are very long lived because their lifetime scales like $e^N$ (with $N\sim
10^{6}$ in a globular cluster)~\cite{lifetime}. Therefore, these gaseous states
can play an important role in the dynamics~\cite{bt}. Indeed, most stellar
systems like globular clusters are in the form of metastable gaseous states.
They are described by the Michie-King model~\cite{michie,king} which is a
truncated Boltzmann distribution taking into account the evaporation of high
energy stars. The thermodynamics of tidally truncated
self-gravitating systems was studied by Katz~\cite{katzking} and Chavanis {\it
et al.}~\cite{clm1}. In practice, a globular cluster relaxes through
gravitational encounters towards a
truncated equilibrium state. As it slowly evaporates, its central density
increases (as a consequence of the virial theorem) and the cluster follows the
King sequence with higher and higher central densities until a point at which is
becomes unstable and undergoes core collapse as described above.

Since the statistical ensembles are not equivalent for self-gravitating systems
(see~\cite{thirring,lblb} for early works on the subject and
\cite{paddy,dvs1,dvs2,katzrevue,ijmpb} for reviews), it can be of interest to
study what happens
 when the system is in contact with a thermal
bath imposing the temperature. In that case, the evolution of the system is
dissipative and it must be studied 
in the canonical ensemble (CE).
If one considers the problem of minimizing
the free energy at fixed mass, one finds that there is no global minimum of
free energy. There is not even an extremum of free energy (canceling
the first variations of free energy at fixed mass) below a
critical temperature $T_{c}=GMm/2.52k_{B}R$ discovered by Emden~\cite{emden}.
The
absence of a minimum of free energy leads to an {\it isothermal
collapse}~\cite{aa1}. From statistical mechanics arguments, the end-product of
this gravitational collapse is expected to be a {\it Dirac peak}
containing all the mass since this structure has an infinite free energy. This
has actually be proven rigorously by Kiessling~\cite{kiessling}.  
Chavanis {\it et al.}~\cite{crs,sc} studied the dynamical
evolution of the isothermal collapse for $T<T_c$ in the canonical ensemble
(fixed
temperature) by solving numerically and analytically the
Smoluchowski-Poisson system describing a gas of self-gravitating
Brownian particles in an overdamped limit. They found that the pre-collapse is self-similar
and that the density profile develops a finite time
singularity. The central density becomes infinite in a
finite time, scaling as $\rho\sim r^{-2}$, but the  core mass $M_0(t)\sim
(t_{\rm coll}-t)^{1/2}$ tends to zero at the collapse time. Therefore, ``the
central singularity contains no mass'' in apparent contradiction with the
thermodynamical expectation. In fact,
the collapse continues after the singularity and a Dirac peak
containing all the mass is finally formed in the post-collapse regime
\cite{post}. On the other hand, at sufficiently high temperatures $T>T_c$, there
exist statistical equilibrium states in the form of local free energy  minima
at fixed mass
(there may also be saddle points of free energy but they are unstable)  
\cite{aa1}. These states are metastable but they are very long lived since their
lifetime scales like $e^N$~\cite{lifetime}. Therefore, these gaseous states can, again, 
play an important role in the dynamics. The self-gravitating Brownian gas may
describe the evolution of planetesimals in the solar nebula in the context of
planet formation. In that case, the particles experience a friction with the
gas and a stochastic force due to Brownian motion or turbulence
\cite{aaplanetes}. If the gas of particles is sufficiently dense (e.g., at
special locations such as large-scale vortices), self-gravity
becomes important leading to gravitational collapse and planet formation.

The caloric curve of classical self-gravitating
systems has the form of a spiral and the stability of the
equilibrium states can be determined by using the Poincar\'e turning point
criterion~\cite{poincare,lbw,katzpoincare1}. It is then found that the
equilibrium states in the
canonical ensemble become unstable after the first turning point of temperature
$T_c$ (corresponding to a density contrast of $32.1$) while the equilibrium
states in the microcanonical ensemble become unstable later, after
the first turning point of energy (corresponding to a density
contrast of $709$). Therefore, the statistical ensembles are inequivalent
\cite{lbw,thirring,lblb}, which is a fundamental feature of systems with
long-range
interactions~\cite{campabook}. For classical self-gravitating systems, the
region of ensemble inequivalence corresponds to a region of negative specific
heats which is allowed (stable) in the microcanonical ensemble but forbidden
(unstable) in the canonical ensemble.

Since the core collapse of classical point masses leads to a singularity
(a binary star in the microcanonical ensemble or a Dirac peak in the canonical
ensemble), some authors have
studied what happens when a physical short-range
regularization is introduced in the problem. In that case, we expect a phase
transition between a gaseous (dilute) phase which is
independent of the small-scale regularization and a
condensed phase in which the small-scale regularization plays a
prevalent  role. Aronson and Hansen~\cite{aronson} considered the case of a
self-gravitating hard-sphere gas in the canonical ensemble,
modeled by
a van der Waals equation of state, and evidenced a first order
canonical phase transition when the filling factor 
$\mu=R/R_*$ (where $R$ is the system size and $R_*\sim N^{1/3}a$ is the radius of a compact 
object in which all the particles of size $a$ are packed together) is
sufficiently
large.  In the condensed phase, the equilibrium state has a
``core-halo'' structure with a dense solid core surrounded by a dilute
atmosphere. Their study was followed by Stahl {\it et al.}~\cite{stahl} who used a more
accurate equation of state and discussed interesting applications to planet formation. These
authors considered both canonical and microcanonical ensembles and
evidenced a first order microcanonical phase transition in the case of
very large filling factors. Instead of
considering a classical
hard-sphere gas, one can also consider a gas of self-gravitating
fermions described by the Fermi-Dirac statistics. This system can have
application in the context of white dwarf stars~\cite{chandrab},
neutron stars~\cite{ov,shapiroteukolsky}, and dark matter halos made of massive
neutrinos~\cite{bmtv,btv,vss,rar,mg16F,predictiveF,arguellerevue}. In that case,
the Pauli exclusion
principle
creates an
``effective'' repulsion between the particles and plays a role similar to
that played by the core radius of classical particles in a hard sphere
gas.\footnote{One can also consider a Fermi-Dirac distribution
in configuration space leading to qualitatively similar results
\cite{exclusion,bbkmm}.} For
self-gravitating fermions, the end-product of the collapse is a
``core-halo'' structure with a degenerate core (equivalent to a
polytrope of index $n=3/2$) surrounded by a dilute atmosphere. In the context
of dark matter halos, the quantum core is called a ``fermion ball''. The
statistical mechanics of self-gravitating fermions was first
considered by Hertel and Thirring~\cite{htf} who showed rigorously (in a
mathematical sense) that the mean field approximation is exact in a
proper thermodynamic limit. In another paper, Hertel and Thirring~\cite{ht2},
and later Bilic and
Viollier~\cite{bvNR}, discussed first
order canonical phase transitions in the self-gravitating Fermi gas but did not
evidence first order microcanonical phase transitions in their
study. Chavanis~\cite{ptfermi} performed an exhaustive study of
phase transitions in
the self-gravitating Fermi gas in both canonical and microcanonical
ensembles and explored the whole range of parameters. He showed the
existence of two critical points, one in each ensemble. The 
control parameter can be written as  $\mu=17.2587 (R/R_*)^{3/2}$, where  $R$ is
the system
size and $R_*=0.181433\, h^2/(Gm^{8/3}g^{2/3}M^{1/3})$ is the radius of a
completely degenerate object of
mass $M$ at $T=0$ (``white dwarf'') determined by the  Planck constant $\hbar$
\cite{ijmpb}. Since
$R_*$  is fixed by quantum mechanics (for a given mass $M$), the 
control parameter measures the size of the system $R$. For large
systems ($\mu>\mu_{\rm MCP}\simeq 2670$), there exist both microcanonical and
canonical phase
transitions (of zeroth and first order), for systems of intermediate size
($\mu_{\rm CCP}<\mu<\mu_{\rm MCP}$) only canonical phase transitions
exist, and for small systems ($\mu<\mu_{\rm CCP}\simeq 83$) there is no phase
transition at
all. Other types of small-scale regularization have been introduced and
lead to a similar phenomenology~\cite{my,fl,ym,ci,dv}. A review of phase
transitions in self-gravitating systems is given in~\cite{ijmpb}. These results
have been extended to self-gravitating fermions in general
relativity~\cite{bvR,rc,calettre,acepjb}. They have also
been discussed in the
context of the fermionic King model (avoiding the artificial box) in
Newtonian gravity~\cite{clm2} and general relativity~\cite{adky}.

In the present work, we study the nature of phase transitions in a
self-gravitating classical gas of point-like particles
in the presence of a central body. The
central body could mimic the effect of a black hole at the center of a
galaxy or at the center of a globular cluster. It could also represent a
rocky core (protoplanet) at the center of a giant gaseous
planet like Jupiter or Saturn. The central body prevents the formation of
singularities resulting from
gravitational collapse and plays a role similar to that of a
short-range regularization. There exists an equilibrium state for all
accessible values of energy and temperature but canonical and
microcanonical phase transitions can take place between a gaseous
phase and a condensed phase. The condensed phase has the
structure of a giant gaseous planet with a solid core surrounded by an
atmosphere.\footnote{It can also describe the structure of a star cluster
(galactic nucleus, globular cluster...) harboring a central black hole.} The
atmosphere has a ``cusp-halo'' structure, where the cusp
corresponds to the rapid increase of the density of the gas at the
contact with the central body. For a fixed density $\rho_{*}$ of the central
body (and a fixed system size $R$),  we show the existence of two critical
points in the phase
diagram, one in each ensemble, depending on the core
radius $R_{*}$. For small radii $R_{*}<R_{*}^{\rm
MCP}$, there exist both
microcanonical and canonical phase transitions (that are zeroth and first order). For intermediate radii
$R_{*}^{\rm MCP}<R_{*}<R_{*}^{\rm CCP}$, only canonical phase transitions are
present. Finally, for large radii $R_{*}>R_{*}^{\rm CCP}$, there is no phase
transition at all.\footnote{One could equivalently fix the radius of the
central body and increase the system size.} This is qualitatively similar to the
results previously
obtained for self-gravitating fermions~\cite{ijmpb}. We also study how the
canonical and
microcanonical critical
points $R_{*}^{\rm CCP}$  and $R_{*}^{\rm MCP}$ depend on the density of the
central body $\rho_*$.

Our results can also have applications for the problem of chemotaxis
in biology~\cite{murray}.
Indeed, there exists a remarkable analogy between self-gravitating
systems and bacterial populations~\cite{crrs}. In this analogy, the density of
the gas $\rho({\bf r},t)$ is the counterpart of the density of bacteria
$\rho({\bf r},t)$, and minus the gravitational potential $-\Phi({\bf r},t)$ is
the counterpart of the concentration $c({\bf r},t)$ of the chemical
(``pheromone'') produced by the bacteria (see Appendix \ref{sec_analogy}). As a
result, the counterpart of a
central mass $M_*$
(black hole, protoplanet,...) giving rise to an external gravitational force
$-\nabla\Phi_{\rm ext}$ is a supply of ``food'' (chemoattractant)
giving
rise to a chemical drift $\nabla c_{\rm ext}({\bf r})$. Therefore, our
study can have applications in biology up to a straightforward change of
notations.

In this article, we will use the notations of astrophysics in order to make the
connection with previous studies. We will essentially consider the spatial dimension
$d=3$ which is the most relevant dimension in astrophysics,
and which leads to the richest variety of phase transitions. 
It corresponds to spherical halos and stars~\cite{chandrab}.
However, the dimension $d=2$ corresponding to cosmic filaments
\cite{stodolkiewicz,ostriker} and the dimension $d=1$ corresponding to
sheets or pancakes
\cite{spitzer,camm} have also been considered in astrophysics.
On the other hand, the dimension
$d=2$ is particularly relevant in biology~\cite{chemo2d} since cells of bacteria
can be
cultured on a petri dish. We will therefore study how the nature of phase
transitions changes as
a function of the dimension of space. A similar study of the effect of the
dimension of space on the nature of gravitational phase transitions has been
performed in
\cite{sc} for classical
self-gravitating systems, in 
\cite{ptd,wdd,kmcfermions,kmcbosons} for self-gravitating
fermions and bosons, and in~\cite{exclusion,bbkmm} for
particles with an exclusion constraint in position space.

The paper is organized as follows. In Sec.~\ref{sec_eq} we
derive the basic equations determining the equilibrium structure of a
self-gravitating isothermal gas around a central body. In Sec.~\ref{sec_pt3} we
plot the caloric curves of this self-gravitating gas in $d=3$ dimensions and
study the corresponding phase transitions. In Sec.~\ref{sec_pt12} we briefly
consider the case of the dimensions $d=1$ and $d=2$. The Appendices regroup
useful formulae that are needed in the theoretical part of the paper.

\section{Equilibrium structure of a self-gravitating  isothermal gas around a central body}
\label{sec_eq}

\subsection{The maximum entropy principle}
\label{ss.var}

We consider a system of $N$ point-like particles of mass $m$ interacting via
Newtonian gravity in a space of dimension $d$. We allow for the
presence of a spherically symmetric central body of mass
$M_{*}$ and radius $R_{*}$. This central body may mimic a black hole at
the center of a galaxy or at the center of a globular cluster. It may also
describe a rocky core
surrounded by a gas (atmosphere) in the
context of planet formation. The particles are enclosed within a
spherical box of radius $R$ so as to prevent the evaporation of the gas
and make a statistical equilibrium state well-defined.\footnote{In $d=3$
dimensions, there is no statistical equilibrium state in an infinite domain
(there are not even extrema of entropy at fixed mass and energy with a
finite mass). In $d=1$ and $d=2$ dimensions, there exist statistical
equilibrium states in an infinite domain that are studied analytically in a
companion paper~\cite{companion}. In the present paper, we shall  only consider
systems
enclosed
within a box of size $R$.} Physically, a gas is never
completely isolated from the surrounding. Therefore, the box can play the
role of a tidal radius in the case of globular
clusters and dark matter halos\footnote{An alternative to the box would be
to consider a truncated Michie-King model~\cite{michie,king} like in
Refs.~\cite{katzking,clm1,clm2}.} or represent the Hill
sphere in the
context of planet
formation. The Hamiltonian of the self-gravitating
system is
\be
H=\sum_i \frac{1}{2}m v_i^2+m^2\sum_{i<j} u(|{\bf r}_i-{\bf r}_j|)+m\sum_i
\Phi_{\rm ext}({\bf r}_i),
\ee
where $i=1,...,N$ runs over the particles of the gas. The first term
is the kinetic energy, the second term represents the interaction
energy of the particles of the gas and the third term takes into account the
interaction between the gas and the central body. The gravitational potential of
interaction in $d$ dimensions  is given by
\be
\label{defu1}
u(|{\bf r}-{\bf r}'|)=-\frac{1}{d-2} \frac{G}{|{\bf r}-{\bf r}'|^{d-2}}\qquad
(d\neq 2),
\ee
\be
\label{defu2}
u(|{\bf r}-{\bf r}'|)=G \ln\frac{|{\bf r}-{\bf r}'|}{R}\qquad (d=2).
\ee
We treat the influence of the
central body as an external potential (see Appendix \ref{sec_vt}):
\be
\label{tpcore1}
\Phi_{\rm ext}({\bf r})=-\frac{1}{d-2}\frac{GM_*}{r^{d-2}}\quad (d\neq 2),
\ee
\be
\label{tpcore2}
\Phi_{\rm ext}({\bf r})=GM_*\ln \left (\frac{r}{R}\right )\quad (d= 2).
\ee

Let
$f({\bf r},{\bf v})$ denote the distribution function of the system,
i.e., $f({\bf r},{\bf v}) d{\bf r}d{\bf v}$ gives the mass of
the particles of the gas whose position and velocity are in the cell $({\bf
r},{\bf
v}; {\bf r}+d{\bf r},{\bf v}+d{\bf v})$. The integral of
$f$ over the velocity determines the spatial density $\rho=\int f d{\bf v}$.
The total mass of the gas is
\be
\label{mass}
M=\int \rho \, d{\bf r}.
\ee
The spatial integral extends only in the
region surrounding the central body, i.e., in the region  $R_{*}<r<R$ covered
by
the gas. We consider a proper thermodynamic limit $N\rightarrow +\infty$ for
self-gravitating systems in such a way that the rescaled energy and the
rescaled
temperature
\be
\label{fr}
\Lambda=-\frac{ER^{d-2}}{GM^{2}}, \qquad \eta=\frac{\beta GMm}{R^{d-2}},
\ee
are independent on $N$. We also introduce the parameters
\be
\label{muzeta}
\mu=\frac{M_{*}}{M},\qquad \zeta=\frac{R_{*}}{R},
\ee
which represent the normalized mass and the normalized radius of the central
body. In the $N\rightarrow +\infty$ limit,\footnote{A relevant scaling
is $N\rightarrow +\infty$ with $m\sim 1/N$, $R\sim 1$, $G\sim 1$, $E\sim 1$ and
$T\sim 1/N$ (see Appendix A in~\cite{aakin}).} the mean field approximation
is exact, except close to a critical point
\cite{paddy,dvs1,dvs2,katzrevue,ijmpb,houches,cdr,campabook}. Therefore,
ignoring the
correlations between the particles of the gas, the total energy of the
system can be expressed as 
\be
\label{eq.energiemoycc}
E=\frac{1}{2}\int f v^{2}\, d{\bf r}d{\bf v}+\frac{1}{2}\int \rho\Phi \, d{\bf
r}+\int \rho\Phi_{\rm ext} \, d{\bf r},
\ee
where $\Phi_{\rm ext}({\bf r})$ is the potential created by the central body and
$\Phi({\bf r})$ is the gravitational potential created by the
gas (see Appendix \ref{sec_vt}). For $R_*\le r\le R$, they are respectively 
the solutions of the Laplace equation
\be
\label{lap}
\Delta\Phi_{\rm ext}=0,
\ee
and the Poisson equation
\be
\label{eq.equationpoissoncc}
\Delta\Phi=S_{d}G\rho,
\ee
where  $S_d=2\pi^{d/2}/\Gamma(d/2)$ is the surface of a unit sphere in $d$
dimensions. The energy $E=E_{\rm kin}+W_{\rm tot}$ of the gas is
the sum of its kinetic energy $E_{\rm kin}=(1/2)\int f v^{2}\, d{\bf
r}d{\bf v}$ and its total potential energy $W_{\rm tot}=W+W_{\rm
ext}$, where $W=(1/2)\int \rho\Phi \, d{\bf r}$ is
the  self-gravitating energy of the gas and $W_{\rm ext}=\int \rho\Phi_{\rm ext}
\, d{\bf r}$ is the gravitational energy of the gas in the potential created by
the central body (see Appendix
\ref{sec_vt}).

For isolated Hamiltonian systems, the mass and the energy are conserved 
and the thermodynamical potential is the Boltzmann entropy\footnote{The entropy
is defined up to an additive constant (not important in our case), which
explains why the argument of the logarithm is not dimensionless.}
\be
\label{eq.entropiemoycc}
S_{B}=-k_{B}\int \frac{f}{m}\ln \frac{f}{m} d{\bf r}d{\bf v}.
\ee
The Boltzmann entropy $S_{B}=k_B \ln {\cal W}$ measures the ``disorder''
of the system. It is proportional to the logarithm of the number
${\cal W}(\lbrace n_{i}\rbrace)$ of {\it microstates} corresponding to a given {\it
macrostate}~\cite{ogorodnikov}.  At statistical equilibrium, the
system
is in the most mixed state consistent with all the constraints of the
dynamics.  Therefore, if the system is isolated, the equilibrium state
maximizes the Boltzmann entropy $S_B$ at fixed mass $M$ and energy $E$
(microcanonical description). We thus have to solve the maximization
problem
\be
\label{maxs}
\max_{f}\quad \lbrace S_{B}[f] \quad | \quad E[f]=E, \quad M[f]=M\rbrace.
\ee
Alternatively, if the system is in contact with a heat bath imposing 
the temperature $T$, only the mass is conserved and the thermodynamical
potential is the Boltzmann free energy $F_{B}=E-TS_{B}$. It is often more
convenient to work
with the Massieu function
\be
\label{massieu}
J_{B}=\frac{S_{B}}{k_B}-\beta E,
\ee
where $\beta=1/(k_B T)$ is the inverse temperature. The Massieu function is the
Legendre transform of the
entropy with respect to the energy. If the system is in
contact with a heat
bath, the equilibrium
state maximizes the Massieu function $J$ at
fixed mass $M$ (canonical description).
We thus have to solve the maximization
problem
\be
\label{maxj}
\max_{f}\quad \lbrace J_{B}[f] \quad | \quad M[f]=M\rbrace.
\ee
The microcanonical ensemble is the proper description for (isolated)  Hamiltonian
systems and the canonical ensemble is the proper description for (dissipative)
Brownian systems~\cite{ijmpb}.

The extrema of entropy at fixed mass and energy (canceling
the first order variations of entropy under constraints) are
determined by the variational principle
\be
\label{eq.principevariationnelmicrocc}
\frac{\delta S_{B}}{k_B}-\beta\delta E-\alpha_0\delta M=0,
\ee
where $\beta$ and $\alpha_0$ are Lagrange multipliers associated with
the conservation of $E$ and $M$. The extrema of Massieu function at
fixed mass (canceling the first order variations of Massieu function
under
constraints) are determined by the variational principle
\be
\label{eq.principevariationnelcanocc}
\delta J_{B}-\alpha_0\delta M=0,
\ee
where $\alpha_0$ is a Lagrange multiplier associated with the
conservation of $M$. Using the identities
\be
\delta S_{B}=-k_{B}\int \frac{\delta f}{m}\left (\ln \frac{f}{m}+1\right ) \, d{\bf r}d{\bf v},
\ee
\be
\delta E=\frac{1}{2}\int \delta f v^{2}\, d{\bf r}d{\bf v}+\int \Phi\delta\rho
\, d{\bf r}+\int \Phi_{\rm ext}\delta\rho
\, d{\bf r},
\ee
\be
\delta M=\int \delta f \, d{\bf r}d{\bf v},
\ee
we find that the variational principles
(\ref{eq.principevariationnelmicrocc}) and
(\ref{eq.principevariationnelcanocc}) lead to the mean field
Maxwell-Boltzmann distribution
\be
\label{eq.distributionmaxboltzcc}
f=A'e^{-\beta m \lbrack v^{2}/2+\Phi({\bf r})+\Phi_{\rm ext}({\bf r})\rbrack},
\ee
where $A'=me^{-\alpha_0 m-1}$. Introducing the local (kinetic) pressure $P({\bf
r})=\frac{1}{d}\int f v^{2}\,
d{\bf v}$ (see Appendix \ref{sec_che}), we find that the barotropic equation of
state corresponding
to the distribution (\ref{eq.distributionmaxboltzcc}) is that of an
isothermal gas
\be
\label{eq.equationetatcc}
P({\bf r})=\rho({\bf r})\frac{k_{B}T}{m}.
\ee
As a result, the velocity dispersion $\langle v^2\rangle=dP/\rho=dk_B T/m$
is constant.
Integrating Eq.~(\ref{eq.distributionmaxboltzcc}) over the velocity, we find that the spatial density is the mean field  Boltzmann distribution
\be
\label{eq.distributionboltzcc}
\rho=Ae^{-\beta m \lbrack \Phi({\bf r})+\Phi_{\rm ext}({\bf r})\rbrack },
\ee
where $A=(2\pi/\beta m)^{d/2}A'$. Therefore, combining Eqs.~(\ref{eq.equationpoissoncc}) and (\ref{eq.distributionboltzcc}), the structure
of the gas around the central body is obtained
by solving the Boltzmann-Poisson equation
\be
\label{eq.equationboltzpoiss}
\Delta\Phi=S_{d}GAe^{-\beta m ( \Phi+\Phi_{\rm ext})}
\ee
with appropriate boundary conditions (see below),
and by relating the Lagrange multipliers to the constraints. We can then plot
the series of equilibria $\beta(E)$ for given values of $M_*$ and $R_*$ (or in
dimensionless form $\eta(\Lambda)$ for given $\mu$ and $\zeta$). The control
parameter is $E$ in the microcanonical ensemble and $\beta$
in the canonical ensemble. The stable region of the series of
equilibria defines the
caloric curve in the considered ensemble. Note that the extrema (regarding the
first variations) of the entropy
$S_B$
at fixed $E$ and $M$, and the extrema of the Massieu function $J_B$ at
fixed
$T$ and $M$ are the same, and both determine a self-gravitating isothermal gas. 
However, the stability of the gas (regarding the
second variations of $S_B$ or $J_B$ with appropriate constraints)
may differ in the microcanonical (fixed
$E$) and canonical (fixed $T$) ensembles. When this happens, this is
referred to a situation of {\it ensemble inequivalence}. It can be shown
that canonical stability (Eq.~(\ref{maxj})) implies microcanonical
stability (Eq.~(\ref{maxs})), but the converse is in general false for systems with
long-range interactions~\cite{ellis,cc}. For example, negative
specific heats are forbidden in the canonical ensemble while they are allowed in
the microcanonical ensemble. Therefore, canonical stability only
provides a
{\it sufficient} condition of microcanonical stability. Ensemble
inequivalence and phase transitions in self-gravitating systems is
well-documented in the absence of a central
body~\cite{paddy,dvs1,dvs2,katzrevue,ijmpb}. We shall study how
the presence of a central body affects these results.

\subsection{The Emden equation}
\label{ssec.equationemdencc}

Introducing the total gravitational potential
\be
\label{ptot}
\Phi_{\rm tot}=\Phi+\Phi_{\rm ext},
\ee
and using Eq.~(\ref{lap}), we
can rewrite the
Boltzmann-Poisson equation
(\ref{eq.equationboltzpoiss}) as
\be
\label{eq.equationboltzpoisstot}
\Delta\Phi_{\rm tot}=S_{d}GAe^{-\beta m \Phi_{\rm tot}}.
\ee
It can be shown that the
maximum entropy state of a non-rotating self-gravitating system is spherically
symmetric~\cite{antonov}. In that case, we can rewrite the foregoing equation as
\be
\label{bptot}
\frac{1}{r^{d-1}}\frac{d}{dr}
\left(
r^{d-1} \frac{d \Phi_{\rm tot}}{dr}
\right)
= S_d G A e^{-\beta m \Phi_{\rm tot}}.
\ee
It has to be solved with the boundary condition
\be
\label{bco}
\frac{d\Phi_{\rm tot}}{dr}(R_*)=\frac{d\Phi_{\rm
ext}}{dr}(R_*)=\frac{GM_*}{R_{*}^{d-1}}
\ee
resulting from Eq.~(\ref{lg}) and the fact that $d\Phi/dr(R_*)=0$, since $\rho=0$,
for $r\le R_*$ (see Appendix \ref{sec_pfd}). Multiplying Eq.~(\ref{bptot}) by
$r^{d-1}$, integrating between $R_*$ and $r$, and using the boundary condition
(\ref{bco}), we
obtain the integrodifferential equation
\begin{equation}
\label{gros}
r^{d-1} \frac{d \Phi_{\rm tot}}{d r} = GM_*
+  G A \int_{R_*}^r  e^{-\beta m \Phi_{\rm tot}(r')} S_d\, {r'}^{d-1} \,dr'.
\end{equation}
This equation can also be written as 
\begin{equation}
\label{newl}
\frac{d \Phi_{\rm tot}}{d r} = \frac{GM_{\rm tot}(r)}{r^{d-1} },
\end{equation}
where
\begin{equation}
M_{\rm tot}(r)=M_*+M(r)=M_*+\int_{R_*}^r \rho(r') S_d\,
{r'}^{d-1} \,dr'
\end{equation}
is the total mass contained within the sphere of radius
$r$. This is Newton's law in $d$ dimensions (see
Appendix \ref{sec_pfd}).

To determine the structure of the isothermal gas, we introduce the function
\begin{equation}
\psi = \beta m(\Phi_{\rm tot} - \Phi_{\rm tot,0}),
\end{equation}
where $\Phi_{\rm tot,0}=\Phi_{\rm tot}(R_{*})$ is the total gravitational
potential at
$r=R_{*}$. Then, the density can be written as
\begin{equation}
\label{eq.definitionrhocc}
\rho=\rho_{0}e^{-\psi},
\end{equation}
where $\rho_{0}=\rho(R_{*})$ is the density of the gas at the contact
with the central body. By an abuse of language, $\rho_0$ will be called
the central density and  $\Phi_{\rm tot,0}$ will be called the central 
total gravitational
potential. Introducing
the dimensionless radius
\begin{equation}
\label{voir}
\xi = (S_d
G \beta m \rho_{0})^{1/2} r,
\end{equation}
the Boltzmann-Poisson equation~(\ref{bptot}) reduces to the form
\be
\label{eq.equationemdencc}
\frac{1}{\xi^{d-1}}\frac{d}{d\xi}
\left(
\xi^{d-1} \frac{d \psi}{d\xi}
\right)
= e^{-\psi},
\ee
which is called the Emden equation~\cite{emden,chandrab}. In the presence of a central body, this equation has to be solved
with the boundary conditions [see Eq.~(\ref{bco})]
\bea
\label{bcemden}
\psi(\xi_0) = 0,\qquad
\psi'(\xi_0)= \frac{\eta_0}{\xi_0},
\eea
where we have defined
\begin{equation}
\label{xietaz}
\xi_0 = (S_d G \beta m\rho_{0})^{1/2} R_*,\qquad  \eta_0 = \beta G M_*
m/R_*^{d-2}.
\end{equation}
In the absence of a central body, the boundary conditions
are  replaced by $\psi(0)=\psi'(0)=0$~\cite{emden,chandrab}. The Emden equation
(\ref{eq.equationemdencc}) can also be written in the form of an
integrodifferential equation [see Eq. (\ref{gros})]
\be
\label{eq.equationintegrodiffpsicc}
\xi^{d-1}\frac{d \psi}{d \xi} = \eta_0 \xi_0^{d-2}
+ \int_{\xi_0}^\xi  e^{-\psi(\xi')} {\xi'}^{d-1}\,d\xi'.
\ee

\begin{figure}
\begin{center}
(a)
\includegraphics[width=0.9\linewidth,angle=0,clip]{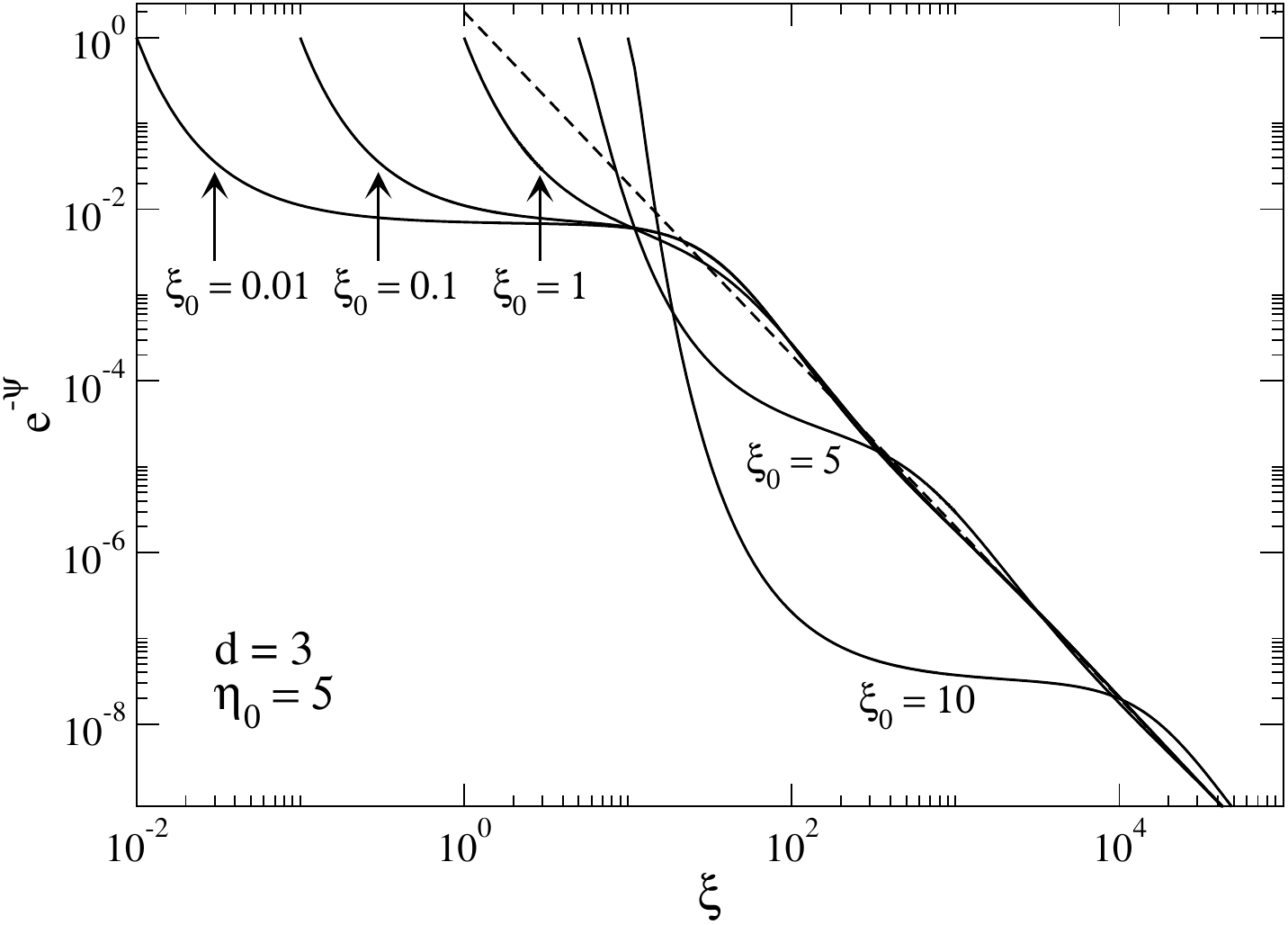}
\hspace{6pt}
(b)
\includegraphics[width=0.9\linewidth,angle=0,clip]{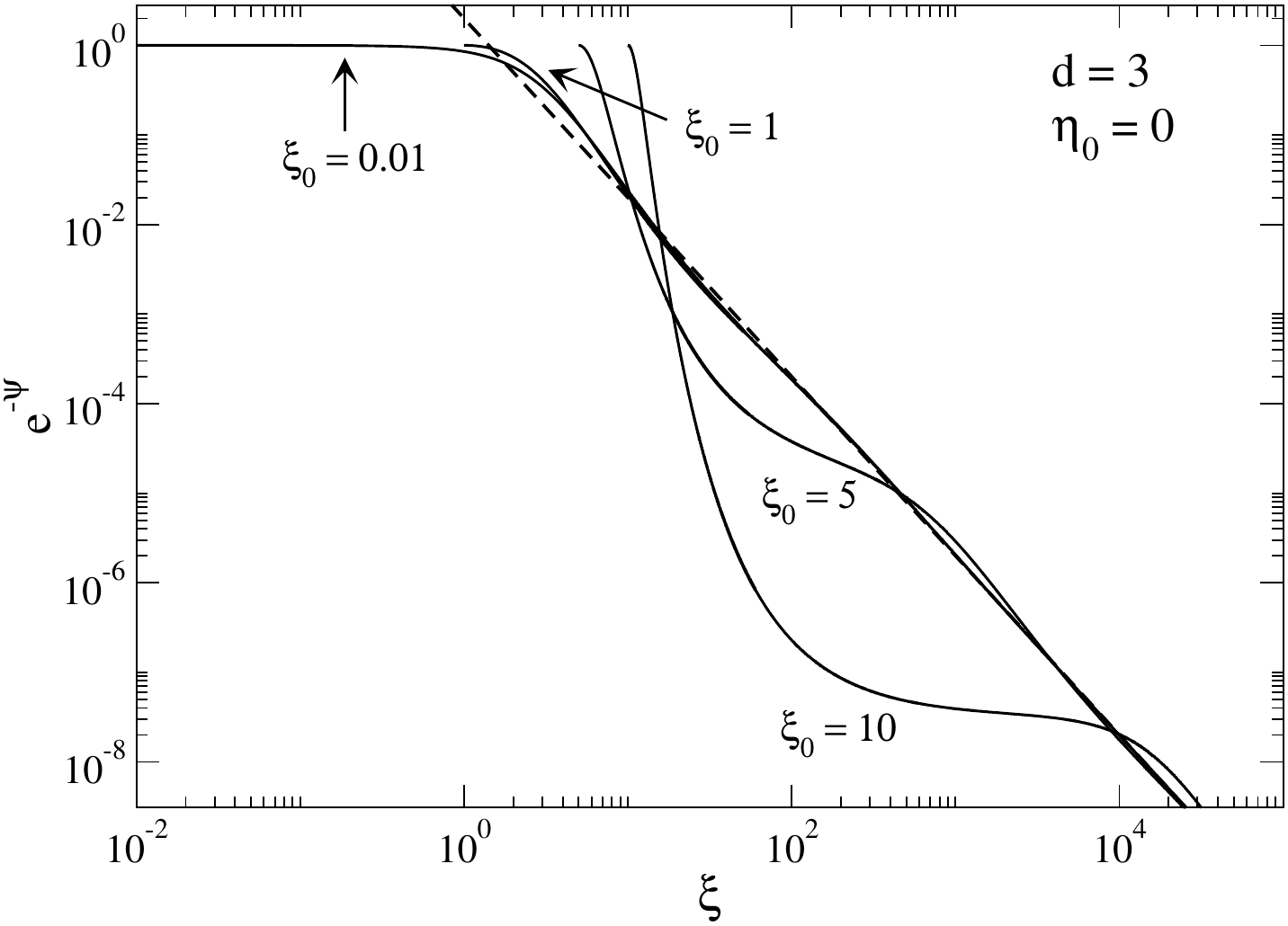}
\end{center}
\caption{\label{fig.profilrhouvcc}Dimensionless density profile $e^{-\psi}$ that
is the solution of the Emden equation with a central body in $d=3$ dimensions.
We have
selected  $\eta_0=5$ and $\eta_0=0$ (hole) for illustration.}
\end{figure}

The Emden equation (\ref{eq.equationemdencc}) must be solved
numerically (see Fig.~\ref{fig.profilrhouvcc}). We can however 
easily determine the asymptotic behaviors
of the solutions. For $\xi\rightarrow \xi_{0}$, we can expand the
solution in Taylor series and we find that
\bea
\label{eq.psixi0}
\psi(\xi) &=& \frac{\eta_0}{\xi_0} \left(\xi-\xi_0\right)
+
\frac{1}{2}\left \lbrack1-\frac{(d-1)\eta_0}{\xi_0^2}\right\rbrack
\left(\xi-\xi_0\right)^2
\nonumber \\
&&
+
\frac{1}{6}\left\lbrack\frac{1-d-\eta_0}{\xi_0}+\frac{d(d-1)\eta_0}{\xi_0^3}\right\rbrack
\left(\xi-\xi_0\right)^3
\nonumber \\
&&
-
\frac{1}{24}\biggl \lbrack 1+\frac{1-d^2-2(d-1)\eta_0-\eta_0^2}{\xi_0^2}
\nonumber \\
&&
+
\frac{(d-2)(d-1)^2\eta_0}{\xi_0^4}\biggr\rbrack
\left(\xi-\xi_0\right)^4 +
{O}\left(\left(\xi-\xi_0\right)^5\right).\nonumber\\
\eea
Therefore, when $\eta_0\neq 0$ and $\xi_0\neq 0$, the density
profile of the gas close to the central body  behaves as
\be
\rho\sim\rho_0 e^{-\frac{\eta_0}{\xi_0}(\xi-\xi_0)}\qquad (\xi\rightarrow
\xi_0).
\ee
It presents a spike which is clearly visible in Fig.~\ref{fig.profilrhouvcc}-a.
By contrast, when $\eta_0=0$ and $\xi_0\neq 0$ (hole), the density
profile of the gas close to the central body  behaves as
\be
\rho\sim\rho_0 e^{-\frac{1}{2}(\xi-\xi_0)^2}\qquad (\xi\rightarrow
\xi_0).
\ee
In that case, it presents a core (see Fig.
\ref{fig.profilrhouvcc}-b). On the other hand, for $d>2$, the asymptotic
behavior of the
solution for $\xi\rightarrow +\infty$ is the same as for the
isothermal self-gravitating gas without a central body~\cite{sc}. The reason is
that an unbounded
isothermal self-gravitating gas carries out an infinite mass so 
the effect of the central body 
becomes negligible at sufficiently
large distances. As a result, the solution behaves as~\cite{sc}
\be
\label{eq.epsixiinf}
e^{-\psi} \sim \frac{2(d-2)}{\xi^2} \qquad {\rm for}\quad \xi\rightarrow
+\infty.
\ee
The particular dimensions $d=1$ and $d=2$ are treated specifically
in a companion paper~\cite{companion}.

\subsection{The fundamental equation of hydrostatic equilibrium}

We can obtain the spatial structure of a self-gravitating gas surrounding a
central
body in a different (but equivalent) manner by starting directly from the condition of
hydrostatic equilibrium (for $R_{*}<r<R$):
\be
\label{eqhydro}
\nabla P+\rho\nabla\Phi_{\rm tot}={\bf 0}.
\ee
Dividing
Eq.~(\ref{eqhydro}) by $\rho$, taking its divergence, and using the Poisson
equation
\be
\Delta\Phi_{\rm tot}=S_d G\rho,
\ee
we obtain 
\be
\nabla\cdot \left (\frac{\nabla P}{\rho}\right )=-S_d G\rho, 
\ee
which is the fundamental equation of hydrostatic equilibrium. 
For the isothermal equation of state (\ref{eq.equationetatcc}), it takes the
form
\be
\frac{k_BT}{m}\Delta\ln\rho=-S_{d} G  \rho.
\ee
For a spherically symmetric distribution, we get
\be
\label{fund2}
\frac{1}{r^{d-1}}\frac{d}{dr}\left ({r^{d-1}}\frac{d\ln\rho}{dr}\right
)=-S_{d}\beta G m \rho
\ee
with the boundary condition 
\be
\left (\frac{d\ln\rho}{dr}\right )(R_*)=-\beta m\frac{GM_*}{R_*^{d-1}}
\ee
obtained from Eqs.~(\ref{eq.equationetatcc}), (\ref{bco}) and (\ref{eqhydro}).
Writing
$\rho(r)=\rho_{0}e^{-\psi(\xi)}$ with the variables $\psi$ and $\xi$ defined
previously, we recover the Emden equation (\ref{eq.equationemdencc}) with the
boundary condition (\ref{bcemden}). The two
descriptions are of course equivalent since the condition of hydrostatic
equilibrium (\ref{eqhydro}) can be obtained by taking the logarithmic derivative
of Eq.
(\ref{eq.distributionboltzcc}) and using Eq.~(\ref{eq.equationetatcc}). More
generally, it is satisfied by any
distribution function that only depends on the individual energy of the
particles: $f({\bf r},{\bf v})=f(\epsilon)$ where $\epsilon=v^2/2+\Phi({\bf
r})$ (see Appendix \ref{sec_che}).

\subsection{The Milne variables}

\begin{figure}
\begin{center}
\includegraphics[width=0.9\linewidth,angle=0,clip]{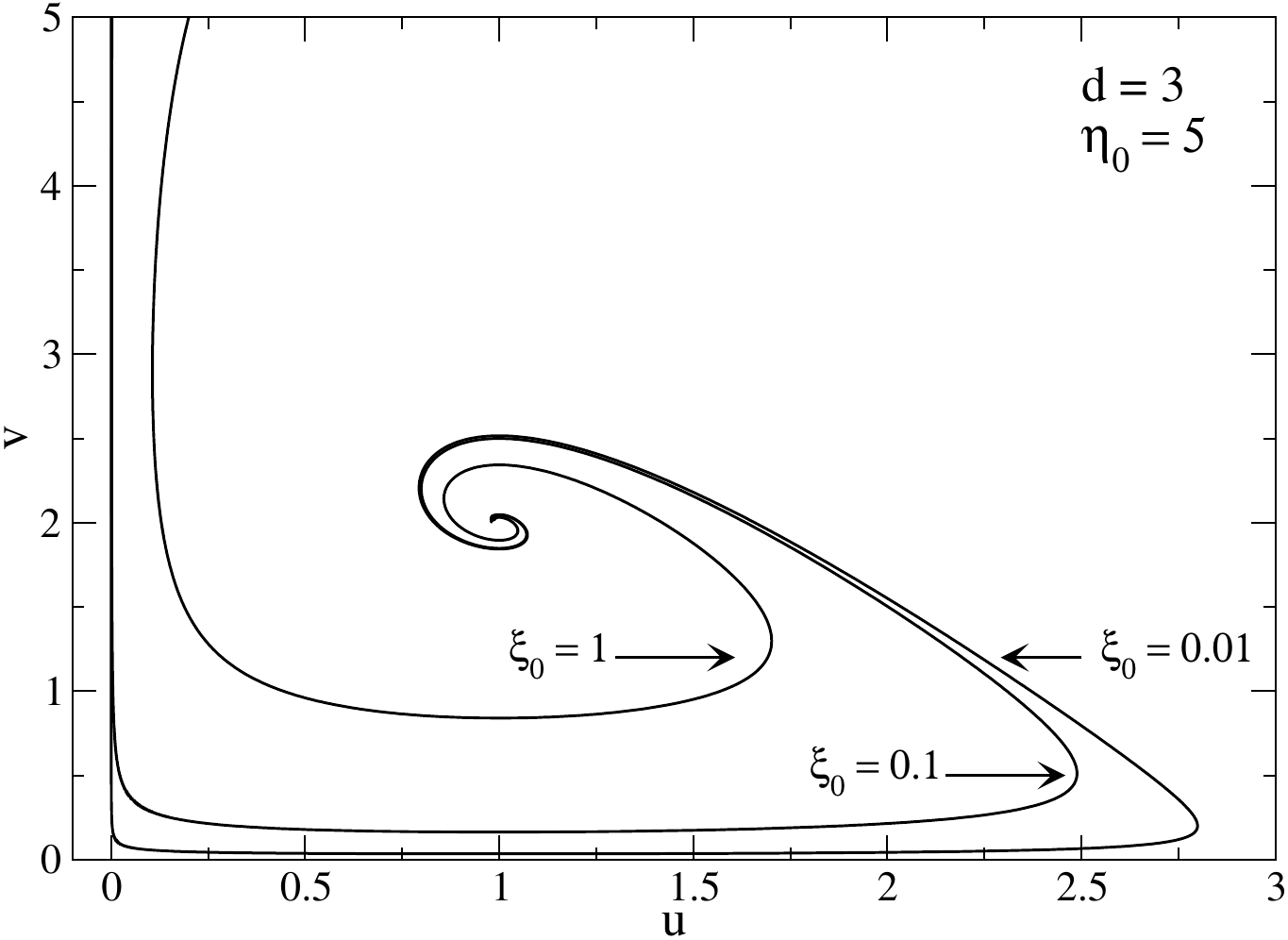}
\end{center}
\caption{\label{fig.uv} Solution of the Emden equation with a central body in
the $(u,v)$ plane in $d=3$ dimensions (we have
selected $\eta_0=5$ for illustration).}
\end{figure}

Let us introduce the analogue of the Milne variables
\bea
u = \frac{d\ln M_{\rm tot}(r)}{d \ln r},\qquad
v = - \frac{d\ln P(r)}{d \ln r},
\eea
where $M_{\rm tot}(r)=M(r)+M_*$ is the total mass enclosed within the
sphere of radius $r$ and $P(r) = \rho(r) k_B T/m$ is the local
pressure. Using the equalities $GM_{\rm tot}(r) = r^{d-1}d\Phi_{\rm tot}/dr$ and
$dM_{\rm tot}/dr = S_d r^{d-1}\rho(r)$ (see
Appendix \ref{sec_pfd}), we easily find that
\bea
u = \frac{\xi e^{-\psi}}{\psi'},\qquad
v = \xi \psi'.
\eea
Therefore, the Milne variables keep the same form as in the absence of
central body~\cite{chandrab}. They satisfy the first order differential
equation
\bea
\frac{u}{v}\frac{dv}{du}=\frac{2-d+u}{d-u-v}.
\eea
The $(u,v)$ curves are parameterized by $\xi$. They start from
$(\xi_0^2/\eta_0,\eta_0)$ for $\xi=\xi_0$ and end at $(d-2,2)$ for $\xi\rightarrow +\infty$ (if $d>2$). An example of $(u,v)$ curve is shown in Fig.~\ref{fig.uv}.

\subsection{The thermodynamical parameters}
\label{ssec.prametresthermo}

For $d>2$, there is no global entropy maximum at fixed mass and energy in an
unbounded
domain. The isothermal gas surrounding the central body has the
tendency to evaporate, leading to higher and higher values of
entropy as the volume that it occupies increases. 
There are not even extrema of entropy at fixed mass and energy in an
unbounded domain
because the solutions of the
Boltzmann-Poisson equation (\ref{eq.equationboltzpoiss}) have infinite mass. As in the case
without central body, we shall confine the gas within a box of radius
$R$. The box delimitates the region where the system is isolated from
the surrounding so that thermodynamical arguments can be applied. It
can mimic the effect of a tidal radius for  globular
clusters and dark matter halos, or represent the Hill sphere in the context of
planet formation. In the
context of the chemotaxis of bacterial populations,
the box has a physical justification as it represents the domain containing
the bacteria.

In a bounded domain, the solutions of the Emden equation 
(\ref{eq.equationemdencc}) are terminated by the box at a
normalized radius given by 
\be
\label{alpha}
\alpha = (S_d G \beta m
\rho_{0})^{1/2}R.
\ee
Using Eqs.~(\ref{muzeta}) and (\ref{alpha}), we can rewrite Eq.~(\ref{xietaz})
as
\be
\label{sn}
\xi_0 = \alpha \zeta,\qquad  \eta_0
= \frac{\eta \mu}{\zeta^{d-2}}. 
\ee
We shall now relate $\alpha$ to the dimensionless temperature $\eta$ and
to the dimensionless
energy $\Lambda$.

\subsubsection{The temperature}

According to the Newton law (\ref{newl})  applied at $r=R$, we
have
\be
\frac{d \Phi_{\rm tot}}{dr}(R)= \frac{G(M_{*}+M)}{R^{d-1}},
\ee
where $M$ is the total mass of the gas. Introducing the variables defined
previously and using
$\xi=r\alpha/R$, the foregoing relation can be rewritten as
\be
\label{eta}
\eta=\frac{\alpha\psi'(\alpha)}{1+\mu}.
\ee
This relation can also be obtained 
by substituting the Boltzmann distribution (\ref{eq.definitionrhocc}) into
the
mass constraint (\ref{mass}) and by using the Emden equation
(\ref{eq.equationemdencc}).

\subsubsection{The energy}
\label{sec_energy}

The computation of the energy is a little more intricate.  
Using the Maxwell-Boltzmann
distribution (\ref{eq.distributionmaxboltzcc}), or the isothermal equation of
state (\ref{eq.equationetatcc}), the kinetic energy 
\be
\label{ny}
E_{\rm kin}=\int f \frac{v^{2}}{2}\, d{\bf r}d{\bf v}=\frac{d}{2}\int P\, d{\bf
r}
\ee
is given by
\be
\label{eq.energiecincc}
E_{\rm kin}=\frac{d}{2}Nk_{B}T,
\ee
just like for a noninteracting (perfect) gas.
Therefore, the
dimensionless kinetic energy is
\be
\label{eq.energiecinadimcc}
-\frac{E_{\rm kin} R^{d-2}}{G M^2} = -\frac{d}{2 \eta}.
\ee
On the other hand, the potential energy $W_{\rm tot}=W+W_{\rm ext}$ is given by
\bea
\label{nrj}
W_{\rm tot} = \frac{1}{2}\int \rho \Phi\,d{\bf r} + \int
\rho \Phi_{\rm ext} \,d{\bf r}.
\eea
Using Eq.~(\ref{ptot}) it can be rewritten as
\bea
\label{eq.energiepotintcc}
W_{\rm tot} = \frac{1}{2}\int \rho \Phi_{\rm tot} \,d{\bf r} + \frac{1}{2}\int
\rho \Phi_{\rm ext} \,d{\bf r}.
\eea
We shall compute the potential energy in
two different manners, either by  using the virial theorem or
by directly evaluating the integral
(\ref{eq.energiepotintcc}). The first
approach based on the virial theorem is only valid for $d \neq 2$ while the
second approach is general.

For $d \neq 2$, it is shown in
Appendix \ref{sec_vt} that the virial theorem for a self-gravitating gas in
hydrostatic
equilibrium surrounding a central body can be written as
\be
\label{eq.equationetatisotherme}
2E_{\rm kin}+(d-2)W_{\rm tot} = d P(R)V-d P(R_{*})V_{*},
\ee
where $V = S_d R^d/d$ is the volume of the box enclosing the gas and
$V_* = S_d R_*^d/d$ is the volume of the central body. This relation is
valid for a general equation of state. For the isothermal equation of
state (\ref{eq.equationetatcc}), we obtain
\be
2E_{\rm kin}+(d-2)W_{\rm tot} = S_d \left (\rho(R) R^d- \rho_0 R_*^d\right
)\frac{k_B T}{m},
\ee
where the first term (kinetic energy) is given by Eq.~(\ref{eq.energiecincc}).
This equation then determines the potential energy $W_{\rm tot}$. Introducing
the dimensionless
variables defined previously, we get
\be
\label{eq.virieladimcc}
-\frac{W_{\rm tot}R^{d-2}}{GM^{2}} = \frac{\alpha^2}{(d-2)\eta^2}
(\zeta^d-e^{-\psi(\alpha)} )+\frac{d}{d-2}\frac{1}{\eta}.
\ee
The total energy is $E=E_{\rm kin}+W_{\rm tot}$.  Adding Eqs.
(\ref{eq.energiecinadimcc}) and (\ref{eq.virieladimcc}), we find that the total
dimensionless energy is given by
\be
\label{eq.energieadimensionneecc}
\Lambda = \frac{4-d}{d-2}\frac{d}{2 \eta}
+ \frac{\alpha^2}{(d-2)\eta^2}(\zeta^d-e^{-\psi(\alpha)}) \quad (d\neq 2).
\ee

It can be useful to obtain another expression of the energy. 
Introducing the dimensionless variables defined previously  in the
expression of the potential energy (\ref{eq.energiepotintcc}), and
using Eq.~(\ref{pcore1}) for $d\neq 2$, we find that the dimensionless
potential energy can be written as
\bea
\label{eq.energiepotadimcc}
-\frac{W_{\rm tot}R^{d-2}}{GM^2} &=& -\frac{1}{2\eta^2\alpha^{d-2}}
\int_{\alpha \zeta}^{\alpha}(\psi+\psi_*) e^{-\psi} \xi^{d-1} \,d\xi
\nonumber\\
&&
+
\frac{\mu}{2(d-2)\eta}\int_{\alpha \zeta}^{\alpha}e^{-\psi} \xi \,d\xi.
\eea
The quantity $\psi_*
\equiv \beta m\Phi_{\rm tot}(R_*)$ is obtained by evaluating  $\psi(\xi)=\beta
m\Phi_{\rm tot}(r)-\psi_{*}$ at $\xi=\alpha$, using $\Phi_{\rm
tot}(R)=-[1/(d-2)]GM_{\rm tot}/R^{d-2}$ (see Appendix \ref{sec_pfd}) for $d\neq
2$. This yields
\be
\psi_* = -\frac{\eta(1+\mu)}{d-2}-\psi(\alpha).
\ee
Adding Eqs.~(\ref{eq.energiecinadimcc}) and
(\ref{eq.energiepotadimcc}), we find that the total dimensionless
energy is given by
\bea
\label{eq.energieadimintcc}
\Lambda &=& -\frac{d}{2\eta}-\frac{1}{2\eta^2\alpha^{d-2}}
\int_{\alpha \zeta}^{\alpha}(\psi+\psi_*) e^{-\psi} \xi^{d-1} \,d\xi\nonumber\\
&&
+ \frac{\mu}{2(d-2)\eta}\int_{\alpha \zeta}^{\alpha}e^{-\psi} \xi \,d\xi\qquad
(d\neq 2).
\eea
This  expression  is equivalent to Eq.~(\ref{eq.energieadimensionneecc}), but it
is more 
complicated since it involves new integrals. However, the present method can be
generalized in $d=2$ dimensions.

Using Eq.~(\ref{pcore2}) for $d=2$, the dimensionless potential
energy (\ref{eq.energiepotintcc}) can be written as
\bea
\label{eq.energiepotadim2dcc}
-\frac{W_{\rm tot}}{GM^2} &=& -\frac{1}{2\eta^2}
\int_{\alpha \zeta}^{\alpha}(\psi+\psi_*) e^{-\psi} \xi \,d\xi
\nonumber\\
&&
- \frac{\mu}{2\eta}\int_{\alpha \zeta}^{\alpha}e^{-\psi}  \ln(\xi/\alpha)\xi \,d\xi.
\eea
The quantity $\psi_* \equiv \beta m\Phi_{\rm tot}(R_*)$ is obtained by
evaluating
$\psi(\xi)=\beta m\Phi_{\rm tot}(r)-\psi_{*}$ at $\xi=\alpha$, using 
$\Phi_{\rm tot}(R)=0$ (see Appendix \ref{sec_pfd}). This yields
\be
\psi_* = -\psi(\alpha).
\ee
Adding Eqs.~(\ref{eq.energiecinadimcc}) and
(\ref{eq.energiepotadim2dcc}), we find that the total dimensionless
energy is given by
\bea
\Lambda &=& -\frac{1}{\eta}-\frac{1}{2\eta^2}
\int_{\alpha \zeta}^{\alpha}(\psi+\psi_{*}) e^{-\psi} \xi \,d\xi
+\frac{1}{2}\mu \ln\alpha\nonumber\\
&&
- \frac{\mu}{2\eta}\int_{\alpha \zeta}^{\alpha}e^{-\psi} \xi \ln\xi
\,d\xi\qquad (d= 2).
\eea
The virial theorem in $d=2$ is discussed in~\cite{companion}.

\subsection{Entropy and free energy}

The entropy and the free energy (or Massieu function) of the equilibrium
state can be calculated as follows. From Eqs. (\ref{eq.entropiemoycc}) and
(\ref{eq.distributionmaxboltzcc}) we get
\be
\label{ef1}
S_B/k_B=\alpha_0 M+\left (\frac{d}{2}+1\right )N+2\beta W+\beta W_{\rm ext},  
\ee
where we have used Eqs. (\ref{ny}) and (\ref{eq.energiecincc}). Applying
Eq. (\ref{eq.distributionboltzcc}) at $r=R$ and using Eqs. (\ref{fr}),
(\ref{eq.definitionrhocc}), (\ref{alpha}) and
$\Phi(R)+\Phi_{\rm ext}(R)=-G(M+M_*)/[(d-2)R^{d-2}]$ for $d\neq 2$ or
$\Phi(R)+\Phi_{\rm ext}(R)=0$ for $d=2$
[see Eqs. (\ref{pcore1}), (\ref{pcore2}), (\ref{nu1}) and (\ref{nu2})], we
find that
\bea
\label{ef2}
\alpha_0 m=-2\ln\alpha-\left (\frac{d}{2}-1\right
)\ln\eta+\psi(\alpha)+\frac{\eta}{d-2}(1+\mu)\nonumber\\
+\ln\left\lbrack
(2\pi)^{d/2}S_d G^{d/2}M^{(d-2)/2}mR^{(4-d)d/2}\right\rbrack-1.\qquad 
\eea
We also have
\be
\label{ef3}
E=\frac{d}{2}Nk_B T+W_{\rm tot}
\ee
with $W_{\rm tot}=W+W_{\rm ext}$. From Eqs. (\ref{fr}) and (\ref{ef3}) we
obtain
\be
\label{ef4}
2\beta W+\beta W_{\rm ext}=-2N\Lambda\eta-dN-\beta W_{\rm ext}.
\ee
Substituting Eqs. (\ref{ef2}) and (\ref{ef4}) into Eq. (\ref{ef1}) we find that
\bea
\label{ef5}
S_B/Nk_B=-\left (\frac{d}{2}-1\right
)\ln\eta-2\ln\alpha+\psi(\alpha)\nonumber\\
+\frac{1}{d-2}(1+\mu)\eta-2\Lambda\eta-\frac{d}{2}-\frac{\beta W_{\rm
ext}}{N}\nonumber\\
+\ln\left\lbrack
(2\pi)^{d/2}S_d G^{d/2}M^{(d-2)/2}mR^{(4-d)d/2}\right\rbrack.
\eea
The term $\beta W_{\rm ext}/N$ can be obtained from Eqs.
(\ref{muzeta}), (\ref{eq.definitionrhocc}),
(\ref{voir}), (\ref{sn}), (\ref{pcore1}) and (\ref{pcore2}) yielding
\be
\label{ef6}
\frac{\beta W_{\rm
ext}}{N}=-\frac{1}{d-2}\mu\int_{\alpha\zeta}^{\alpha}e^{-\psi}\xi\,
d\xi\qquad (d\neq 2),
\ee
\be
\label{ef7}
\frac{\beta W_{\rm
ext}}{N}=\mu\int_{\alpha\zeta}^{\alpha}e^{-\psi}\ln
({\xi}/{\alpha})\xi\, d\xi\qquad (d=2).
\ee
The Massieu function (\ref{massieu}) is then given by
\be
\label{ef8}
\frac{J_{B}}{N}=\frac{S_B}{Nk_B}+\Lambda\eta.
\ee

\subsection{The structure of the gas profile close to the central body}
\label{sec_st}

In this section, we  determine the profile of the gas in the vicinity of the
central body.
The density profile of the isothermal gas is
given by the Boltzmann distribution
\be
\rho(r)=A e^{-\beta m \Phi_{\rm tot}(r)},
\ee
where $\Phi_{\rm tot}=\Phi+\Phi_{\rm ext}$ is the total gravitational potential.
For
$r\rightarrow R_{*}$, the gravitational potential is dominated by the
contribution of the central body $\Phi_{\rm ext}(r)$ so
that\footnote{A better approximation may consist in replacing
$M_*$ by $M(r)$ in the following formulae.}
\be
\Phi_{\rm tot}({r})\simeq -\frac{1}{d-2}\frac{GM_{*}}{r^{d-2}} \qquad (d\neq 2),
\ee
\be
\Phi_{\rm tot}(r)\simeq GM_{*}\ln (r/R) \qquad (d=2).
\ee

For $d=1$, we find that the density profile close to the central body
increases exponentially rapidly as
\be
\rho(x)=\rho_{0}e^{-({x}-{R_{*}})/\epsilon}
\ee
on a typical length scale
\be
\epsilon=\frac{1}{\beta G m M_{*}}.
\ee
This corresponds to a ``cusp''  since
the first derivative $\rho'(R_*)=-\rho_0/\epsilon$ of the density profile is
nonzero at $x=R_*$. This solution remains valid if the central body is a Dirac
mass  ($R_{*}=0$) in which case
$\rho=\rho_0 e^{-x/\epsilon}$ for $x\rightarrow
0$~\cite{companion}.

For $d=2$, we find that the density profile close to
the central body increases like a power law: 
\be
\rho(r)=\rho_{0}\left (\frac{R_{*}}{r}\right )^{\beta GM_{*}m}.
\ee
Again, we have a cusp since $\rho'(R_*)=-\beta GM_*m\rho_0/R_*\neq 0$.
This power-law behavior remains valid if the central body is a Dirac
mass ($R_{*}=0$) in which case $\rho(r)=Ar^{-\beta GM_* m}$.
The density diverges at $r=0$ and the profile is
normalizable provided that $\beta GM_{*}m<2$~\cite{companion}.

For $d=3$, we find that the density profile close to the central body is
\be
\label{eq.densitetoymodel}
\rho(r)=\rho_{0}e^{\beta GM_{*}m \left (\frac{1}{r}-\frac{1}{R_{*}}\right )}.
\ee
We note that the density profile is not normalizable in the case where
the central body is a Dirac mass ($R_{*}=0$).  For $r\rightarrow R_{*}$, we find
that the density
profile increases exponentially rapidly as
\be
\rho(r)=\rho_{0}e^{-(r-R_{*})/\epsilon}
\ee
on a typical length scale
\be
\epsilon=\frac{R_{*}^{2}}{\beta GM_{*}m}.
\ee
Therefore, at the contact with the central body, there is a density
{\it spike} (or cusp) of the gas on a
typical length
$\epsilon$ since $\rho'(R_*)=-\rho_0/\epsilon\neq 0$.  Then, taking
$r\rightarrow +\infty$ in Eq.~(\ref{eq.densitetoymodel}), we find that
the spike is followed by a {\it plateau} where the density is nearly constant:
\be
\rho=\rho_{0}e^{-\frac{\beta GM_{*}m}{R_{*}}}.
\ee
This plateau extends on a typical length $L$ such that the mass of gas
contained in this region becomes comparable to the mass of the central
body. This corresponds to the condition
\be
\rho_{0}e^{-\frac{\beta GM_{*}m}{R_{*}}}\frac{4}{3}\pi (L^{3}-R_{*}^{3})=M_{*}.
\ee
If $R_{*}\ll L$, we get the estimate
\be
L\simeq \left (\frac{3M_{*}}{4\pi \rho_{0}}\right )^{1/3}e^{\frac{\beta
GM_{*}m}{3R_{*}}}.
\ee
For $r>L$, the self-gravity of the gas must be taken into account and, at large distances,
the density decreases with a typical $r^{-2}$ behavior corresponding to the
standard self-gravitating
isothermal sphere~\cite{chandrab}. This ``cusp $+$ plateau $+$
halo'' structure
(or simply
``cusp-halo'' structure) is clearly visible in Fig.~\ref{fig.profilrhouvcc}. 
These results are similar to those found in the 
case of self-gravitating fermions~\cite{mnras,ptfermi} where the role of
the central body is played by the completely degenerate fermion ball (quantum
core).

From these considerations, we can distinguish four types of
configurations which are reminiscent of the morphology of certain objects
in planetology:

(i) If the mass of the central body is small (or if there is no
central body) and if the gas is not too dense we have a
non-self-gravitating homogeneous gas that could represent a diffuse nebula.

(ii) If the mass of the central body is large and if the gas is
not too  dense we have a non-self-gravitating gas experiencing only the
gravitational potential of the central body (solid planetary core). The
density
of the gas increases rapidly at the contact of the solid core and forms a spike.
A massive solid core surrounded
by a tiny atmosphere corresponds to the structure of telluric planets like the
Earth.

(iii) If the mass of the central body is small (or if there is
no central body) and if the gas is sufficiently dense we have a standard
isothermal self-gravitating gas (isothermal sphere) with a finite central
density. It could
represent a protoplanet.

(iv) If the mass of the central body is large 
and if the gas is sufficiently dense we have a self-gravitating gas
experiencing the gravitational potential of the central body (solid
planetary core). The density of
the gas increases rapidly at the contact of the solid core and forms a spike.
This thin layer is then followed by an envelope held by its self-gravity. This
corresponds to the structure of giant planets like Jupiter and Saturn.

Cases (i) and (iii) correspond to the ``gaseous phase'' where
the influence of the central object is weak. Cases (ii) and (iv) correspond to
the ``condensed phase'' where the influence of the central object is strong.
In that case, the density of the gas is enhanced close to the solid core and
forms a spike. The
gaseous phase and the condensed phase can themselves be divided in two
categories depending on whether the envelope is self-gravitating or not.

The case of globular clusters and dark matter halos, which are
self-gravitating, corresponds to points (iii) and (iv) depending whether they
contain a central black hole or not.

\subsection{The numerical procedure}

In order to plot the series of equilibria $\eta(\Lambda)$ for 
fixed external parameters $\zeta$ and $\mu$, we proceed as follows: (i)
We fix $\xi_0$ and  make a guess $\eta_0^{\rm guess}$ for $\eta_0$; (ii) we
solve the Emden equation  (\ref{eq.equationemdencc}) with the boundary
condition from Eq. (\ref{bcemden}) until $\alpha=\xi_0/\zeta$.
This gives $\eta$ and $\Lambda$ according to the formulae of Sec.
\ref{ssec.prametresthermo} and  $\eta_0=\eta\mu/\zeta^{d-2}$. We iterate
this procedure until
the values of $\eta_0^{\rm guess}$ and $\eta_0$ coincide; (iii) in that case,
the chosen value of $\xi_0$ determines $\alpha$, $\eta$ and $\Lambda$. By
varying $\xi_0$, we can obtain the whole series of equilibria  $\eta(\Lambda)$
for given values of $\zeta$ and $\mu$.

\section{Caloric curves and phase transitions in the presence of a central body
in $d=3$
dimensions}
\label{sec_pt3}

In this section, we describe phase transitions which
take place in a self-gravitating system with a central body in $d=3$ dimensions.
We fix the typical density of the central body, $\rho_*=M_*/R_*^{3}$,
and plot the series of equilibria $\beta(E)$ for different
values of the radius $R_*$ of the central body.  We could also fix the mass 
$M_*$ of the central body instead of its average density, but we think that
fixing the average density is more relevant if we want to apply our results, for
example, to the problem of planet formation.\footnote{Rocky
cores have a given typical density $\rho_*=M_*/R_*^3$ determined by their
composition. By contrast, a central black hole has a constant mass to radius
ratio $M_*/R_*=c^2/2G$ determined by the Schwarzschild relation.}
Furthermore, the
classical isothermal gas without central body~\cite{antonov,lbw} is recovered
for
$R_*\rightarrow 0$ with fixed $\rho_*$ while the limit $R_*\rightarrow 0$ with
fixed $M_*$ corresponds to a central Dirac mass. This corresponds to a very
different situation. 

The dimensionless density of the central body is
\be
\kappa\equiv \frac{M_* R^3}{M R_*^3} = \frac{\mu}{\zeta^3}.
\ee
We shall work with the dimensionless variables defined previously.
However, it may be useful in the discussion to take $M=R=G=m=1$ (we can
always introduce appropriate scales to be in this situation). 
In that case, $\zeta=R_{*}$ represents the radius of the central body, 
$\kappa=\rho_{*}$ the density of the central body, $\eta=\beta$ the 
inverse temperature of the gas and $\Lambda=-E$ the energy of the gas (with the opposite sign). In the discussion, we shall use
the physical variables $R_{*}$, $\rho_{*}$, $\beta$ and $E$, and in the figures, we shall
use the dimensionless variables $\zeta$, $\kappa$, $\eta$ and $\Lambda$.

\begin{figure}
\begin{center}
\includegraphics[width=0.9\linewidth,angle=0,clip]{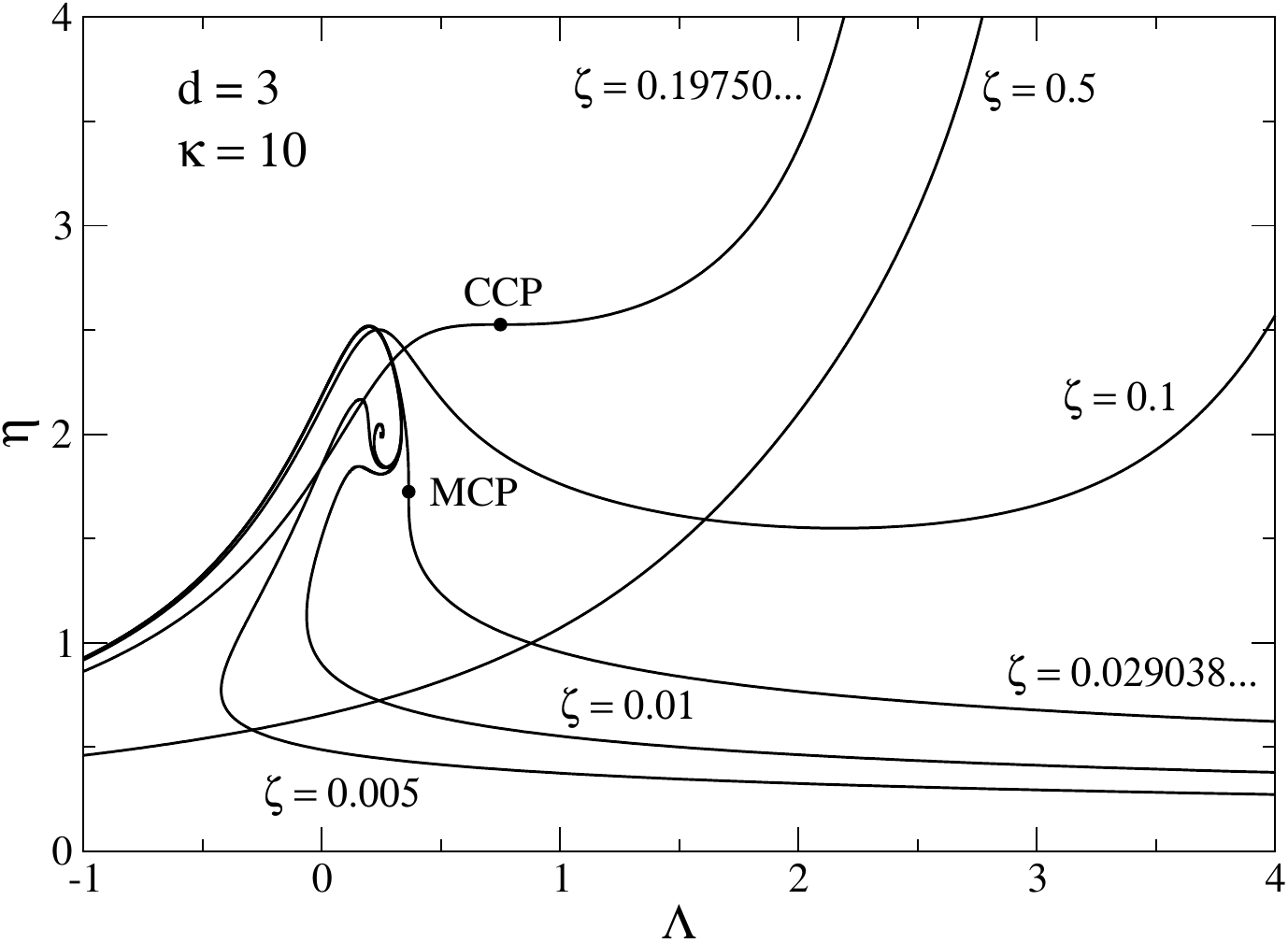}
\end{center}
\caption{\label{fig.deroulcompletcc1}Series of equilibria at
fixed central body density $\kappa$ (we have taken
$\kappa=10$ for
illustration) for several values of the central body radius
$\zeta$. As $\zeta$ increases, the series of
equilibria unwinds.
There exists a microcanonical critical point (MCP) and
a canonical critical point (CCP) above which the microcanonical
 and the canonical phase transitions respectively disappear. }
\end{figure}

\begin{figure}
\begin{center}
\includegraphics[width=0.9\linewidth,angle=0,clip]{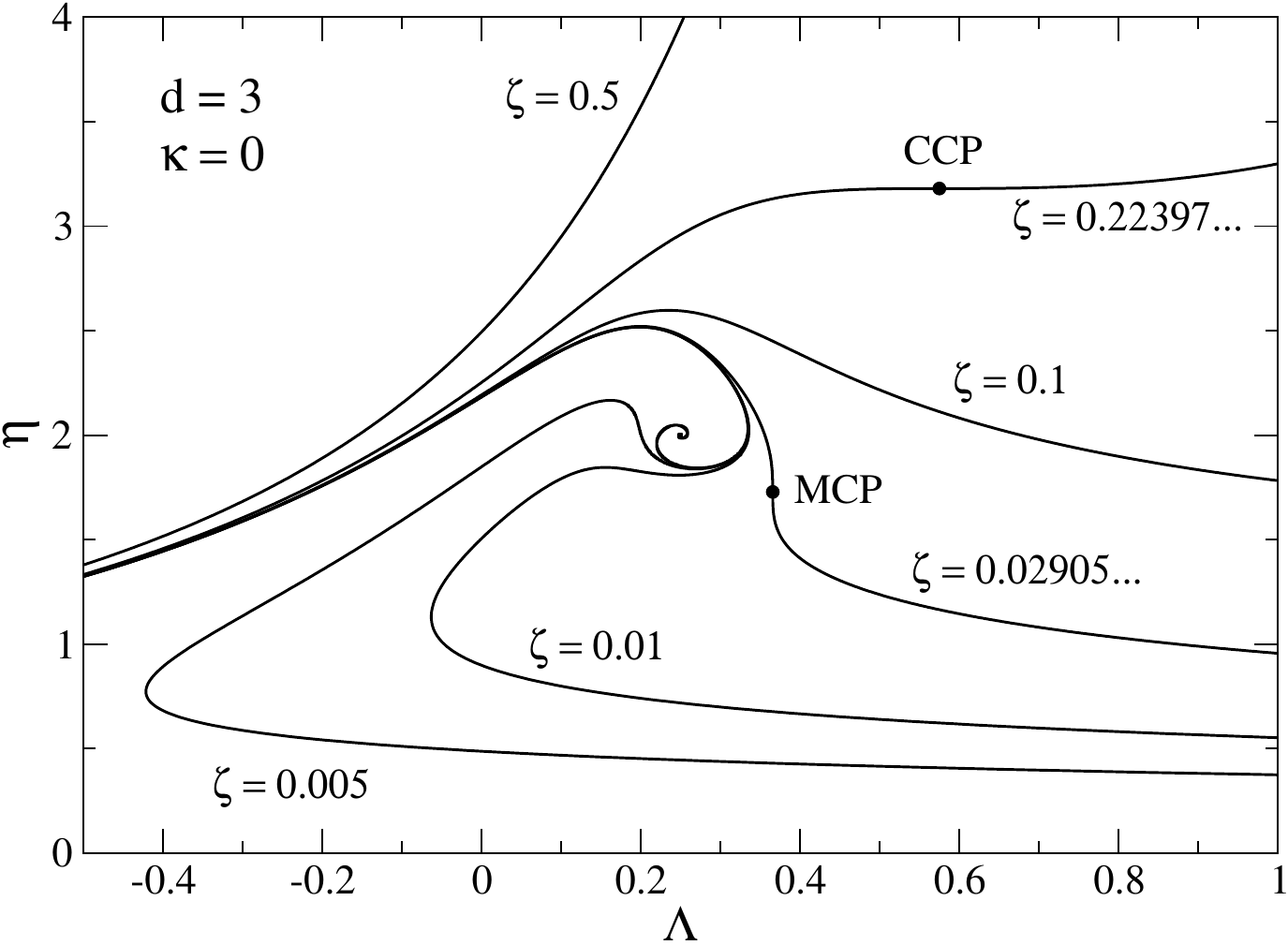}
\end{center}
\caption{\label{fig.deroulcompletcc2} Same as Fig.~\ref{fig.deroulcompletcc1} 
for $\kappa = 0$ and $\zeta\neq 0$ (hole).  This corresponds to a system of
self-gravitating particles contained in a volume delimited by two shells at
radii $r=R_*$ and $r=R$.}
\end{figure}

In Figs.~\ref{fig.deroulcompletcc1} and \ref{fig.deroulcompletcc2}, 
we plot the curve $\beta(E)$ for a fixed central body density
$\rho_{*}$ and for  different values
of the central body radius $R_{*}$.  This curve contains all the extrema of
entropy
(resp. free energy) at fixed mass and energy
(resp. temperature).  For that reason, it is called
the {\it series of equilibria}. The series of equilibria 
is parameterized by the density contrast ${\cal R}=\rho_0/\rho(R)=e^{\psi(\alpha)}$
between the central body and the box. The stable part of this curve in each
ensemble defines the {\it caloric curve}. The series of equilibria (extremal
states)
is the same in the canonical and microcanonical ensembles while the caloric
curve (stable
 states) can be different in case of ensemble inequivalence. For $R_{*}\rightarrow 0$, we
recover the classical spiral of a self-gravitating isothermal gas
without central body~\cite{paddy,dvs1,dvs2,katzrevue,ijmpb}.
For finite $R_{*}$, the spiral unwinds and
different shapes are possible depending on the value of $R_{*}$. For
$R_{*}<R_{*}^{\rm MCP}$, the curve has a $Z$-shape structure leading to
zeroth and first order microcanonical and canonical phase transitions. At the
microcanonical
critical point $R_{*}=R_{*}^{\rm MCP}$, the microcanonical phase
transitions disappear. For $R_{*}^{\rm MCP}<R_{*}<R_{*}^{\rm CCP}$, the curve
has an $N$-shape structure leading to zeroth and first order canonical phase
transitions. At
the canonical critical point $R_{*}=R_{*}^{\rm CCP}$, the canonical phase
transitions disappear. Finally, for $R_{*}>R_{*}^{\rm CCP}$, the curve is
monotonic so that there is no phase transition anymore. In that case,
the statistical ensembles are equivalent. 

We see that the description
of phase transitions in a self-gravitating classical gas with a
central body is very similar to that already given for
self-gravitating fermions~\cite{ijmpb}. The central body provides a small scale regularization 
of the gravitational force that prevents complete collapse and the formation of
singularities. In that sense, it has an effect similar to the Pauli exclusion
principle in quantum mechanics 
or to the core radius of the particles in a hard spheres gas. Since the
phenomenology is the same in all these situations, we shall be relatively brief
and refer to
the review~\cite{ijmpb} for more details about phase transitions in self-gravitating systems. We note, however, that in the
present case, the structure of the series equilibria $\beta(E)$
depends on two parameters $\rho_{*}$ and $R_{*}$ characterizing the
central body while in the case of self-gravitating fermions there is only one
parameter $\mu=(gm^4/h^3)\sqrt{512\pi^4G^3MR^3}$ related to the inverse of the
Planck constant ($\mu\sim
1/\hbar^3$) or to the ratio $\mu\sim (R/R_{*})^{3/2}$ between the system size
$R$ and the size $R_*$
of a fermion ball (a self-gravitating Fermi gas at $T=0$) with mass
$M$. Furthermore, in the present case, the central body is an
external object distinct from the gas while the fermion ball (or the solid
core) is created by the
system itself. What plays the role of the condensed object in the present
context
is the cusp at the contact with the central body (there is also a cusp in the
case of self-gravitating fermions \cite{mnras}).

\subsection{The case of a small central body $R_{*}<R_{*}^{\rm MCP}$ in the
microcanonical ensemble: $Z$-shape structure}
\label{sec_pt3a}
\begin{figure}
\begin{center}
\includegraphics[width=0.9\linewidth,angle=0,clip]{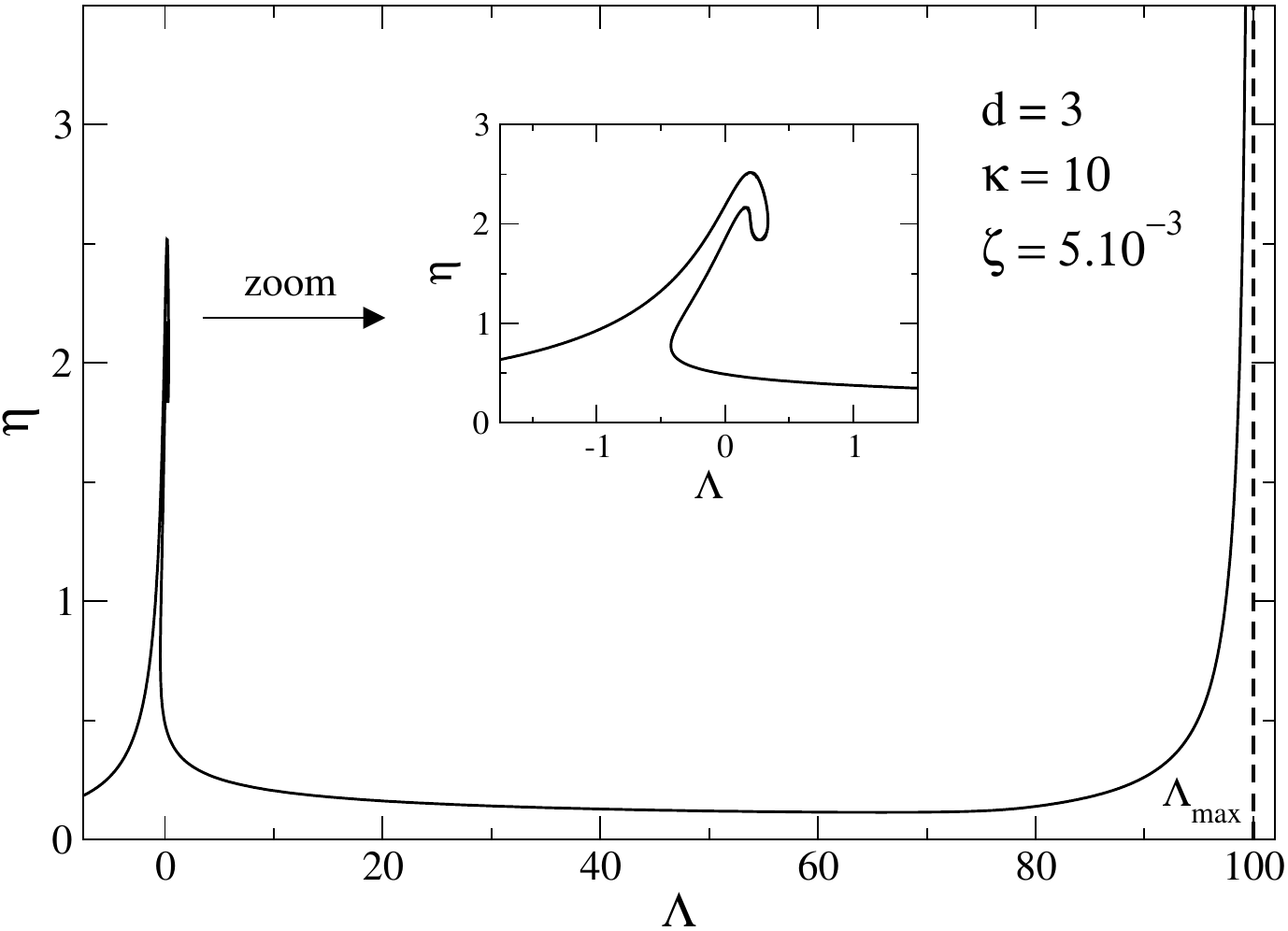}
\end{center}
\caption{\label{neck1}Series of equilibria for $\kappa=10$ and $\zeta =
5\times 10^{-3}<\zeta_{\rm MCP}$. It has a characteristic
$Z$-shape structure (``dinosaur's
neck''~\cite{ijmpb}). There exists an equilibrium state for all accessible
values
of the energy $\Lambda\le\Lambda_{\rm max}(\zeta,\kappa)$ in the microcanonical
ensemble, and for all
values of temperature $\eta\ge 0$ in the canonical ensemble. However,
interesting phase transitions, described in the text, appear for
energies around $\Lambda_{c}$ and temperatures around $\eta_c$ (see the
following figures).}
\end{figure}

\begin{figure}
\begin{center}
\includegraphics[width=0.9\linewidth,angle=0,clip]
{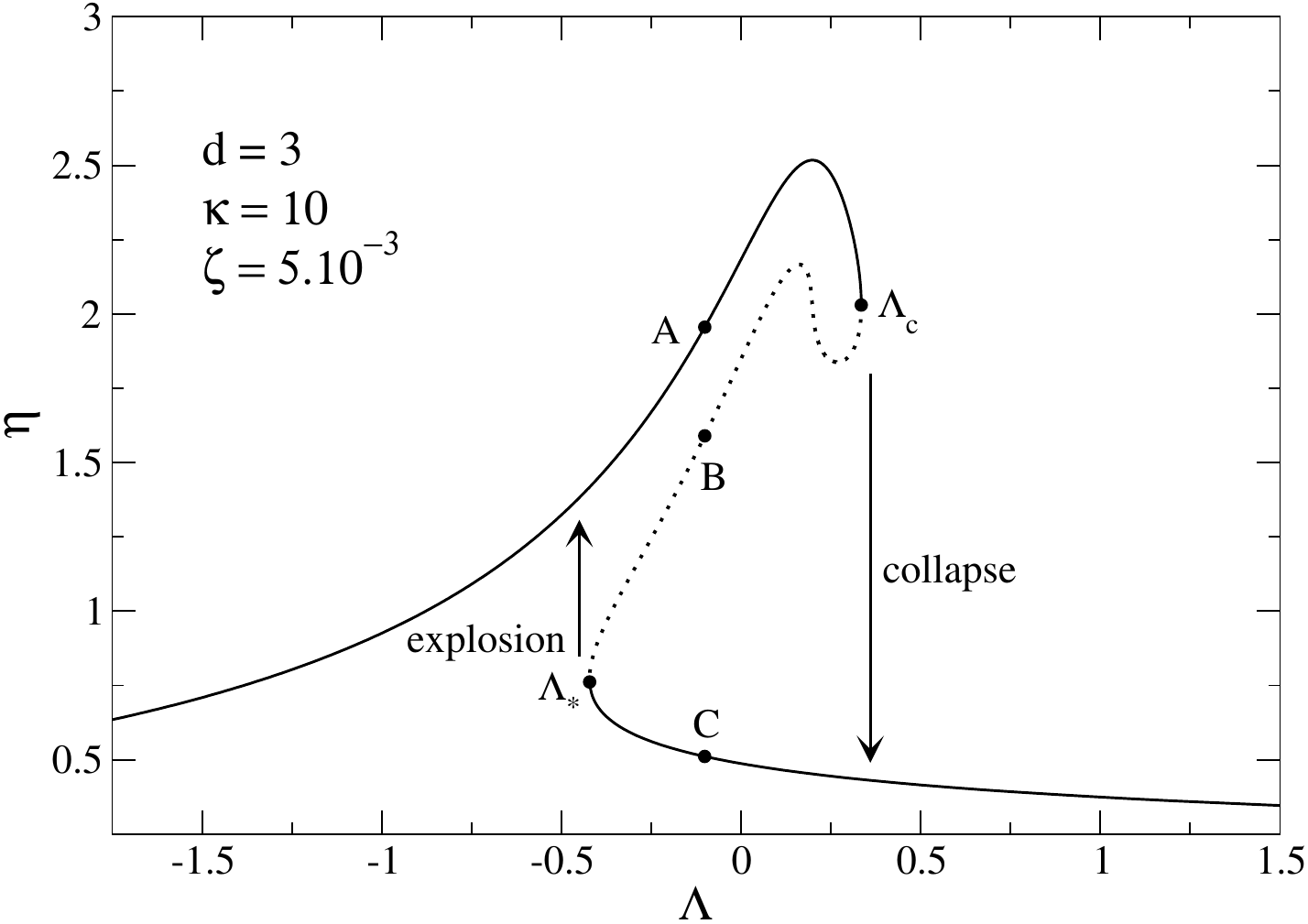}
\end{center}
\caption{\label{neck2} Physical caloric curve in the microcanonical ensemble
(solid lines) containing stable (global entropy maxima) and metastable (local
entropy maxima) equilibrium states. Unstable saddle points of entropy are
represented by dotted lines. }
\end{figure}

We first consider the case of a small central body $R_{*}<R_{*}^{\rm MCP}$
so that the series of equilibria has a $Z$-shape structure resembling
a ``dinosaur's neck''~\cite{ijmpb} (see Figs.~\ref{neck1} and \ref{neck2}).

In this section, we consider the case of an isolated system so that
the control parameter is the energy and the relevant statistical
ensemble is the microcanonical ensemble. As explained previously, the
series of equilibria contains all the extrema of entropy at
fixed mass and energy. The thermodynamically  stable states in the
microcanonical ensemble correspond to entropy {\it maxima} at fixed mass
and energy (minima or saddle points of entropy must be discarded). They
can be determined by  applying the turning point method of 
Poincar\'e (see, e.g.~\cite{katzrevue,ijmpb}). At very
high energies, self-gravity is negligible with respect to thermal
motion (velocity dispersion) and the system is equivalent to a
non-interacting gas in a box. We know from ordinary thermodynamics
that this gas is stable (entropy maximum). Therefore, using the Poincar\'e
criterion, we conclude that the upper branch of the series of
equilibria is stable (entropy maxima at fixed mass and energy) until
the first turning point of energy $E_{c}$ where the tangent is
vertical. For sufficiently small values of $R_{*}$, this is close to
the Antonov energy $-0.335$~\cite{antonov,paddy}. At that point, the curve $\beta(-E)$
rotates clockwise so that a mode of stability is
lost.\footnote{We note that the region of negative specific heat
just before the first turning point of energy is stable in the microcanonical
ensemble. The system
becomes unstable in the microcanonical ensemble after the first turning point
of
energy when the specific heat becomes positive. The regions of negative
specific heat and the situations of ensemble inequivalence will be discussed in
more detail in the following sections.} Therefore, the
configurations in the intermediate branch are unstable saddle points of
entropy at fixed mass and energy. However, at the second turning point
of energy $E_{*}$, the curve $\beta(-E)$ rotates anti-clockwise so
that the mode of stability is regained. Therefore, the lower branch
is stable (entropy maxima at fixed mass and energy). The energy
$E_{*}$ depends on $R_{*}$ and tends to $E_{*}(R_{*})\rightarrow
+\infty$ when $R_{*}\rightarrow 0$.

\begin{figure}
\begin{center}
\includegraphics[width=0.9\linewidth,angle=0,clip]{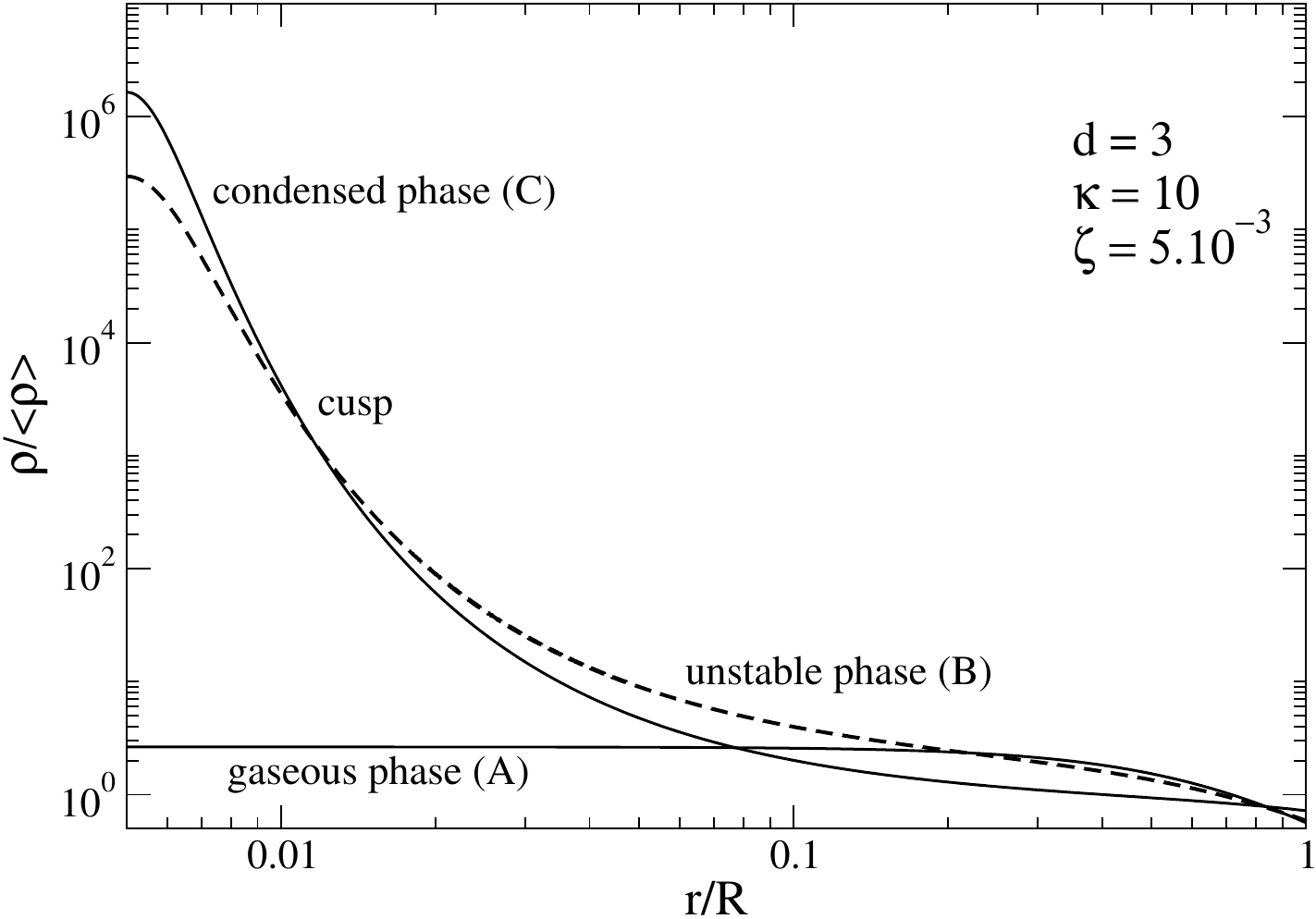}
\end{center}
\caption{\label{profile}Typical normalized density profiles
$\rho/\langle{\rho}\rangle$ with $\langle{\rho}\rangle=3M/4\pi R^3$
corresponding to gaseous (points $A$), condensed
(points $C$) and unstable (points $B$) states. The condensed and
unstable states have a cusp-halo structure.}
\end{figure}

Typical density profiles of the series of equilibria are shown in
Fig.~\ref{profile}. The states (A) on the upper branch have an almost
uniform density with a small density contrast ${\cal
R}=\rho(R_{*})/\rho(R)$. They form the {\it gaseous phase}. At large
distances, their density profile decreases approximately like
$r^{-2}$. The states (C) on the lower curve are highly inhomogeneous.
They present a high-density cusp at the contact with the central body
surrounded by a dilute atmosphere. Therefore the density contrast
${\cal R}=\rho(R_{*})/\rho(R)$ is large.  They form the {\it condensed
phase}. They have a ``cusp-halo'' structure which is the counterpart
of the ``core-halo'' structure of self-gravitating fermions (in that case 
the core is a degenerate quantum body). The cusp
contains a lot of potential energy. By forming the cusp,
some energy is released in the form of kinetic energy in the
halo which  is very hot and almost uniform. This
explains the plateau following the cusp (see Sec.~\ref{sec_st}). Finally, the
states (B) on
the intermediate (unstable) branch are similar to the gaseous states
(the density profile decreases approximately like $r^{-2}$ at large
distances) except that they form an embryonic cusp  at the contact with
the
core. This is the equivalent of a ``germ'' in the language of phase
transitions (see below).

\begin{figure}
\begin{center}
(a)
\includegraphics[width=0.9\linewidth,angle=0,clip]{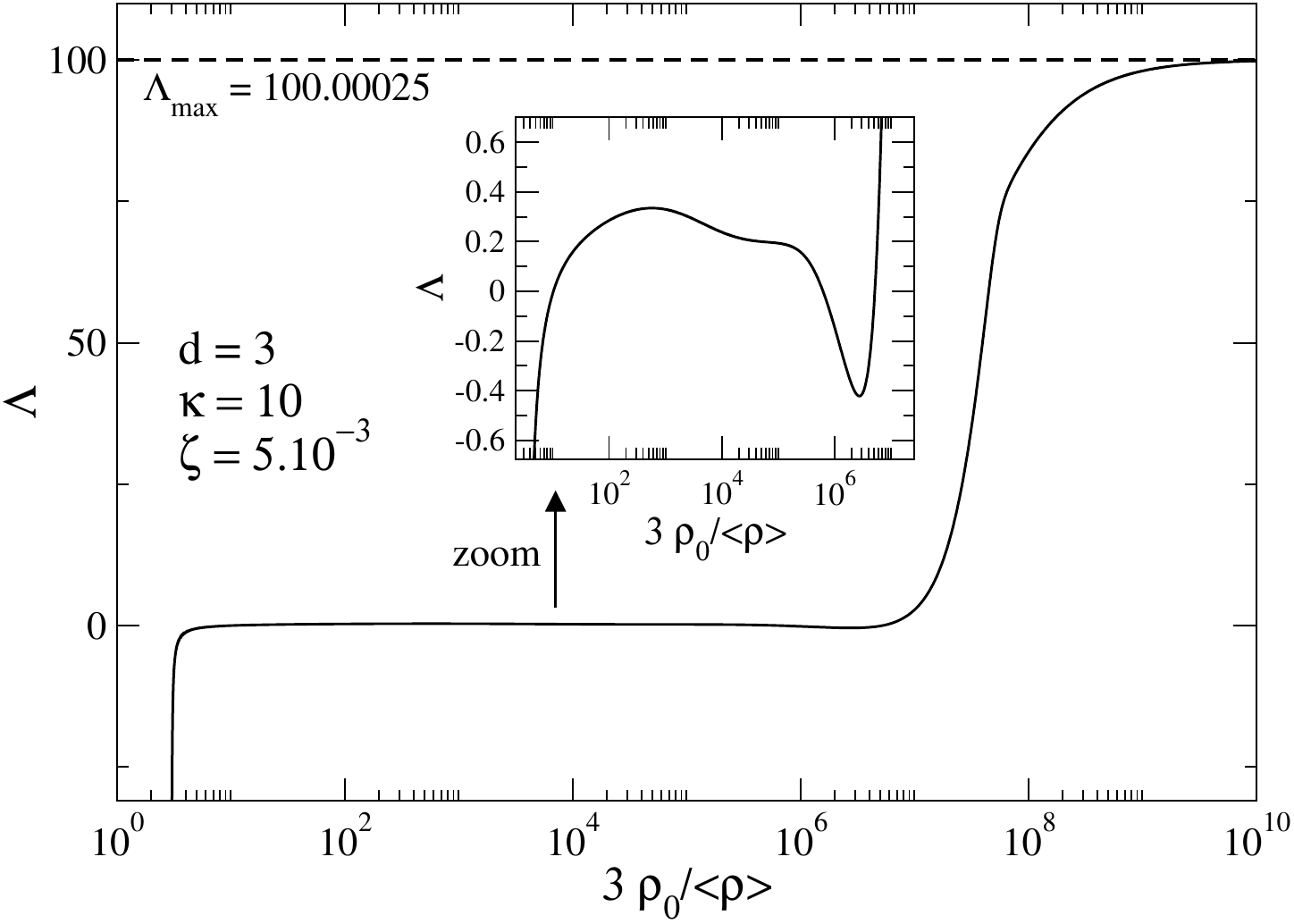}
\hspace{6pt}
(b) \includegraphics[width=0.9\linewidth,angle=0,clip]{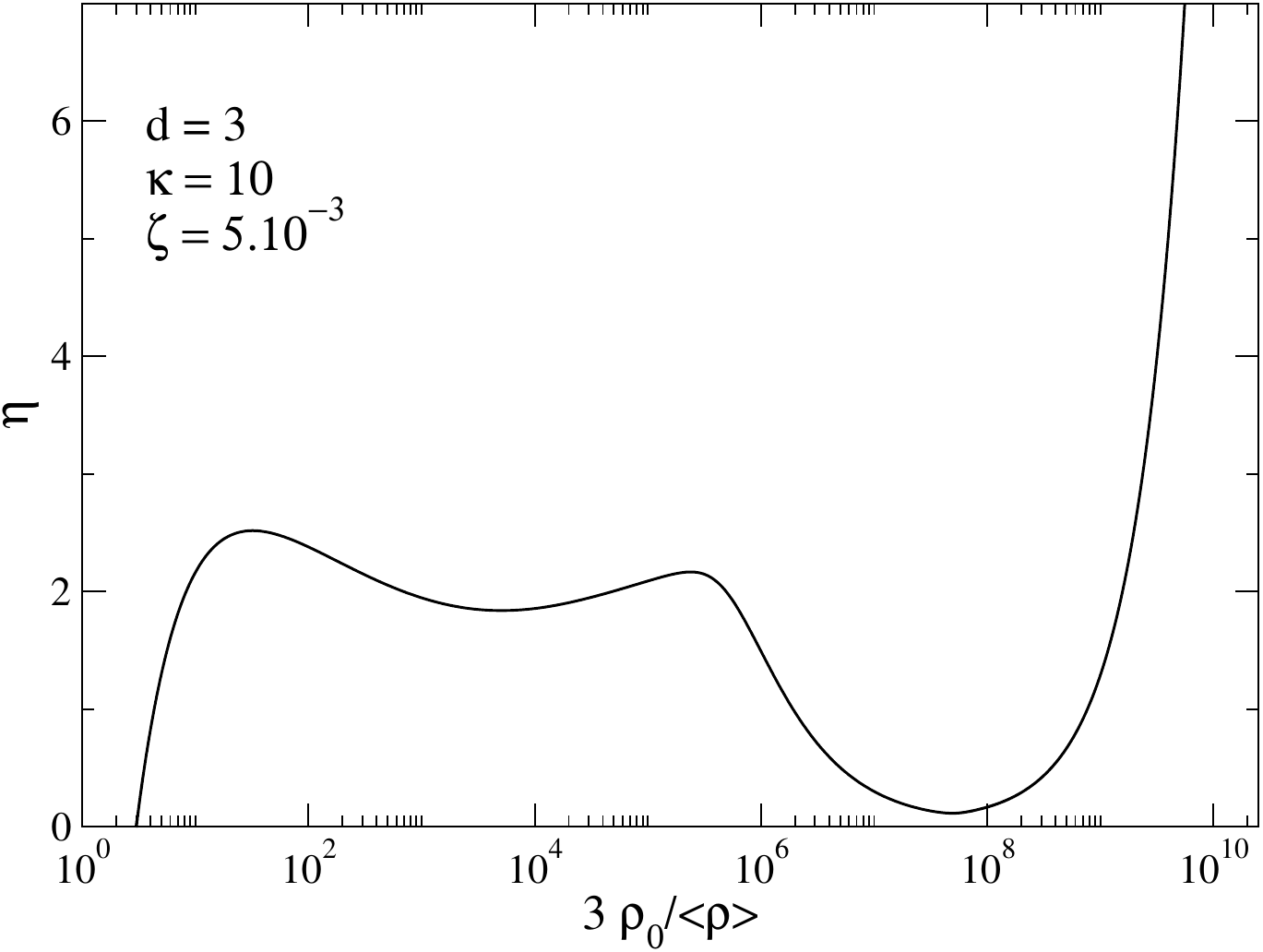}
\end{center}
\caption{\label{osc1}Evolution of the energy $\Lambda$ and the inverse
temperature $\eta$ as a function of the normalized
central density $\rho_0/\langle{\rho}\rangle$.}
\end{figure}

In Fig.~\ref{osc1}, we have plotted the inverse temperature $\beta$ and minus
the energy $-E$ as a function of the central density $\rho_0=\rho(R_*)$ (more
precisely
the value of the density of the gas at the contact with the central
body). The central density parameterizes the series of equilibria. We
clearly see the appearance of oscillations giving rise to the spiral
on the series of equilibria. They lead to microcanonical instability
at the first turning point of energy and to canonical instability at 
the first turning point of temperature (stability is regained at the last
turning points).

\begin{figure}
\begin{center}
\includegraphics[width=0.9\linewidth,angle=0,clip]
{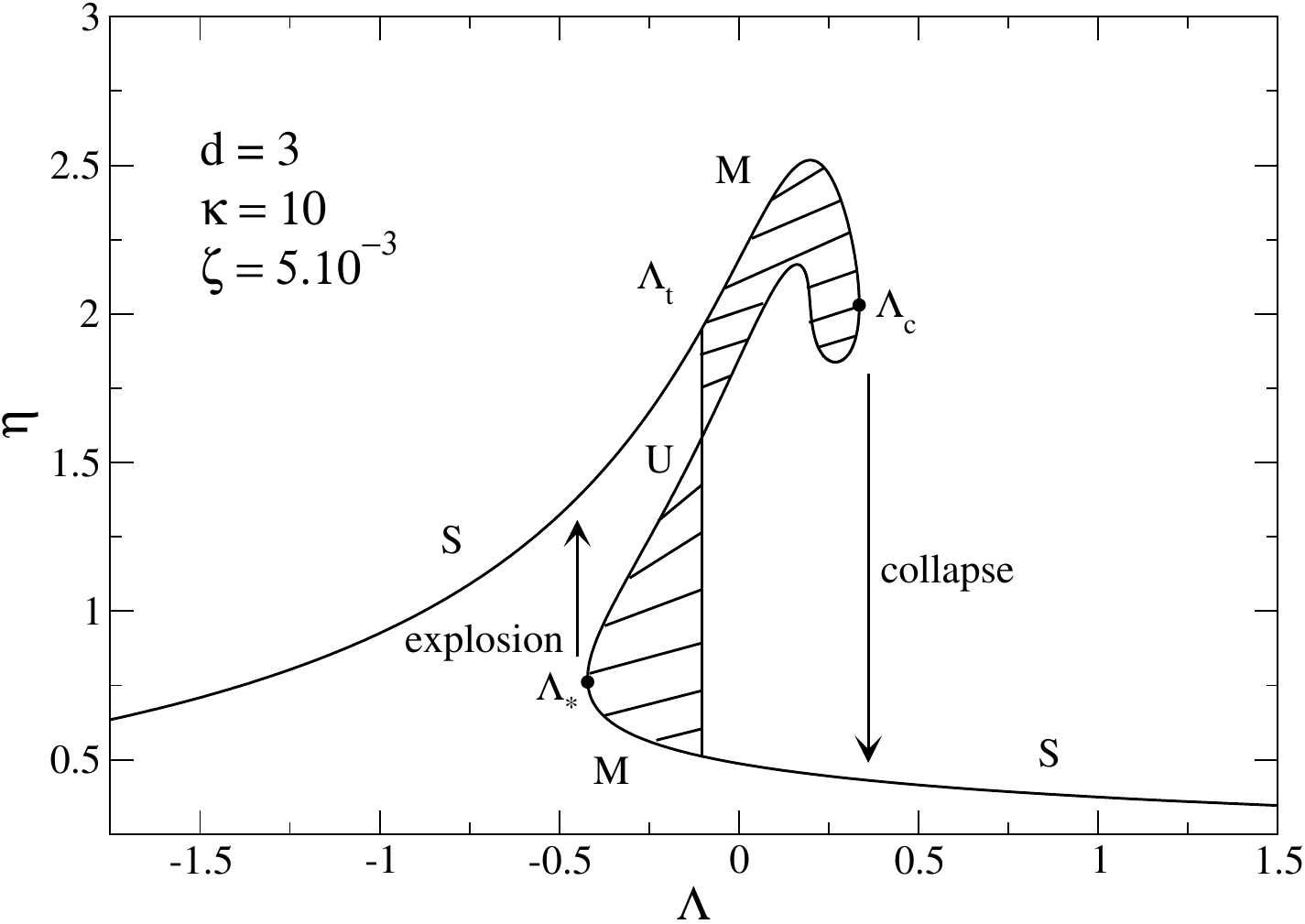}
\end{center}
\caption{\label{neck2b} Caloric curve in the microcanonical ensemble. 
We have indicated the fully stable states (global entropy maxima S), the
metastable states (local entropy maxima M) and the unstable states (saddle
points U). The strict caloric curve is composed by the fully stable states and
exhibits a first order phase transition at $\Lambda_t$. In practice, this phase
transition does not take place because the metastable states have a very long
lifetime. The physical caloric curve is composed of the fully stable and
metastable states (see Fig.~\ref{neck2}). The ends of the metastable branches
are called spinodal points. At these points, the system exhibits zeroth order
phase transitions associated with a collapse and an explosion.}
\end{figure}

In principle, to obtain the caloric curve, we must determine which
states are local entropy maxima and which states are global entropy
maxima. This can be done by performing a vertical Maxwell construction
or by plotting the entropy of the two phases as a function of the energy and determining the
transition energy $E_{t}$ at which they become equal~\cite{ijmpb}. The {\it strict
caloric curve} in the microcanonical ensemble is obtained by keeping
only fully stable states (global entropy maxima), as in Fig.~\ref{neck2b}. From this curve, we
would expect a first order microcanonical phase transition to occur at
$E=E_{t}$, connecting the gaseous phase to the condensed phase. It
would be accompanied by a discontinuity of temperature and specific
heat. However, for self-gravitating systems, the metastable states are
long-lived because the probability of a fluctuation able to trigger
the phase transition is extremely weak.  Indeed, in order to trigger a phase
transition between a gaseous state and a condensed state, the system has to
overcome the entropic barrier played by the solution of the
intermediate branch (points B) with a ``germ''.  Now, the
height of the entropic barrier scales like $N$ so that the probability
of transition scales like $e^{-N}$.  Therefore, metastable 
states (local entropy maxima) are very robust~\cite{lifetime} because their lifetime
scales like the exponential  of the number of particles, $t_{\rm life}\sim
e^{N}$, and it
becomes infinite at the thermodynamic limit $N\rightarrow +\infty$.  In practice, the first order
phase transition at $E_{t}$ does {\it not} take place and, for
sufficiently large $N$, the system remains ``frozen'' in the
metastable phase. Therefore, the {\it physical caloric curve} in the
microcanonical ensemble must take into account the metastable states
(local entropy maxima) which are as much relevant as the fully stable
equilibrium states (global entropy maxima). This physical caloric
curve is multi-valued and corresponds to the solid lines in
Fig.~\ref{neck2}. It is obtained from the series of equilibria by
discarding the unstable saddle points (dashed line).

Reducing progressively the energy from high values (for unbounded systems, the
mechanism by which energy decreases 
 may be due physically to a slow
evaporation), the system remains in the gaseous phase (points $A$)
until the critical value $E_{c}$ at which the gaseous phase ceases to
exist. At that point, called a spinodal point, the system undergoes a gravothermal
catastrophe~\cite{lbw}. However, in the present case, complete collapse is
prevented by the central body and the systems ends up in the condensed
phase (points $C$). Since the gravitational collapse is accompanied by
a discontinuous jump of entropy this phase transition is of zeroth
order. The resulting equilibrium state has a ``cusp-halo''
structure. The cusp contains a small fraction of the mass and this
fraction decreases as $R_{*}\rightarrow 0$ (in the absence of a central body,
the gravothermal catastrophe leads to a {\it binary star} with a
small mass but a large binding energy~\cite{ijmpb}). The rest of the mass is
diluted in a hot and massive envelope held by the box. In an open system (i.e.,
if
the box is removed) the halo would be dispersed at infinity so that
only the cusp (thin atmosphere) would remain. If we now increase 
the energy of the gas (for unbounded systems, the mechanism to supply energy
could be due to an accretion), the system
remains in the condensed phase until the critical value $E_*$ at which the
condensed phase ceases to exist. At that second spinodal point, the system
undergoes an explosion, reverse to the collapse, and returns to the gaseous
phase. Since the collapse and the explosion occur at different values of the
energy (due to the presence of metastable states), we can generate an hysteretic
cycle in the microcanonical ensemble by varying the energy between $E_c$ and
$E_{*}$. This hysteretic cycle has been followed numerically by
Ispolatov and Karttunen~\cite{ik} for particles interacting via a softened
gravitational potential (the softening regularizes the singularity of the
gravitational potential and plays a role similar to that of the central body in
our case).

\subsection{The case of a moderate central body $R_{*}^{\rm
MCP}<R_{*}<R_{*}^{\rm CCP}$ in the microcanonical ensemble: $N$-shape structure}
\label{sec_pt3b}

\begin{figure}
\begin{center}
\includegraphics[width=0.9\linewidth,angle=0,clip]{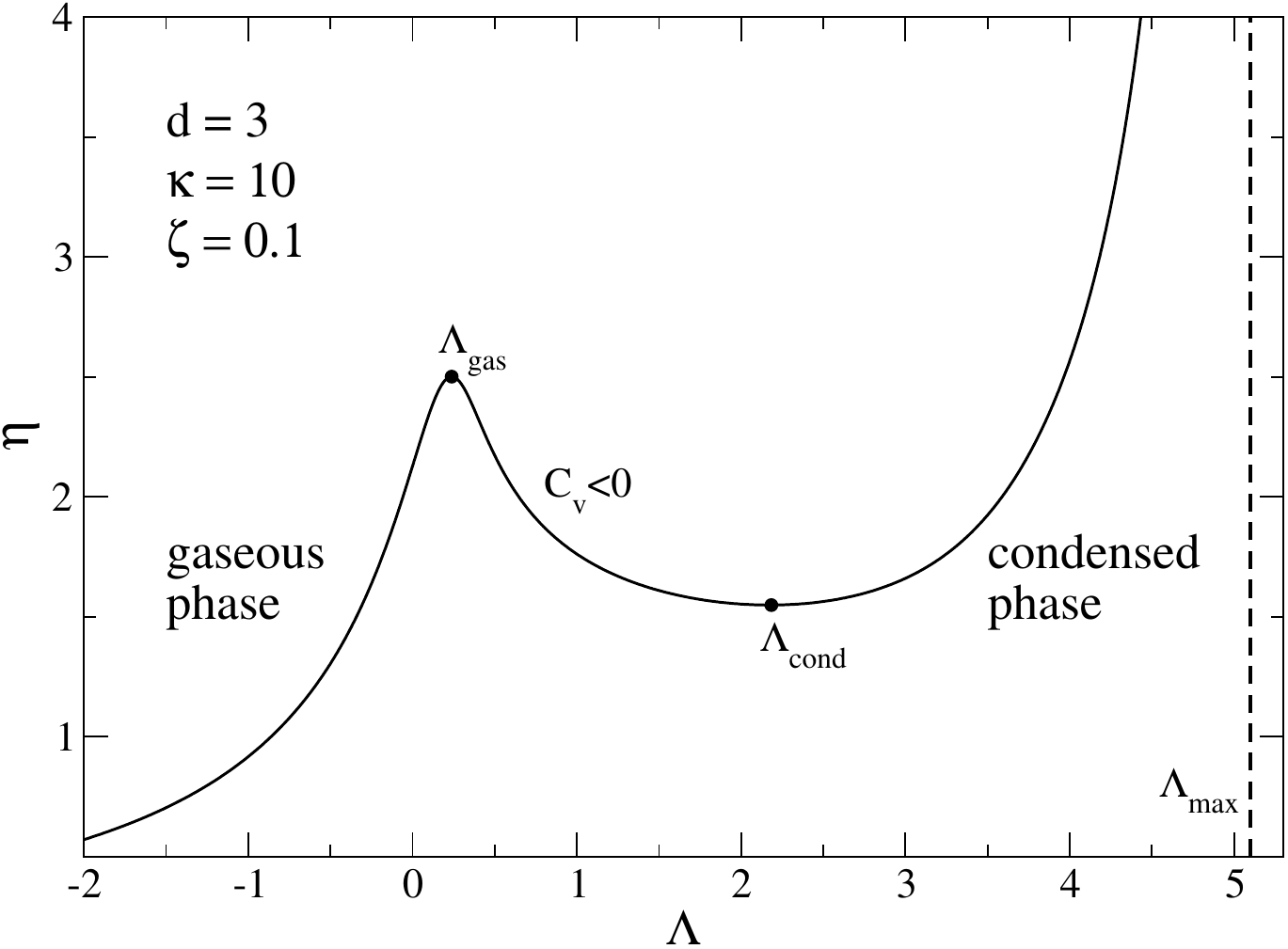}
\end{center}
\caption{\label{nshape} Caloric curve in the microcanonical ensemble  for
$\kappa = 10$ and $\zeta_{\rm MCP}<\zeta =
0.1<\zeta_{\rm CCP}$. It
has a characteristic $N$-shape structure. All the solutions are global
entropy maxima at fixed mass and energy (fully stable). The solutions
between $\Lambda_{\rm gas}$ and $\Lambda_{\rm cond}$ have negative specific
heats. There is no phase transition in the microcanonical ensemble, only a concentration of the gas near the central body as energy is progressively decreased. }
\end{figure}

We now consider the case of a  central body of intermediate core radius,
$R_{*}^{\rm MCP}<R_{*}<R_{*}^{\rm CCP}$, so that the series of equilibria has an
$N$-shape structure (see
Fig.~\ref{nshape}).

We first describe the structure of the caloric curve in the
microcanonical ensemble. In that case, the control parameter is the
energy $E$. For a moderate value of the core radius, the trace of the
classical spiral has almost disappeared and the $\beta(E)$ curve is univalued.
Since the solutions are entropy maxima for $E\rightarrow +\infty$, and
since there is no turning point of energy (no vertical tangent) in the
series of equilibria, we conclude from the Poincar\'e criterion that all the
states are stable and
correspond to global entropy maxima at fixed mass and
energy. Therefore, for sufficiently large values of the core radius,
there is no phase transition in the microcanonical ensemble. The
gravothermal catastrophe at $E_{c}$ is suppressed. However, there is a
sort of condensation as the energy is progressively decreased. At
high energies, the density profiles are almost homogeneous and they are held by
the box. In that case, the specific heat is positive. At
smaller energies, the density contrast increases and the system has a
``cusp-halo'' structure. Between $E_{\rm gas}$ and $E_{\rm cond}$, the influence
of the central body and of the box are weak and these
states have negative specific heats. Finally, at low energies, the
cusp becomes more and more massive and thin until the minimum energy
$E_{\rm min}$ at which all the mass is in contact with the central body (see
Appendix \ref{sec_lmax}). In that
case, the specific heat is positive again.

\subsection{The case of a moderate central body $R_{*}^{\rm
MCP}<R_{*}<R_{*}^{\rm CCP}$ in the canonical ensemble: $N$-shape structure}
\label{sec_pt3c}

\begin{figure}
\begin{center}
\includegraphics[width=0.9\linewidth,angle=0,clip]{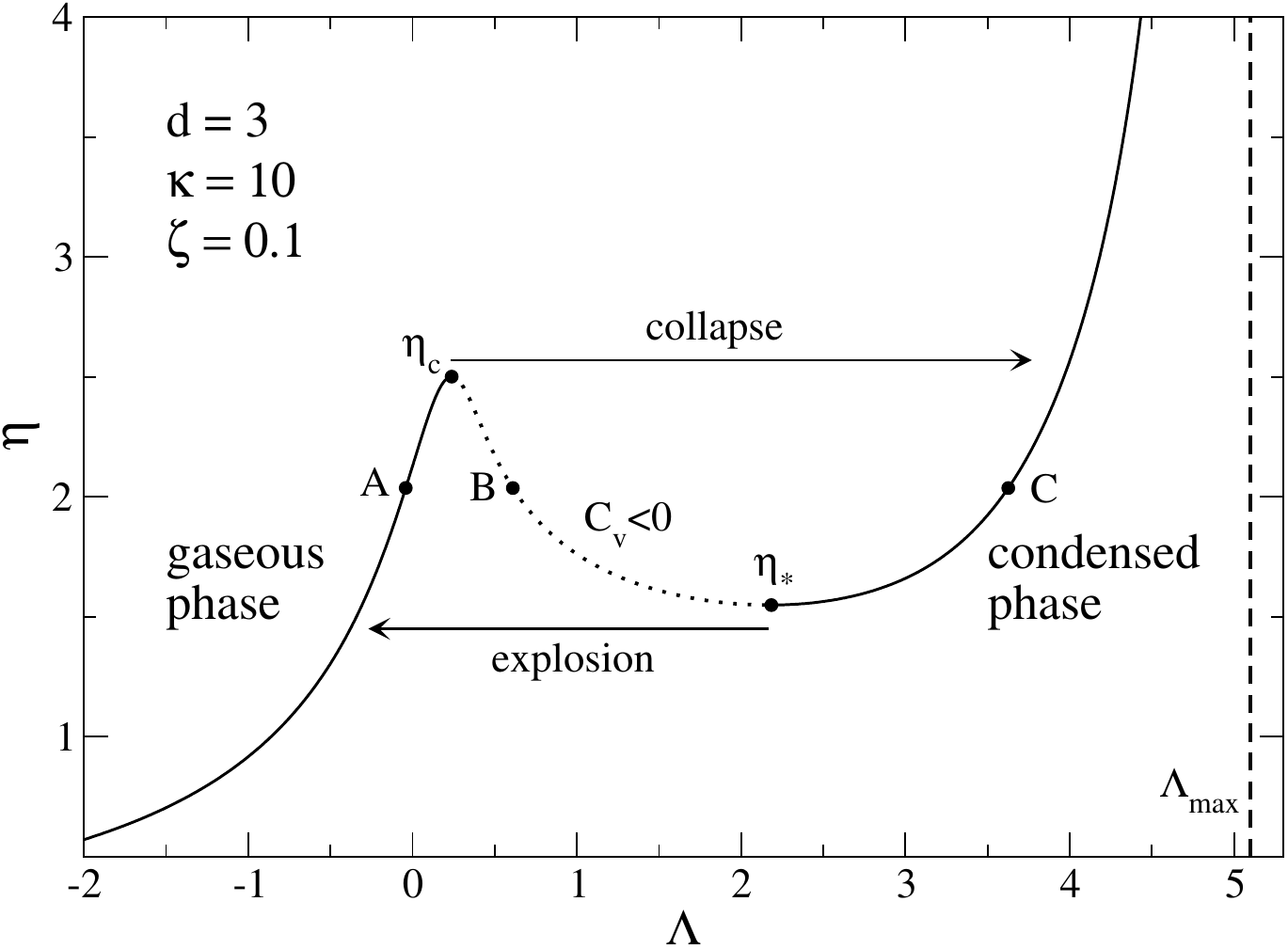}
\end{center}
\caption{\label{nez}
Physical caloric curve in the canonical ensemble (solid lines) 
containing stable (global free energy minima) as well as metastable (local free
energy minima) equilibrium states. Unstable saddle points of free energy at
fixed mass are represented by dotted lines and they lie in the region of
negative specific heats.}
\end{figure}

We now describe the structure of the caloric curve in the
canonical ensemble for the same values of the parameters as in the
previous section. In that case, the control parameter is the
temperature $T$ and the reader may find useful to rotate
Fig.~\ref{nshape} by $90^{\rm o}$ to have the control parameter $\eta\sim 1/T$ in
abscissa. The series of
equilibria $E(T)$ is multi-valued and this gives rise to canonical
phase transitions (see Fig.~\ref{nez}). In fact, we are in a situation
parallel to the one
described in Sec.~\ref{sec_pt3a} in the microcanonical ensemble, provided that we
interchange the role of $E$ and $T$. The series of equilibria contains
all the extrema of free energy at fixed mass. The thermodynamically
stable
states  in the canonical ensemble correspond to free
energy {\it minima}  at fixed mass (maxima or saddle
points of free energy must be
discarded). They can be determined by applying the turning point method of
Poincar\'e. At very high temperatures, self-gravity is negligible
with respect to thermal motion and the system is equivalent to a
non-interacting gas in a box. We know from ordinary thermodynamics that this
gas is stable in every ensemble. Therefore, using
the Poincar\'e criterion, we conclude that the left branch is stable
(free energy minima at fixed mass) until the first turning point of
temperature $T_{c}$ where the tangent is horizontal. For
sufficiently small values of $R_{*}$, this is close to the Emden
inverse temperature $2.52$~\cite{emden,aa1}. At that point, the curve $-E(\beta)$
rotates clockwise so that a mode of stability is lost. Therefore, the
configurations in the intermediate branch are unstable saddle points
of free energy. They lie in the region of negative specific heats
which is forbidden (unstable) in the canonical
ensemble.\footnote{Negative specific heat is a sufficient (but
not necessary) condition of thermodynamical instability in the canonical
ensemble~\cite{cc}.} However, at
the second turning point of temperature $T_{*}$, the curve
$-E(\beta)$ rotates anti-clockwise so that the mode of stability is
regained. Therefore, the right branch is stable (free energy minima
at fixed mass). The temperature $T_{*}$ depends on $R_{*}$ and tends
to $T_{*}(R_{*})\rightarrow +\infty$ when $R_{*}\rightarrow 0$.

\begin{figure}
\begin{center}
\includegraphics[width=0.9\linewidth,angle=0,clip]{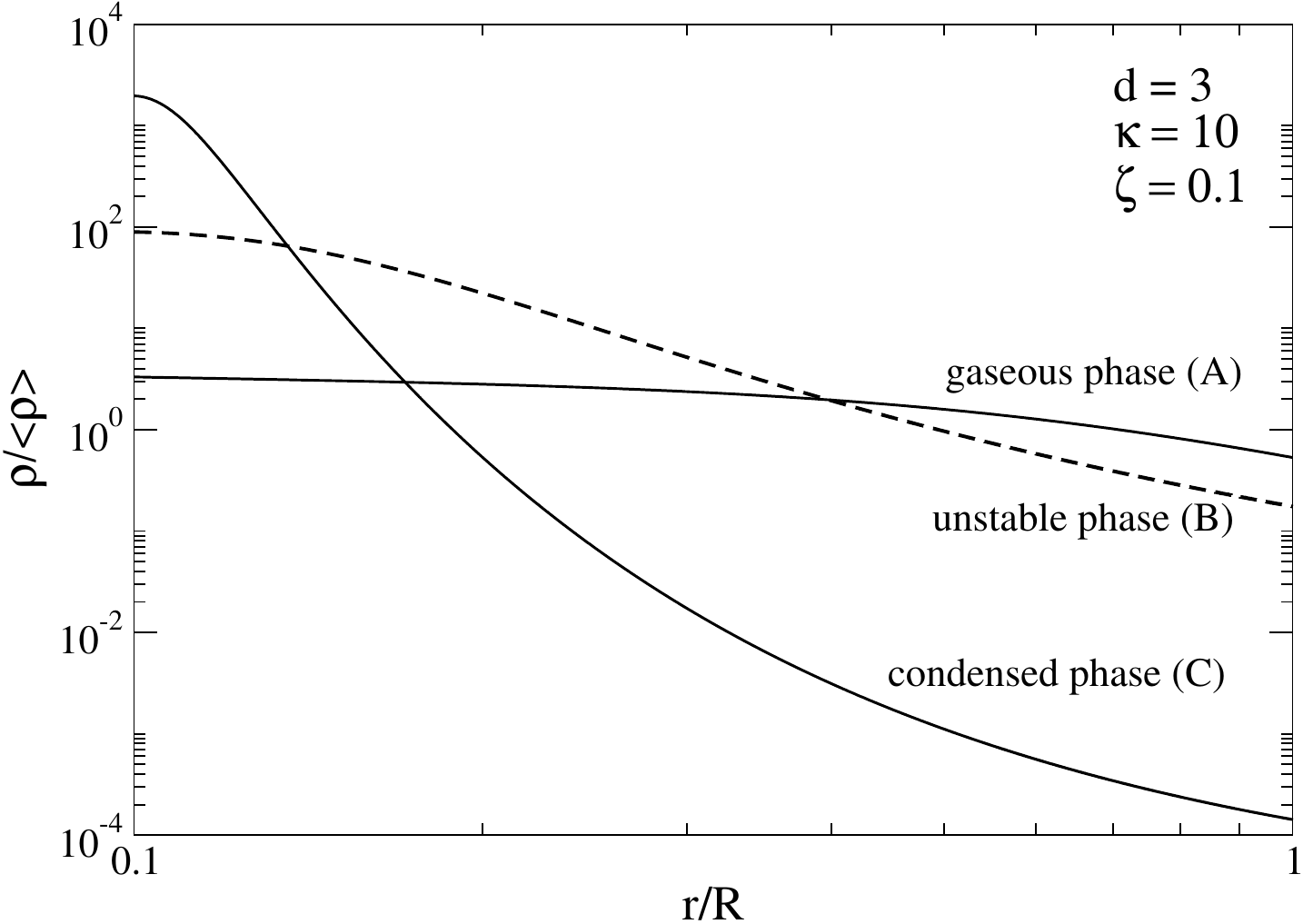}
\end{center}
\caption{\label{prof}Typical normalized density profiles
$\rho/\langle{\rho}\rangle$ corresponding to gaseous (points $A$), condensed (points $C$) and unstable (points $B$) states.}
\end{figure}

Typical density profiles of the series of equilibria
are shown in Fig.~\ref{prof}. The states (A) on the left branch have almost
uniform density profiles and they form the {\it gaseous phase}.  The
states (C) on the right branch are highly inhomogeneous with a cusp-halo
structure. They form the {\it condensed phase}.  Finally, the states (B) on
the intermediate (unstable) branch are similar to the gaseous states
 except that they contain an embryonic cusp at the contact with the central
body.
This
is the equivalent of a ``germ'' in the language of phase
transitions.

\begin{figure}
\begin{center}
(a)\includegraphics[width=0.9\linewidth,angle=0,clip]{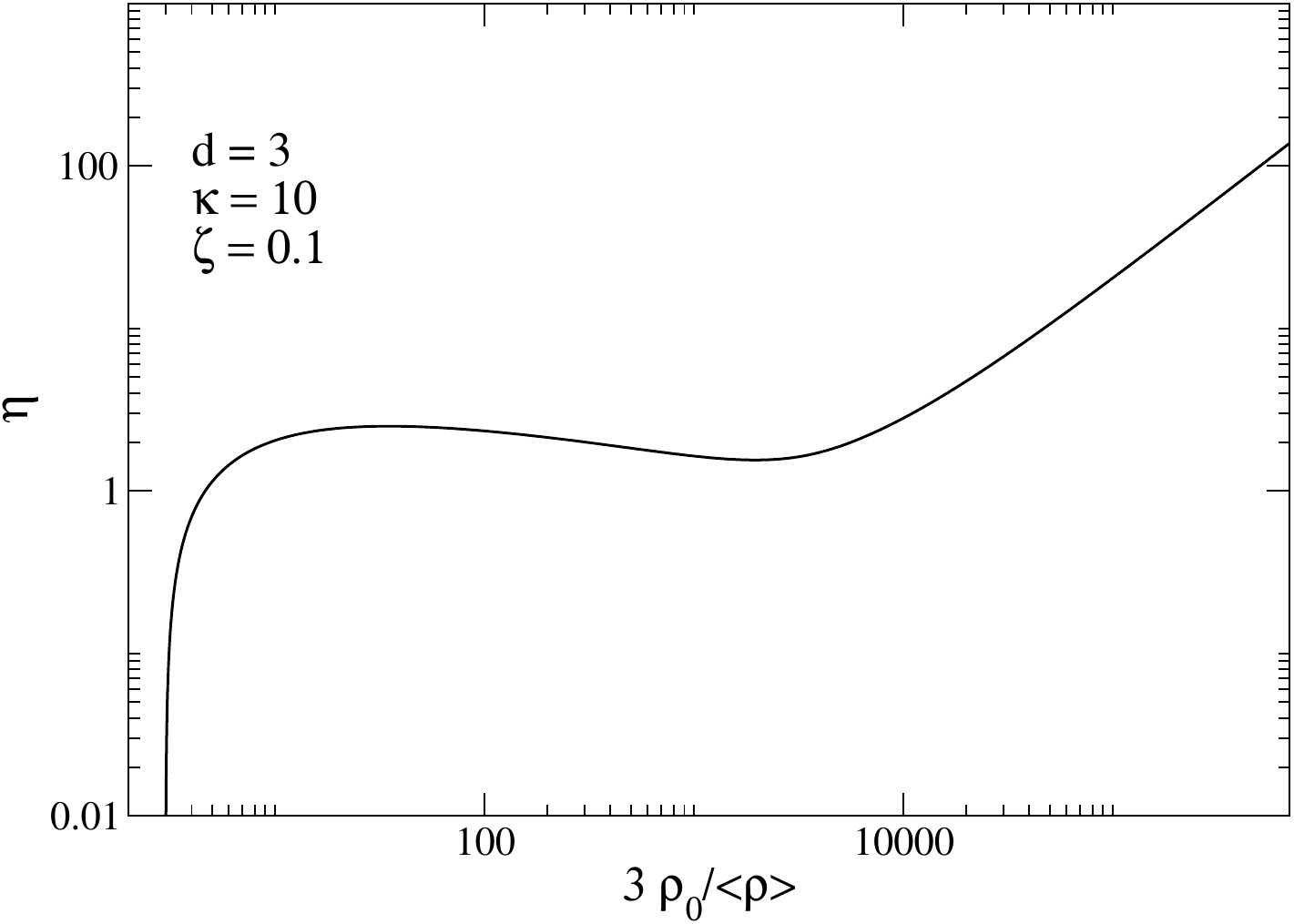}
\hspace{6pt}
(b) 
\includegraphics[width=0.9\linewidth,angle=0,clip]{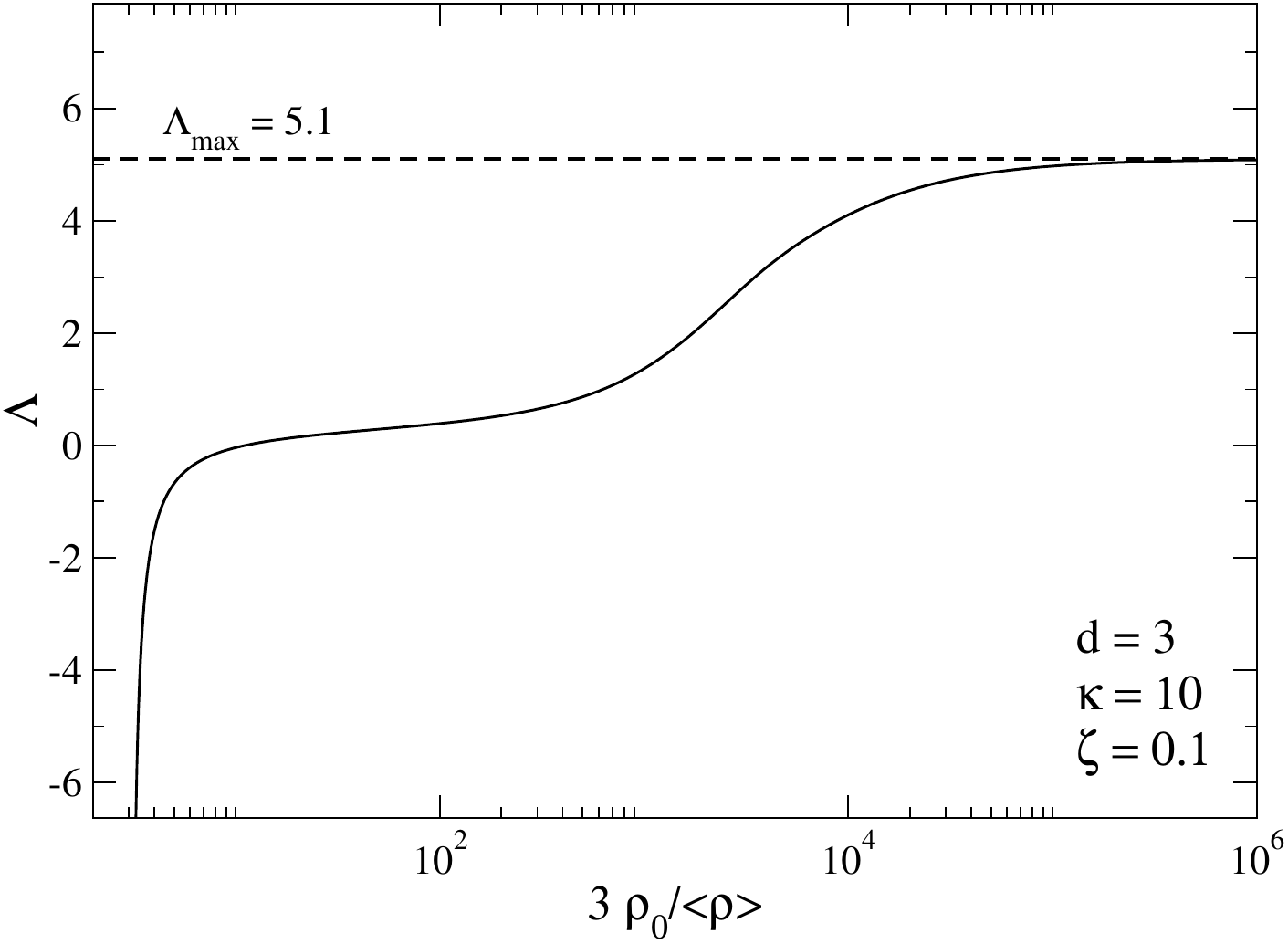}
\end{center}
\caption{\label{osc2}Evolution of the dimensionless energy $\Lambda$ and
inverse
temperature $\eta$ as a function of the normalized
central density $\rho_0/\langle{\rho}\rangle$.}
\end{figure}

In Fig.~\ref{osc2}, we have plotted the inverse temperature $\beta$
and minus the energy $-E$ as a a function of the central density $\rho_0=\rho(R_*)$ (more
precisely the value of the density of the gas at the contact with the
central body). The central density parameterizes the series of
equilibria. We clearly see the oscillations of $T(\rho_0)$ giving rise to
canonical instability at the first turning point of temperature (stability is
regained at the last turning point of temperature). On the other hand, the
absence of oscillations of  $E(\rho_0)$ is associated  to full stability in the
microcanonical ensemble (see Sec.~\ref{sec_pt3c}).

\begin{figure}
\begin{center}
\includegraphics[width=0.9\linewidth,angle=0,clip]
{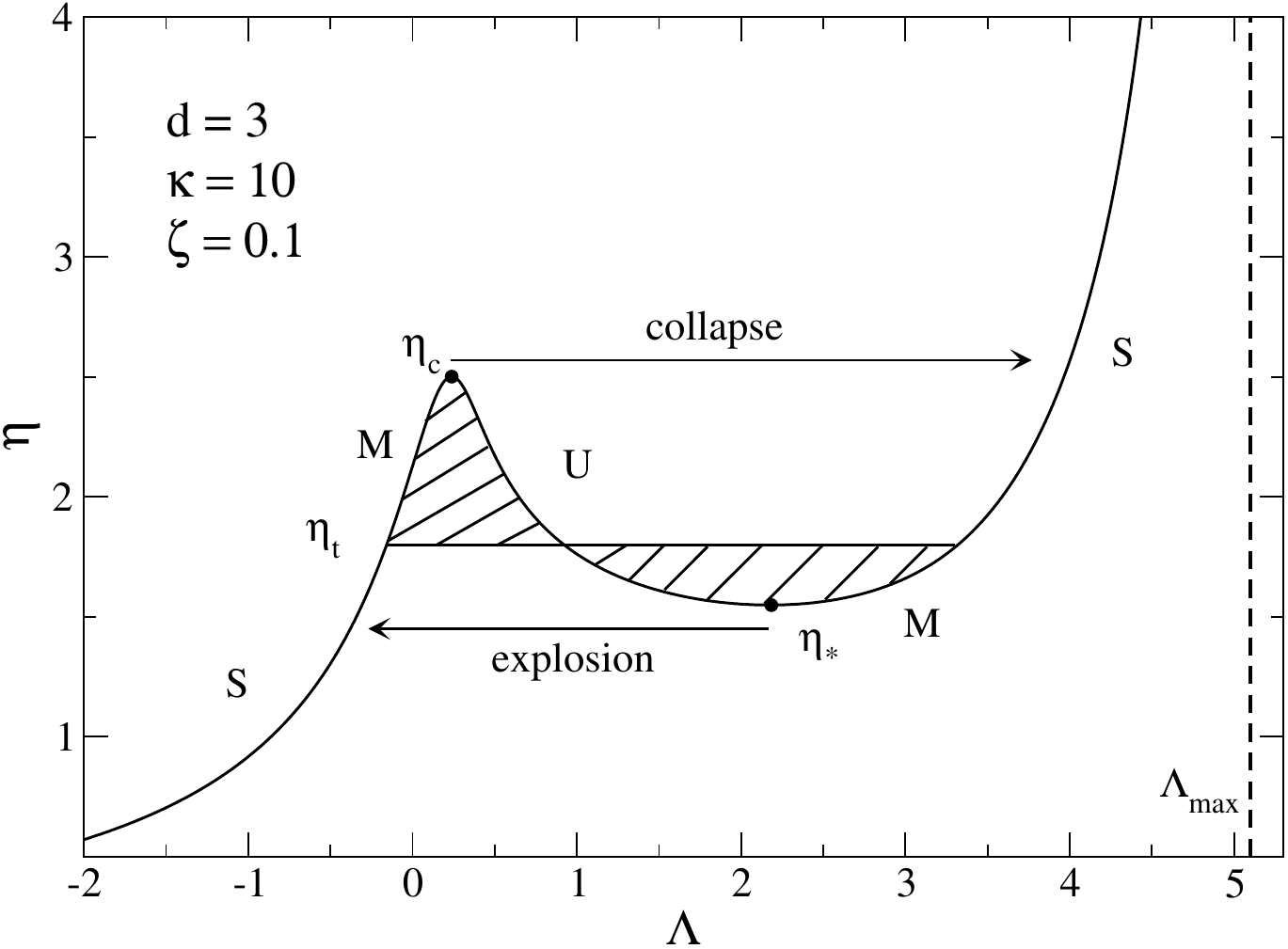}
\end{center}
\caption{\label{nezma} Caloric curve in the canonical ensemble. 
We have indicated the fully stable states (global free energy minima S), the
metastable states (local free energy minima M) and the unstable states
(saddle points U). The strict caloric curve is composed by the fully stable
states and exhibits a first order phase transition at $\eta_t$ which replaces
the
region of negative specific heats allowed in the microcanonical ensemble (see
Fig.~\ref{nshape}). In practice, this phase transition does not take
place because the metastable states have a very long lifetime. The physical
caloric curve is composed of the fully stable and metastable states (see Fig.
\ref{nez}). The ends of the metastable branches are called spinodal points. At
these points, the system exhibits zeroth order phase transitions associated with
a collapse and an explosion. }
\end{figure}

In principle, to obtain the caloric curve, we must determine which
states are local  free energy minima and which states are global
 free energy minima. This can be done by performing a horizontal
Maxwell construction or by plotting the free energy of the two phases
as a function of the temperature and determining the transition temperature $T_{t}$ at which they
become equal. The {\it strict caloric curve} in the canonical ensemble is
obtained by keeping only fully stable states (global free
energy minima) as in Fig.~\ref{nezma}. From this curve, we would expect a first order canonical
phase transition to occur at $T=T_{t}$, connecting the gaseous phase
to the condensed phase. It would be accompanied by a discontinuity of
energy and specific heat. Therefore, the region of negative specific heats in the
microcanonical ensemble (see Fig.~\ref{nshape}) would be replaced by a
phase transition
(plateau) in the canonical ensemble (see Fig.~\ref{nezma}).   However, for
self-gravitating systems, the metastable states are long-lived because
the probability of a fluctuation able to trigger the phase transition
is extremely weak. Indeed, in order to trigger a phase transition between
a gaseous state and a
condensed state, the system has to overcome the
barrier of free energy played by the solutions of the intermediate branch (points B)
with a ``germ''. Yet, the height of the  barrier of free energy scales like
$N$ so that the probability of transition scales like $e^{-N}$.
Therefore, metastable equilibrium states (local free energy minima) are
very robust~\cite{lifetime} because their lifetime scales like the exponential
of the number of particles,  $t_{\rm life}\sim e^{N}$, and it becomes infinite
at the thermodynamic limit $N\rightarrow +\infty$.
In practice, the first order phase transition at $T_{t}$ does {\it
not} take place and, for sufficiently large $N$, the system 
remains ``frozen'' in the metastable phase. Therefore, the {\it physical caloric
curve} in the canonical ensemble must take into account the
metastable states (local free energy minima) which are as much relevant as
the fully stable equilibrium states (global free energy minima). This
physical caloric curve is multi-valued and corresponds to the solid
lines in Fig.~\ref{nez}. It is obtained from the series of
equilibria by discarding the unstable saddle points (dashed line).

Reducing progressively the temperature from high values, the system
remains in the gaseous phase (points $A$) until the critical value
$T_{c}$ at which the gaseous phase ceases to exist. At that point, called a spinodal point, the
system undergoes an isothermal collapse~\cite{aa1}. However, in the present case,
complete collapse is prevented by the central body and the systems ends
up in the condensed phase (points $C$). Since the gravitational
collapse is accompanied by a discontinuous jump of free energy, this phase
transition is of zeroth order. The resulting equilibrium state has a
``cusp-halo'' structure. The cusp contains a large fraction of the
mass and this fraction increases as $R_{*}\rightarrow 0$ (in the
absence of a central body, the isothermal collapse leads to a {\it Dirac peak}
containing all the mass~\cite{ijmpb}). The rest of
the mass is diluted in a light envelope held by the box. In an open
system (i.e., if the box is removed) the halo would be dispersed to
infinity so that only the cusp (thin atmosphere) would remain. If we
now increase the temperature of the gas, the system remains in the condensed phase until the
critical value $T_*$ at which the condensed phase ceases to exist. At
that second spinodal point, the system undergoes an explosion, reverse to the
collapse, and returns to the gaseous phase. Since the collapse and the
explosion occur at different values of the temperature (due to the presence
of metastable states), one can generate an hysteretic cycle in the
canonical ensemble by varying the temperature between $T_c$ and
$T_{*}$. This hysteretic cycle has been followed numerically by
Chavanis {\it et al.}~\cite{crrs} for a self-gravitating Fermi gas (as we have
already indicated, the Pauli exclusion principle plays a role similar to that of
the central body in our case).

\subsection{The case of a large  central body $R_{*}>R_{*}^{\rm CCP}$:
monotonicity and equivalence of statistical ensembles}
\label{sec_pt3d}

\begin{figure}
\begin{center}
\includegraphics[width=0.9\linewidth,angle=0,clip]{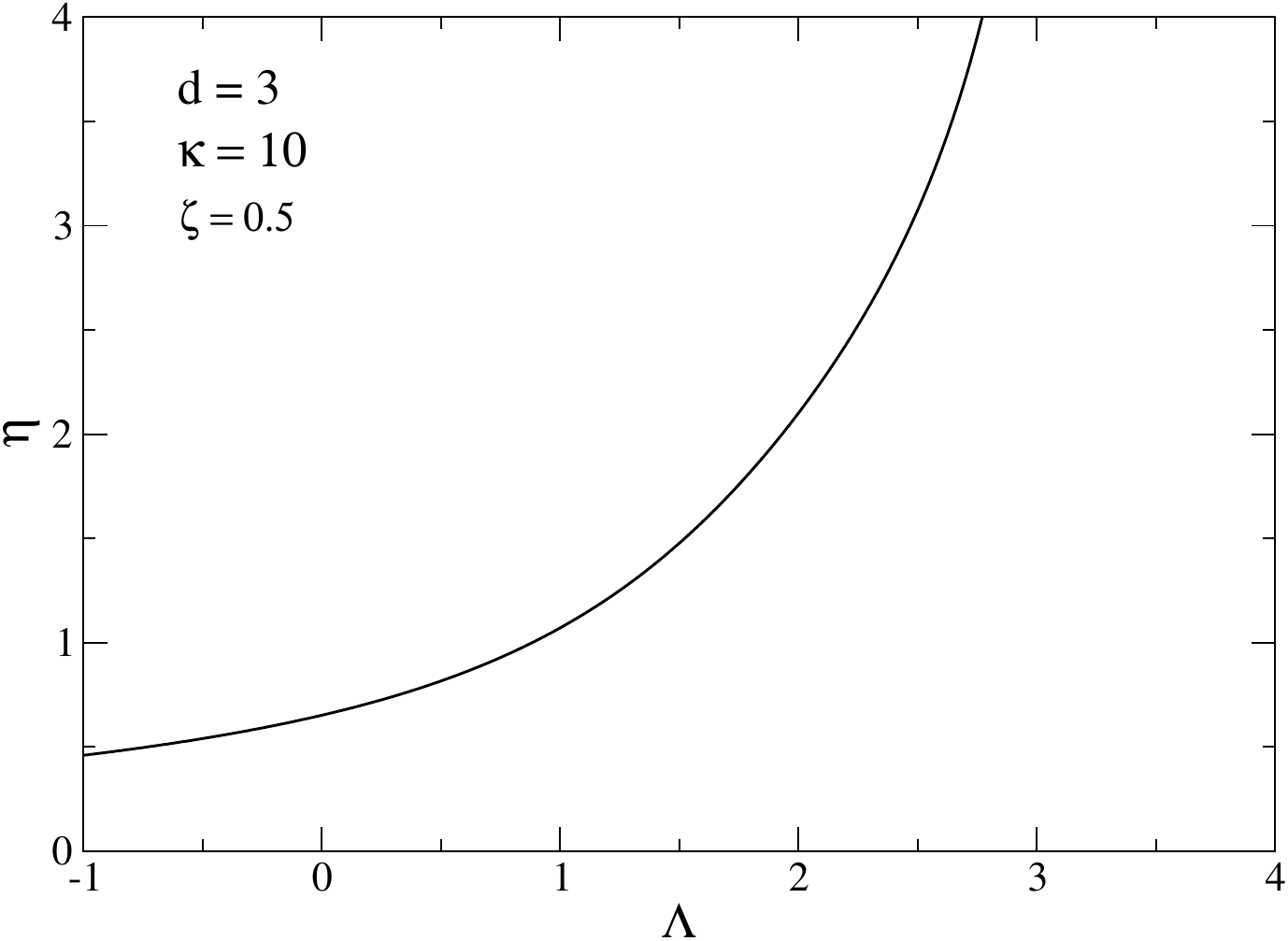}
\end{center}
\caption{\label{monotone} Caloric curve for $\kappa=10$ and
$\zeta=0.5>\zeta_{\rm CCP}$. It is monotonic and the ensembles are equivalent. }
\end{figure}

We now consider the case of a large central body
$R_{*}>R_{*}^{\rm CCP}$ so that the series of equilibria is monotonic (see
Fig.~\ref{monotone}). In that case, the system does not present any phase transition and the ensembles are equivalent.

\subsection{The case of a small central body $R_{*}<R_{*}^{\rm MCP}$ in the
canonical ensemble: multi-rotations}
\label{sec_pt3e}

\begin{figure}
\begin{center}
\includegraphics[width=0.9\linewidth,angle=0,clip]
{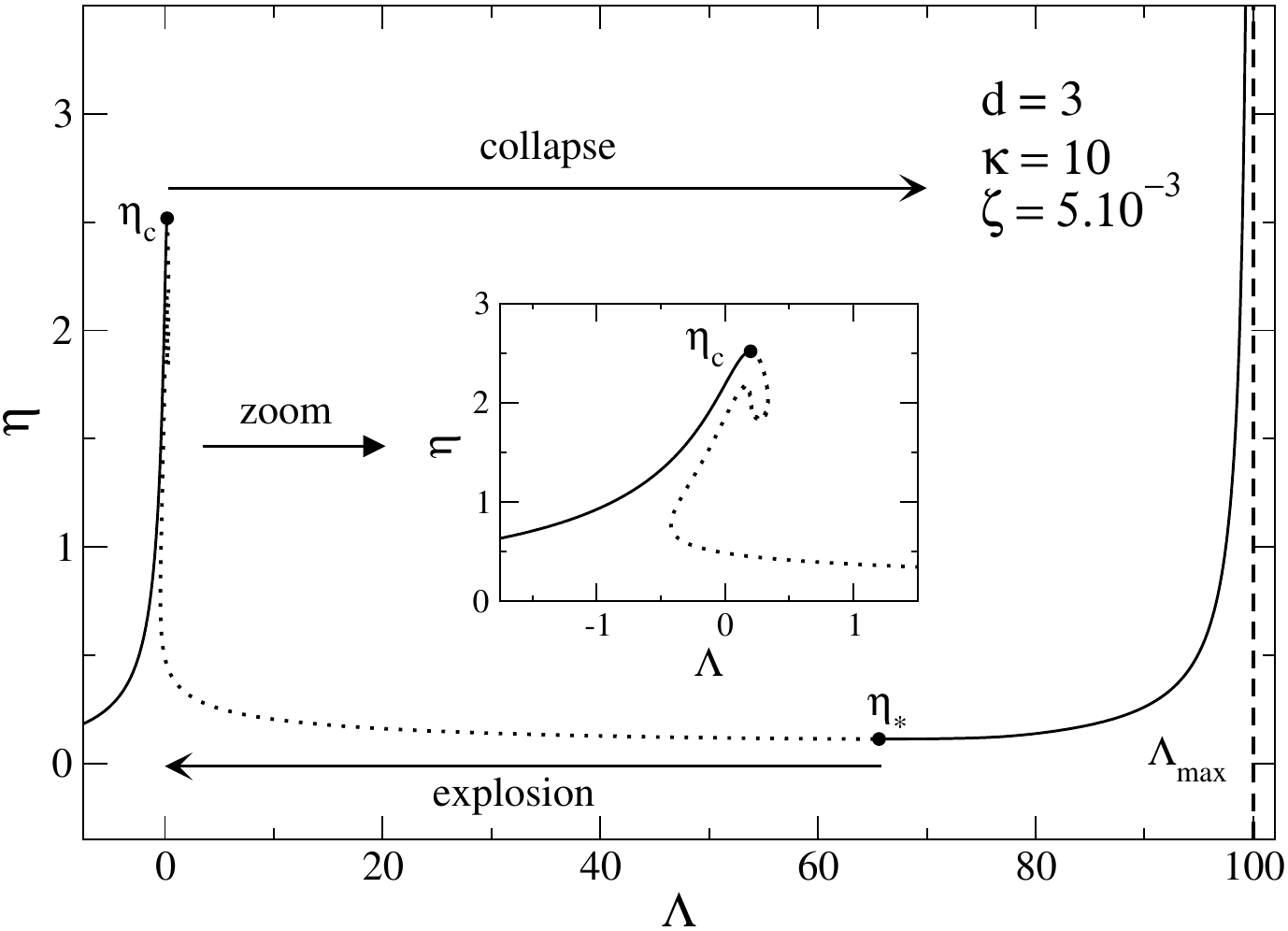}
\end{center}
\caption{\label{fig.elz0.005cc}Physical canonical caloric curve  for $\zeta =
5\times 10^{-3}<\zeta_{\rm MCP}$ and $\kappa =
10$. The dotted lines represent the
unstable states (saddle points).}
\end{figure}

For a small central body $R_{*}<R_{*}^{\rm MCP}$, the system can display
both microcanonical and canonical phase transitions (zeroth and first order). The
microcanonical caloric curve has been described in Sec.~\ref{sec_pt3a}. Let us
describe the corresponding caloric curve in the canonical
ensemble. The series of equilibria displays several turning points of
temperature (see Fig.~\ref{fig.elz0.005cc}). At the first turning point $T_c$,
the curve rotates clockwise so that a mode of
 stability is lost. At the second turning point of
temperature, another mode of stability is lost. At the third turning
point of temperature, the curve rotates anti-clockwise so that the
second mode is regained and at the fourth turning point of temperature
$T_{*}$ the first mode is regained.\footnote{We note that
certain equilibrium states, deep in the spiral, are unstable in
the canonical ensemble (and
also in the microcanonical ensemble) although they
have a positive specific heat (see footnote 14).} Therefore, the left branch
(before
$T_c$) and the right branch (after $T_{*}$) are canonically stable
(free energy minima) and form the physical canonical caloric
curve. Although the structure of the series of equilibria in the
unstable domain is more complex than in Sec.~\ref{sec_pt3b}, because it displays
several rotations, the caloric curve corresponding to the stable part
of the series of equilibria is simple and similar to Fig.~\ref{nez} obtained for
a larger value of the core radius. The strict canonical caloric curve containing
only fully stable states (global free energy minima) would also be similar to
Fig.~\ref{nezma}.

\subsection{The limit  of a vanishing central body $R_{*}\rightarrow 0$}
\label{sec_pt3f}

It is of interest to study the limit of a vanishing central body
$R_{*}\rightarrow 0$ in order to make the connection with the
classical studies of Antonov~\cite{antonov}, Lynden-Bell and Wood~\cite{lbw},
and Katz~\cite{katzpoincare1}.

\begin{figure}
\begin{center}
\includegraphics[width=0.9\linewidth,angle=0,clip]{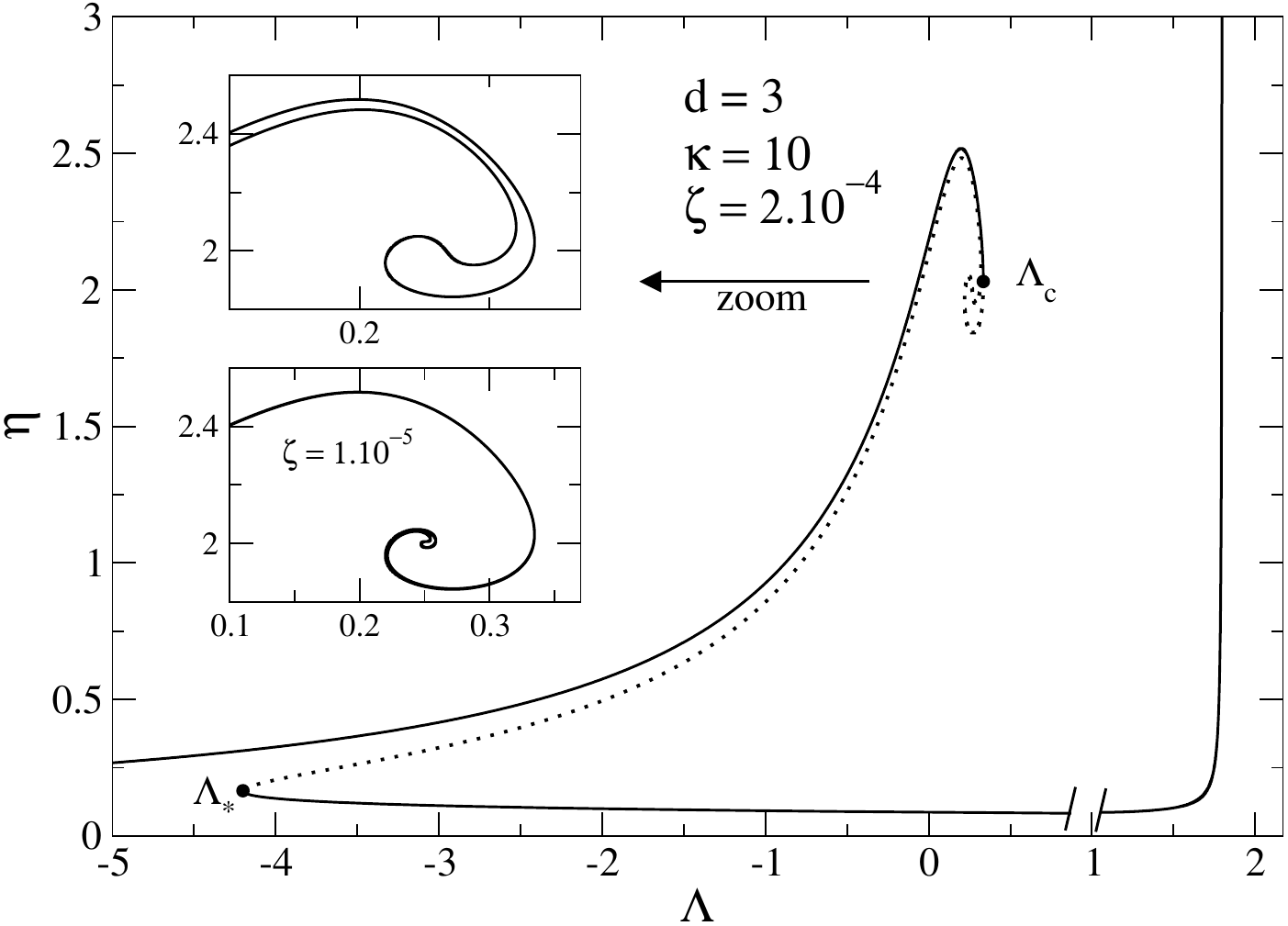}
\end{center}
\caption{\label{fig.elz0.0002cc}Series of equilibria at fixed
$\kappa = 10$ for several small values of
$\zeta$. As $\zeta \rightarrow 0$, the series of
equilibria makes more and more rotations around the singular sphere
$(\Lambda,\eta)=(1/4,2)$ 
before unwinding (see inset). When $\zeta = 0$, we recover the classical spiral
of a spherical isothermal self-gravitating system.}
\end{figure}

\begin{figure}
\begin{center}
(a)
\includegraphics[width=0.9\linewidth,angle=0,clip]{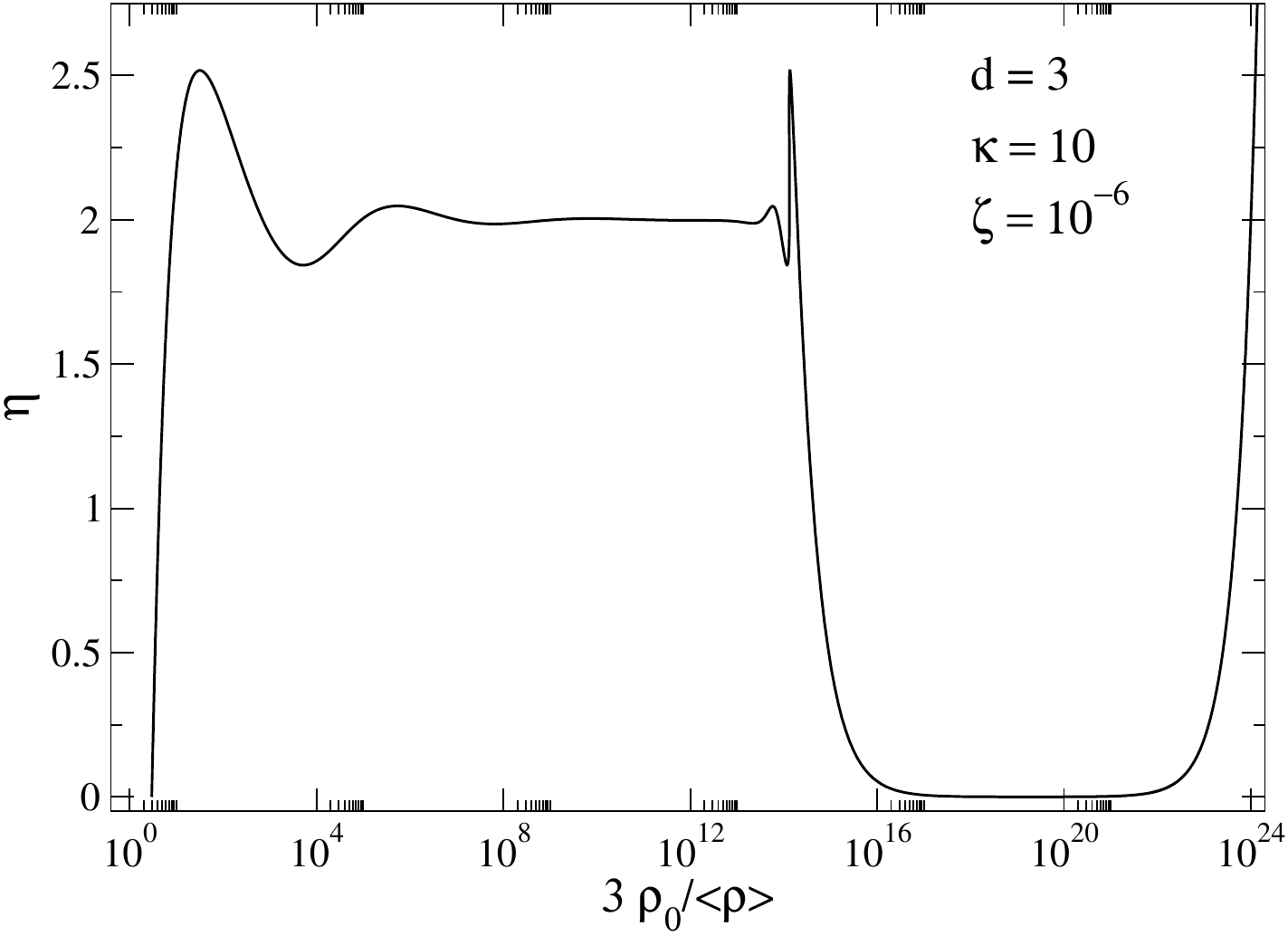}
\hspace{6pt}
(b)
\includegraphics[width=0.9\linewidth,angle=0,clip]{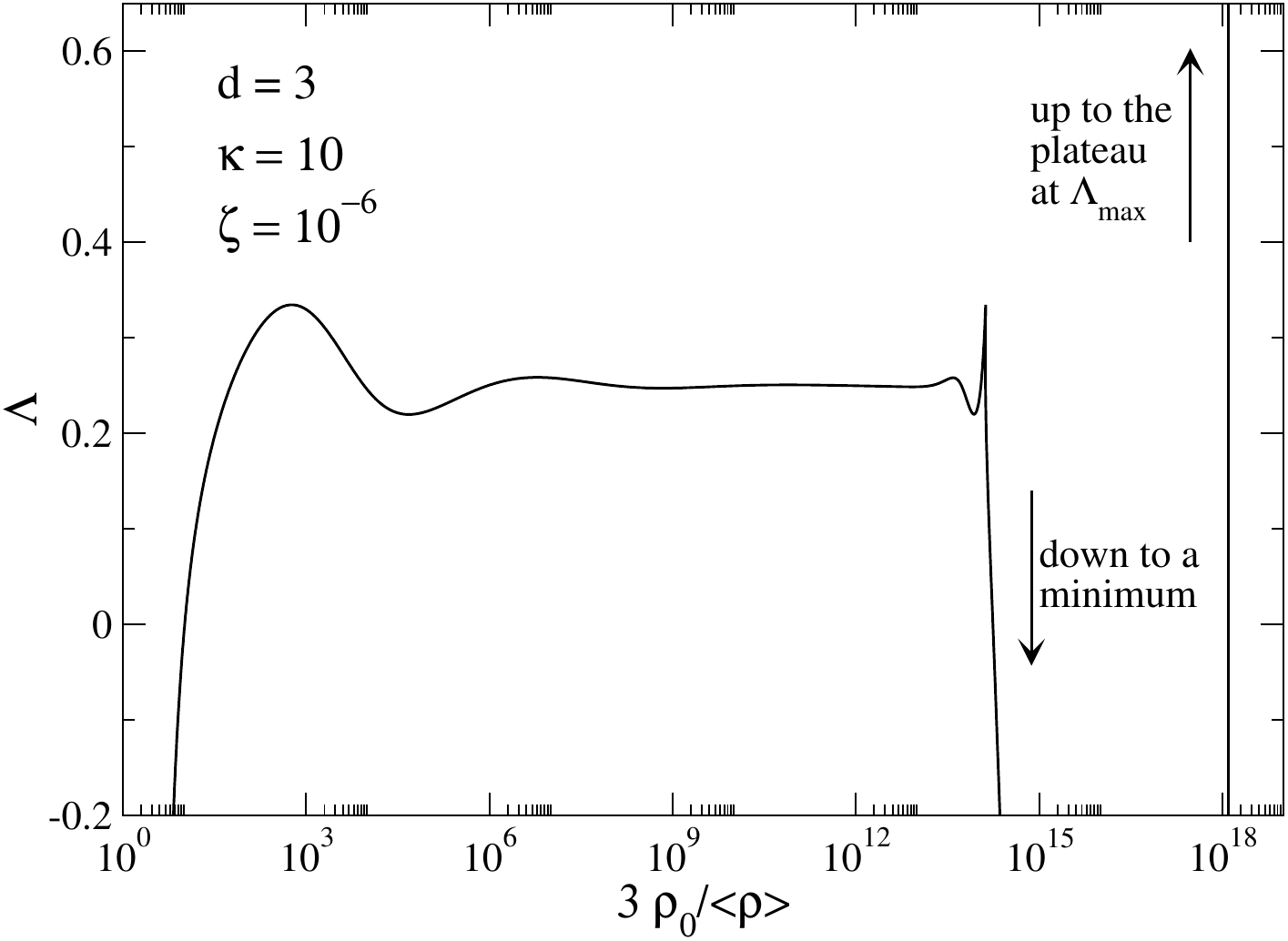}
\end{center}
\caption{\label{fig.erhocc}Evolution of the energy $\Lambda$ and inverse
temperature $\eta$ as a function of the normalized
central density $\rho_0/\langle{\rho}\rangle$.}
\end{figure}

As the radius of the central body decreases, the series of equilibria
winds up several times before unwinding so that more and more turning
points of energy and temperature appear. This is illustrated in Figs.~\ref{fig.elz0.0002cc}
and  \ref{fig.erhocc}. However, the discussion concerning the phase transitions remains
essentially unchanged. In the microcanonical ensemble, the gaseous phase (upper
branch) is stable until
the energy $E_c$ and the condensed phase (lower branch) is stable from the
energy
$E_{*}$. In the canonical ensemble, the gaseous phase (left branch) is stable
until the
temperature $T_c$ and the condensed phase (right branch) is stable from the
temperature $T_{*}$. When $R_{*}\rightarrow 0$, the energy
$E_{*}\rightarrow -\infty$ and the temperature $T_{*}\rightarrow +\infty$.
Similarly, $E_{t}\rightarrow -\infty$ and $T_{t}\rightarrow +\infty$. Therefore,
the gaseous
branch only contains metastable states. The condensed branch in
the microcanonical ensemble approaches the $\beta=0$ axis and is
formed by configurations presenting a thin cusp containing a weak mass but a
large potential energy
surrounded by a hot and massive halo (in the absence of a central body, the
gravothermal catastrophe leads to a tight binary star surrounded by a very hot
halo with a very large entropy). The
condensed branch in the canonical ensemble is rejected to $E\rightarrow
-\infty$ and is formed by configurations presenting a thin cusp containing most
of the mass (in the
absence of a central body, the isothermal collapse leads to a Dirac peak
containing all the mass and having a very negative free energy). The unstable
branch makes
several oscillations and leads to the classical spiral discussed by
Antonov~\cite{antonov}, Lynden-Bell and Wood~\cite{lbw}, and Katz
\cite{katzpoincare1}.

\subsection{Critical points}

\begin{figure}
\begin{center}
\includegraphics[width=0.9\linewidth,angle=0,clip]
{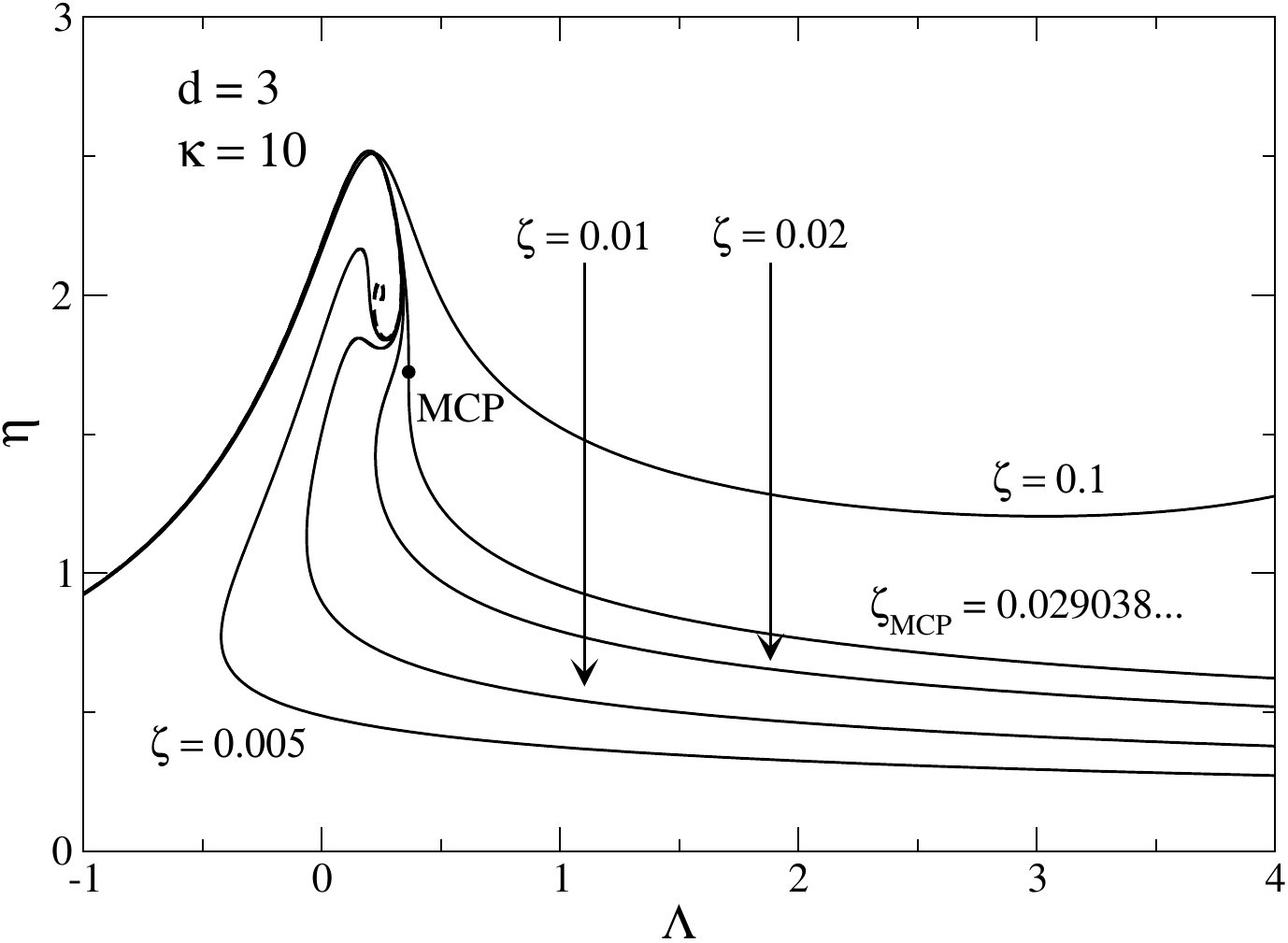}
\end{center}
\caption{\label{microcrit}Enlargement of the series of equilibria near the
microcanonical critical point. When $\zeta=\zeta_{\rm MCP}$, the curve
$\eta(\Lambda)$
presents an inflexion point and the microcanonical phase transition
(gravothermal catastrophe) is suppressed. }
\end{figure}

\begin{figure}
\begin{center}
\includegraphics[width=0.9\linewidth,angle=0,clip]
{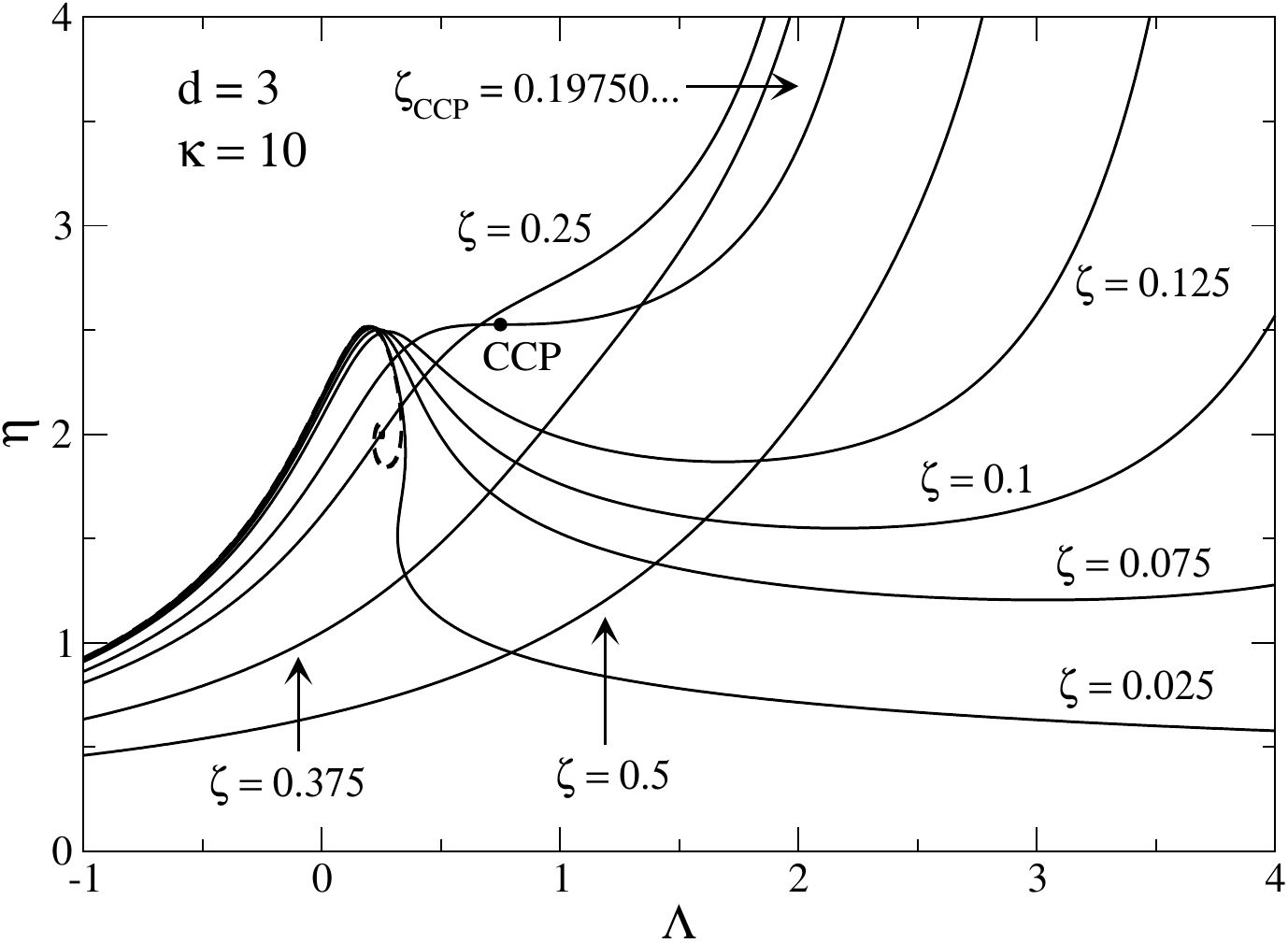}
\end{center}
\caption{\label{canocrit}Enlargement of the series of equilibria near the
canonical critical point. When $\zeta=\zeta_{\rm CCP}$, the curve
$\Lambda(\eta)$
presents an inflexion point and the canonical phase transition (isothermal
collapse) is suppressed.}
\end{figure}

The deformation of the series of equilibria as a function of the core
radius $R_{*}$ (for a fixed density $\rho_{*}$ of the central body) is
represented in Figs.~\ref{microcrit} and \ref{canocrit}. There are two critical
points in the
problem, one in each ensemble. When $R_{*}<R_{*}^{\rm MCP}$, the series of
equilibria $\beta(E)$ presents turning points of energy and
temperature so that there exist phase transitions in the
microcanonical and canonical ensembles. We have seen that the first order phase
transitions may not take place in practice due to the long lifetime of
the metastable
states. However, there remains zeroth order phase transitions associated with the
gravothermal catastrophe in the microcanonical ensemble and the isothermal
collapse in the canonical ensemble. Since the domains
of stability differ in the canonical and microcanonical ensembles, the ensembles
are
not equivalent. Indeed, some states are stable (i.e., accessible) in
the microcanonical ensemble while they are unstable  (i.e., inaccessible) in the canonical ensemble. Since canonical stability implies
microcanonical stability~\cite{cc}, the microcanonical ensemble contains
more stable states than the canonical one. At the microcanonical critical
point $R_{*}=R_{*}^{\rm MCP}$, the series of equilibria $\beta(E)$
presents an inflexion point so that the microcanonical phase
transition (gravothermal catastrophe) is suppressed (see Fig.~\ref{microcrit}).
When
$R_{*}^{\rm MCP}<R_{*}<R_{*}^{\rm CCP}$, the series of equilibria $\beta(E)$
presents turning points of temperature but no turning point of
energy. Therefore, there exist phase transitions in the canonical
ensemble but not in the microcanonical ensemble. The ensembles are not
equivalent as revealed by the region of negative specific heat in
the microcanonical ensemble that is
replaced by a phase transition in the canonical ensemble. At the canonical
critical
point $R_{*}=R_{*}^{\rm CCP}$, the series of equilibria $E(\beta)$
presents an inflexion point so that the canonical phase
transition (isothermal collapse) is suppressed (see Fig.~\ref{canocrit}). When
$R_{*}>R_{*}^{\rm CCP}$, the series of equilibria $\beta(E)$ is monotonic so
that there is no phase transition. In that case, the statistical ensembles are
equivalent.

\begin{figure}
\begin{center}
\includegraphics[width=0.9\linewidth,angle=0,clip]{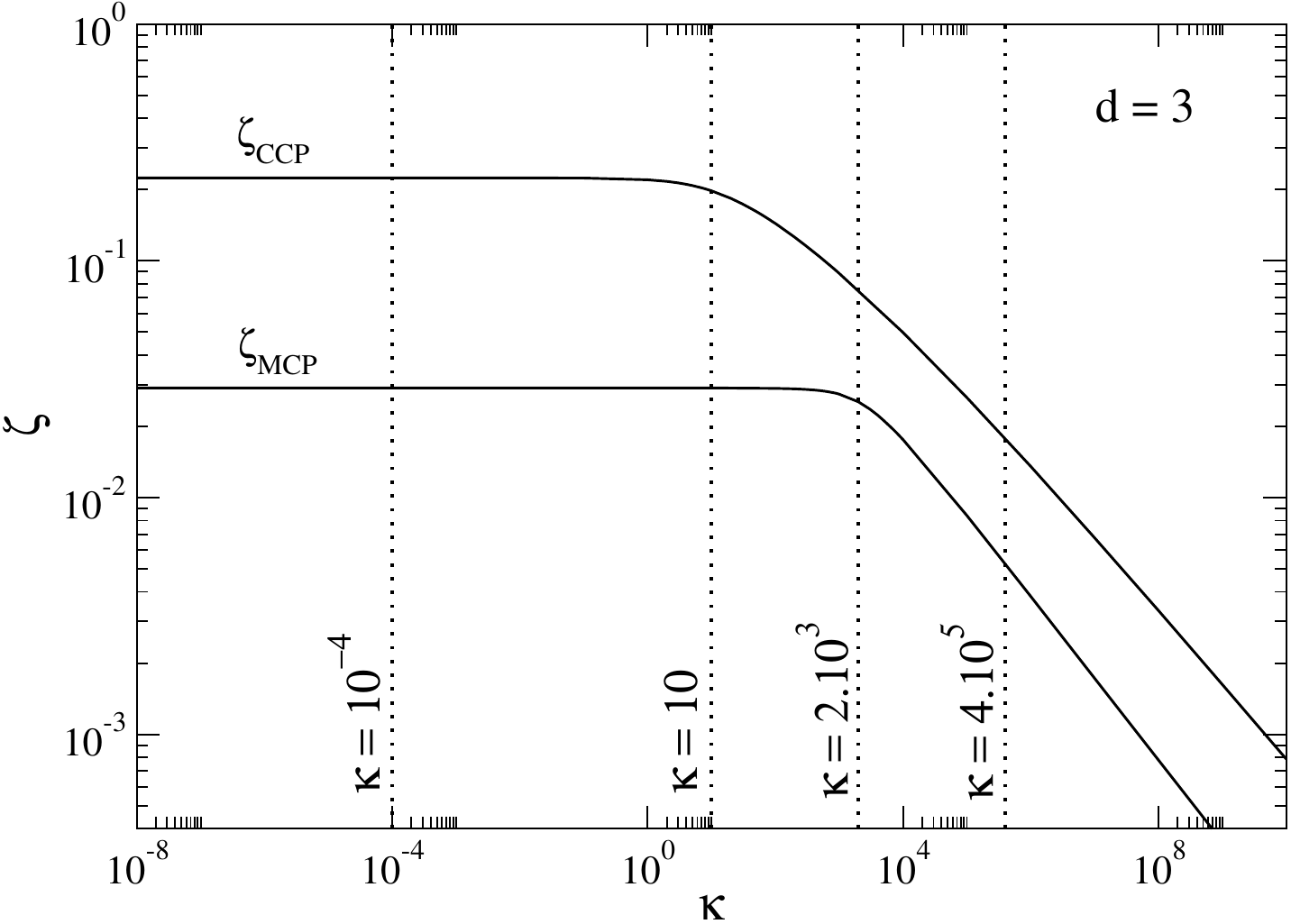}
\end{center}
\caption{\label{fig.deroulzetacc}Dependence of the
microcanonical and canonical
critical points $\zeta_{\rm MCP}$ and $\zeta_{\rm CCP}$ with
the density of the central body
$\kappa$. We have indicated for reference the values of $\kappa$ that have 
been selected to construct  the phase diagrams of Sec.~\ref{sec_cmpd} (see Figs.
\ref{fig.phaseccc} and \ref{fig.phasemccc} below).}
\end{figure}

The values of the microcanonical and canonical critical points
$R_{*}^{\rm MCP}(\rho_{*})$ and $R_{*}^{\rm CCP}(\rho_{*})$ depend on the
density $\rho_{*}$ of the central body. This dependence is represented
in Fig.~\ref{fig.deroulzetacc}. At sufficiently large densities,
the critical radii $R_{*}^{\rm MCP}(\rho_{*})$ and $R_{*}^{\rm CCP}(\rho_{*})$ 
decrease algebraically as $\rho_{*}^{-\alpha}$ with an exponent
$\alpha\sim 0.3$.

\subsection{Canonical and microcanonical  phase diagram}
\label{sec_cmpd}

Typical curves illustrating canonical and microcanonical phase 
transitions are represented in Figs.~\ref{nez} and  \ref{neck2}  respectively.
The phase diagrams of an isothermal gas with a central body can be directly
obtained from these curves by identifying characteristic energies
and characteristic
temperatures.

\begin{figure}
\begin{center}
(a)
\includegraphics[width=0.9\linewidth,angle=0,clip]
{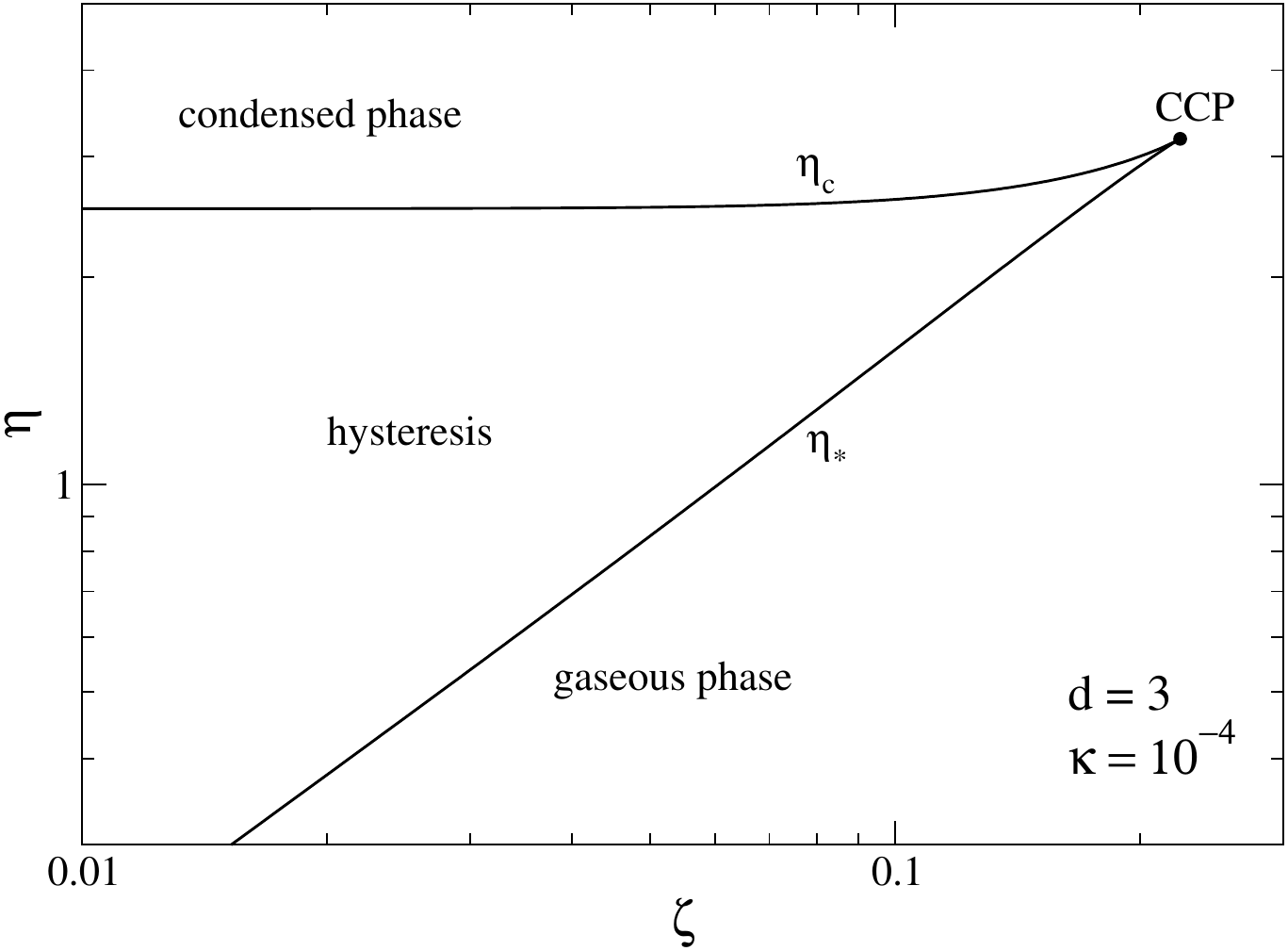}
\hspace{6pt}
(b)
\includegraphics[width=0.9\linewidth,angle=0,clip]
{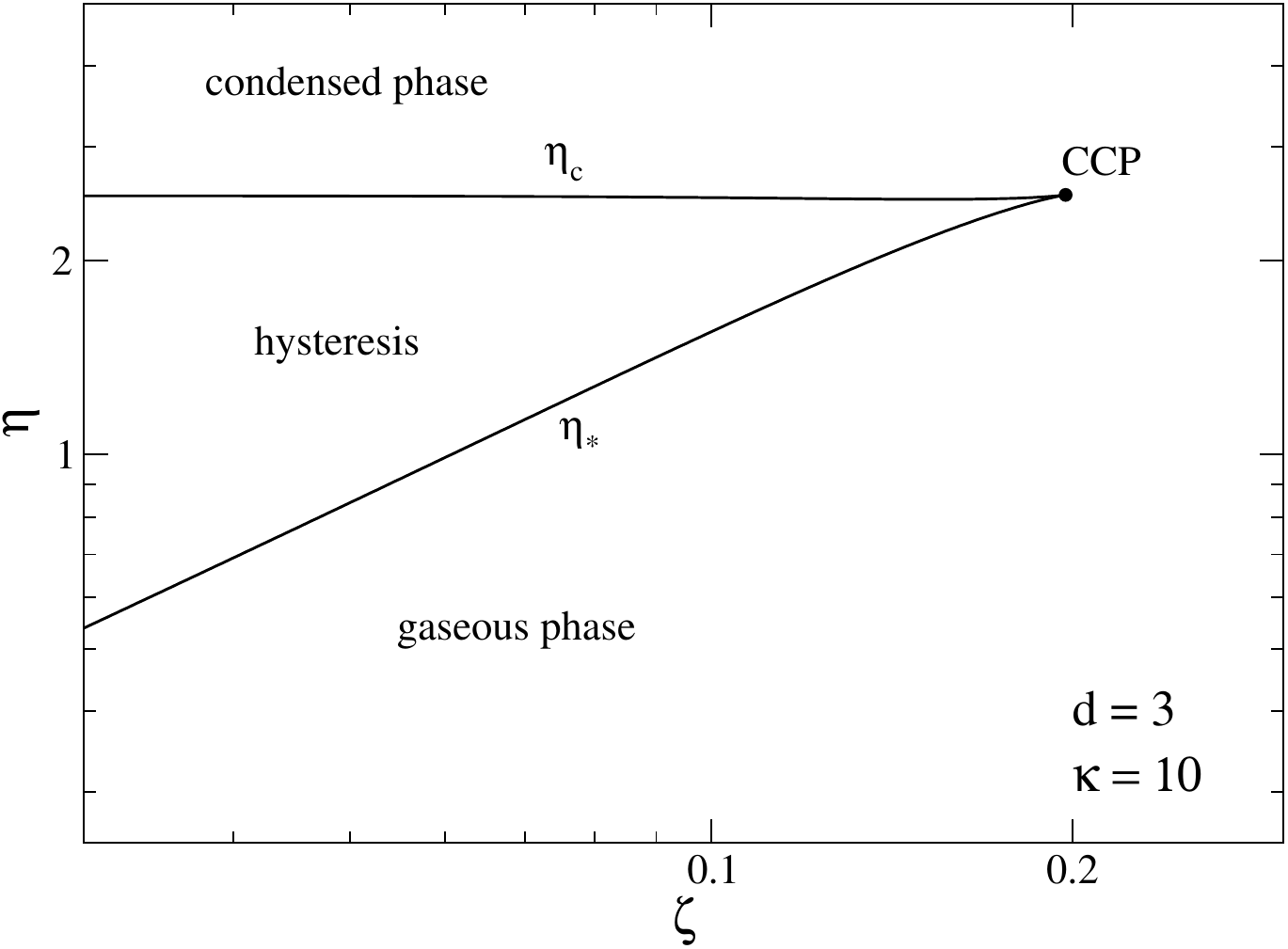}
\vspace{12pt} \\
(c)
\includegraphics[width=0.9\linewidth,angle=0,clip]
{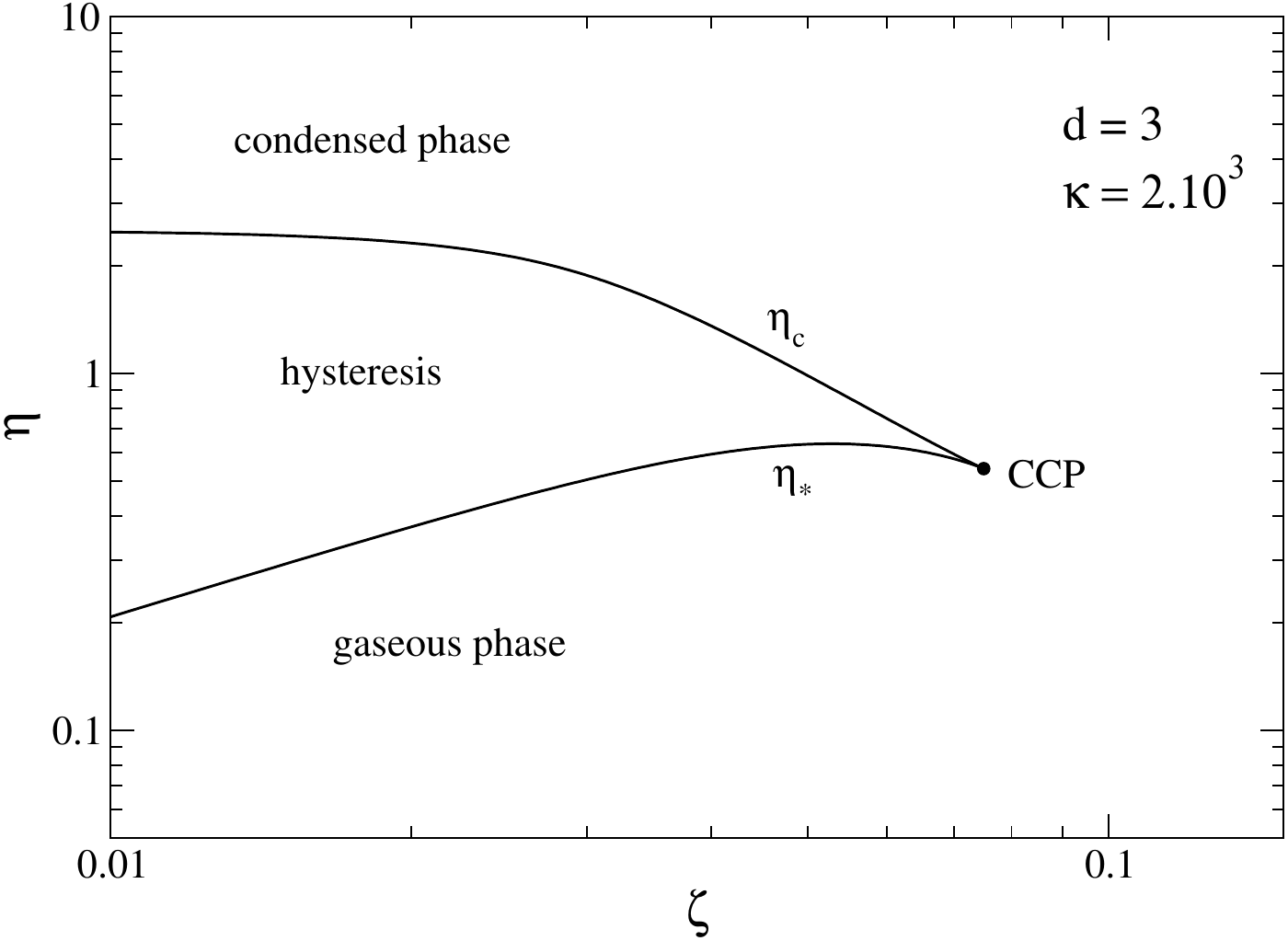}
\end{center}
\caption{\label{fig.phaseccc}Canonical phase diagrams 
in the $(\zeta,\eta)$ plane for different values of the parameter $\kappa$
indicated in Fig.~\ref{fig.deroulzetacc} (the canonical phase diagram for
$\kappa
= 4\times 10^5$ is similar to the one for $\kappa =
2\times 10^3$ so it is not represented). The region delimited by $\eta_{c}$ and
$\eta_{*}$ corresponds to an hysteretic zone where the actual phase of the
system depends on its history. If the system is initially prepared in a
gaseous state, it will remain gaseous until the minimum temperature $1/\eta_{c}$,
at which it will collapse and become condensed. Inversely, if the system is
initially prepared in a condensed state, it will remain condensed until the
maximum temperature $1/\eta_{*}$, at which it will explode and become
gaseous.}
\end{figure}

In the canonical ensemble (see Fig.~\ref{nez}), where $T$ is the control parameter, we note
$T_{c}$ (collapse temperature) the terminal point of the metastable gaseous
phase (first turning point of temperature), and $T_{*}$ (temperature of
explosion) the terminal point of the metastable condensed phase (last
turning point of temperature). These are the canonical
spinodal points. The canonical phase diagram $(R_*,T)$
is represented in Fig.~\ref{fig.phaseccc} for different values of the density
$\rho_{*}$ of the central body. For $T>T_{*}$, the system is in the
gaseous phase and for $T<T_{c}$ the system is in the condensed phase. For
$T_{*}<T<T_{c}$ it can be in one of these two phases (as a stable or a
metastable state) depending on the way it has been prepared. This is
the region where hysteretic effects take place. The two
characteristic temperatures $T_{c}$ and $T_{*}$ merge at the canonical
critical point $R_{*}=R_{*}^{\rm CCP}$. The strict phase diagram should exhibit
the transition temperature $T_t$ corresponding to the first order phase
transition (see~\cite{ijmpb} for self-gravitating fermions). However, since this
phase transition does not take place in practice, we have only represented the
physical phase diagram.

\begin{figure}
\begin{center}
(a)
\includegraphics[width=0.9\linewidth,angle=0,clip]
{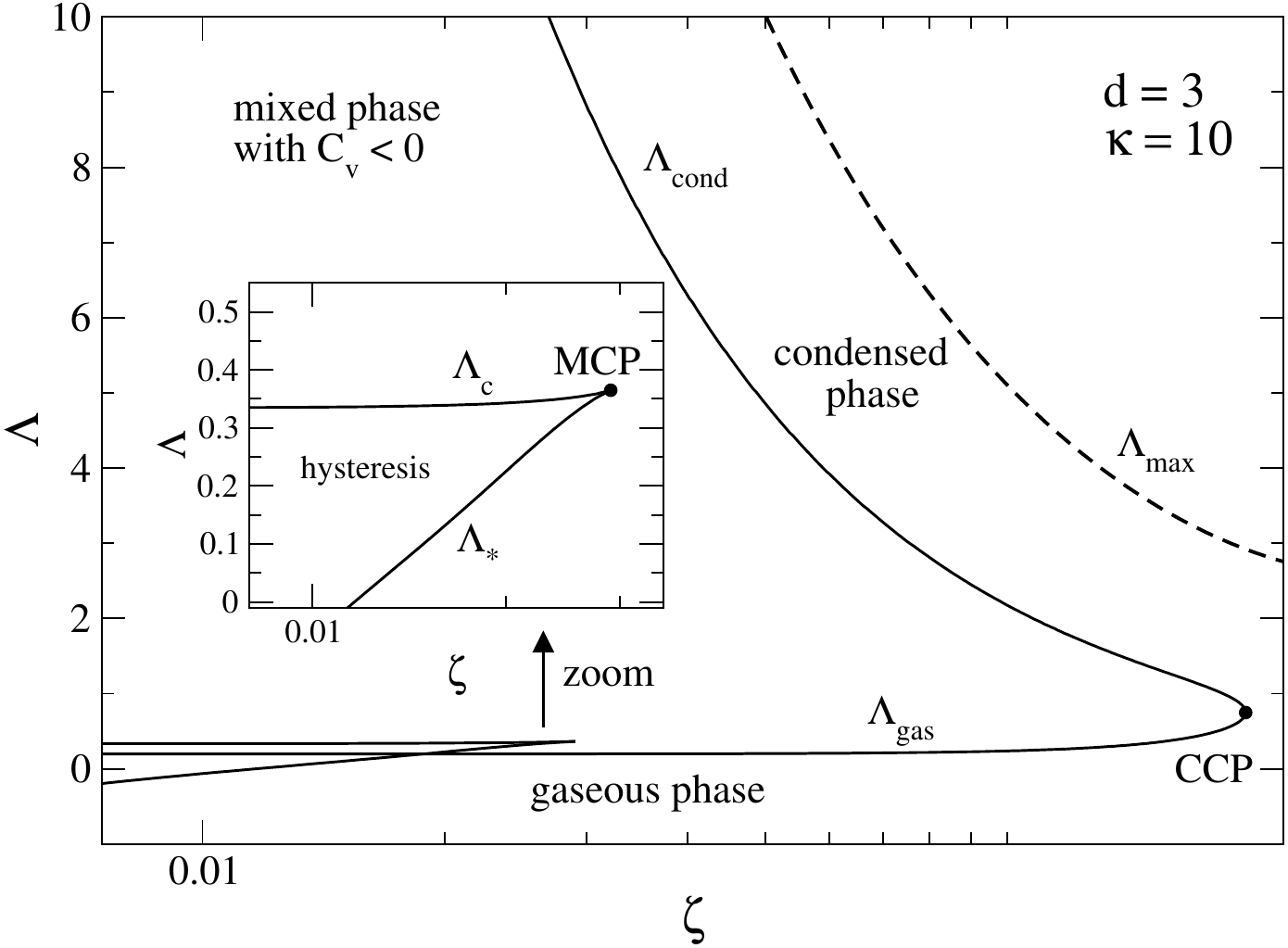}
\hspace{6pt}
(b)
\includegraphics[width=0.9\linewidth,angle=0,clip]
{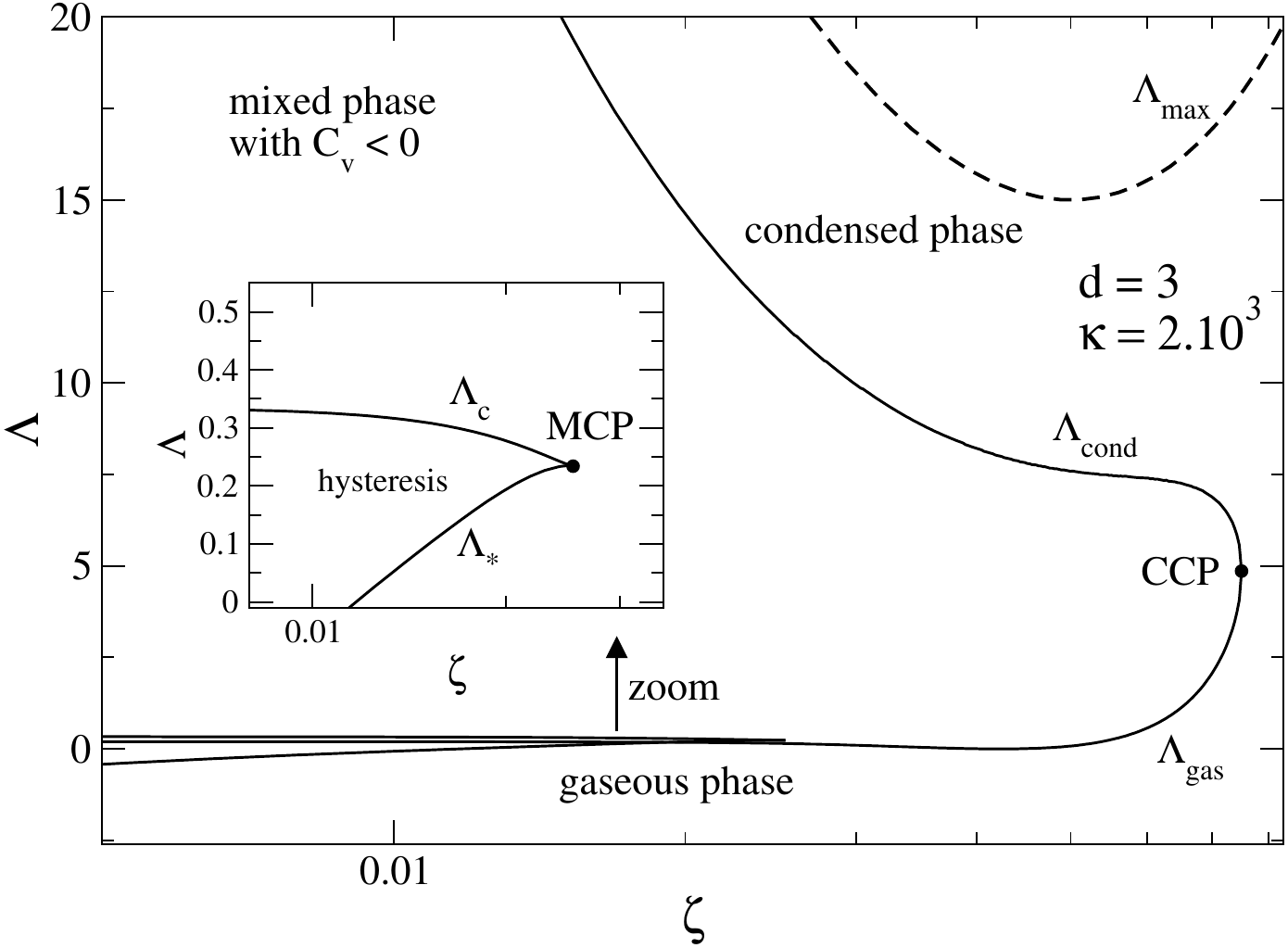}
\vspace{12pt} \\
(c)
\includegraphics[width=0.9\linewidth,angle=0,clip]
{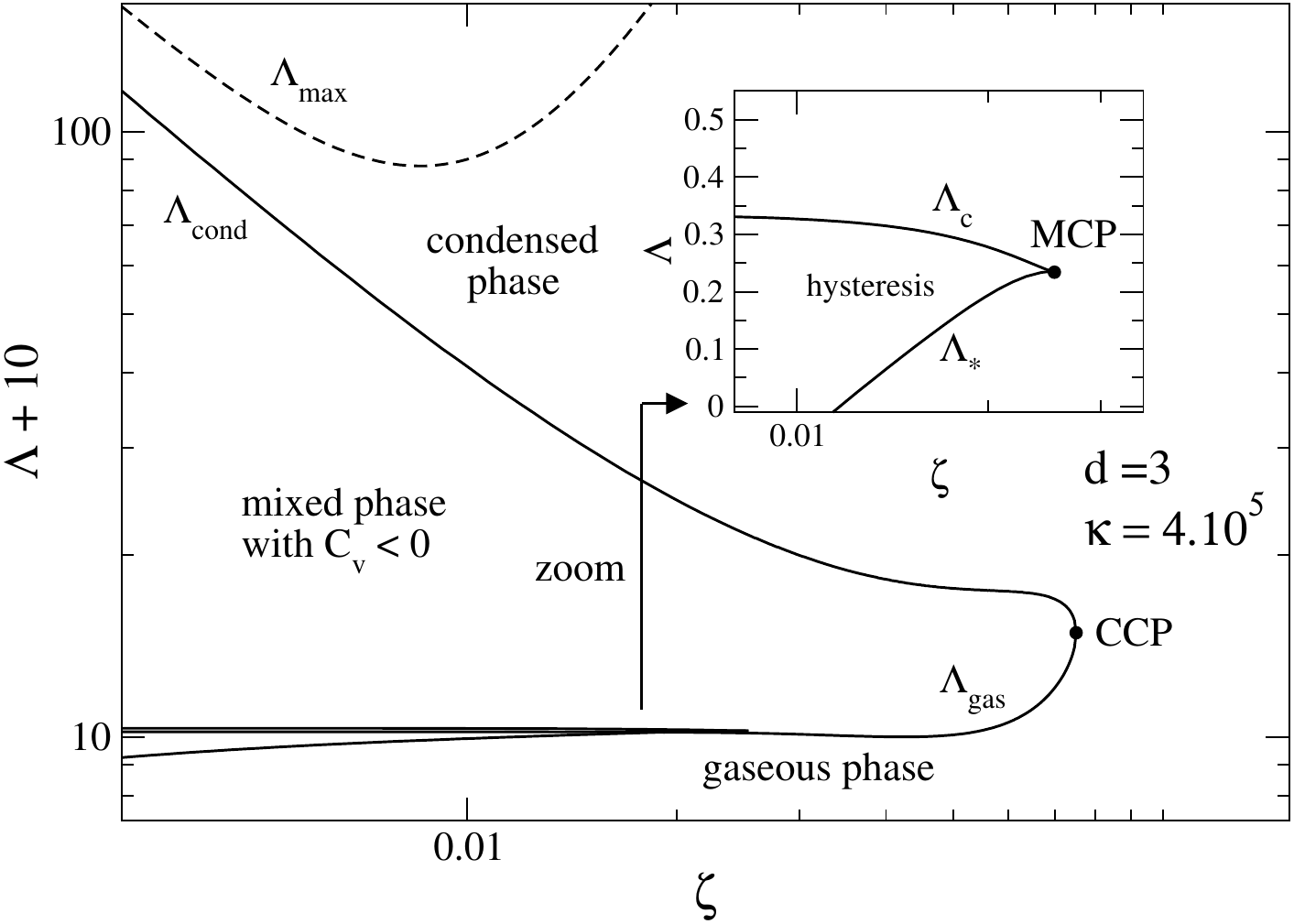}
\end{center}
\caption{\label{fig.phasemccc} Microcanonical phase diagrams in the
$(\zeta,\Lambda)$ 
plane for different values of the parameter $\kappa$ indicated in Fig.
\ref{fig.deroulzetacc} (the microcanonical phase diagram for
$\kappa = 10^{-4}$ is similar to the one for $\kappa =10$ so it is not
represented). The region delimited by $\Lambda_{c}$ and $\Lambda_{*}$
corresponds to an hysteretic zone where the actual phase of the system depends
on its history. If the system is initially prepared in a gaseous state, it
will remain gaseous until the minimum energy $-\Lambda_{c}$, at
which it will collapse and become condensed. Inversely, if the system is
initially prepared in a condensed state, it will remain condensed until the
maximum energy $-\Lambda_{*}$, at which it will explode and
become gaseous.}
\end{figure}

In the microcanonical ensemble (see Fig.~\ref{neck2}), where $E$ is the control parameter, we
note $E_{c}$ (collapse energy) the end point of the metastable gaseous
phase (first turning point of energy), and $E_{*}$ (energy of
explosion) the end point of the metastable condensed phase (last
turning point of energy). These are the microcanonical
spinodal points. We also note $E_{\rm min}$ the
minimum energy.
The microcanonical phase diagram $(R_*,E)$ is represented in Fig.~\ref{fig.phasemccc}
for different values of the density $\rho_{*}$ of the central
body. For $E>E_{*}$, the system is in the gaseous phase and for
$E_{\rm min}<E<E_{c}$ the system is in the condensed phase. For $E_{*}<E<E_{c}$
it can
be in one of these two phases (as a stable or a metastable state)
depending on the way it has been prepared. This is the region where
hysteretic effects take place. The two characteristic energies
$E_{c}$ and $E_{*}$ merge at the microcanonical critical point
$R_{*}=R_{*}^{\rm MCP}$. To complete the diagram, we have also denoted
$E_{\rm gas}$ the energy at which we enter in the region of negative
specific heats (first turning point of temperature) and $E_{\rm cond}$ the
energy at which we leave the region of negative specific heats (last
turning point of temperature), see Fig.~\ref{nshape}. These two characteristic energies
$E_{\rm gas}$ and $E_{\rm cond}$ merge at the canonical critical point
$R_{*}=R_{*}^{\rm CCP}$. The strict phase diagram should exhibit the transition
energy $E_t$ corresponding to the first order phase transition (see~\cite{ijmpb}
for self-gravitating fermions). However, since this phase transition does not
take place in practice, we have only represented the physical phase diagram.

We note that the microcanonical phase diagram is richer than the canonical phase
diagram due to the existence of a negative specific heat region. We recall that
canonical stability implies microcanonical stability but the converse is false
in case of ensemble inequivalence~\cite{cc}.
Therefore, canonical stability is a
sufficient but not necessary condition of microcanonical stability.  Since
canonical equilibria are always realized as microcanonical equilibria, they
constitute a sub-domain of the microcanonical phase diagram. In particular, the
states with energy between $E_{\rm gas}$ and $E_{\rm cond}$, that
have negative specific heats, are not accessible  in the canonical ensemble
(they are unstable
saddle points of free energy) while they are accessible in the microcanonical
ensemble (they are entropy
maxima at fixed mass and energy). Therefore, this region corresponds to a domain
of ensemble inequivalence.

\subsection{The case of a denser and denser central body and the case of a
central Dirac mass}

\begin{figure}
\begin{center}
\includegraphics[width=0.9\linewidth,angle=0,clip]{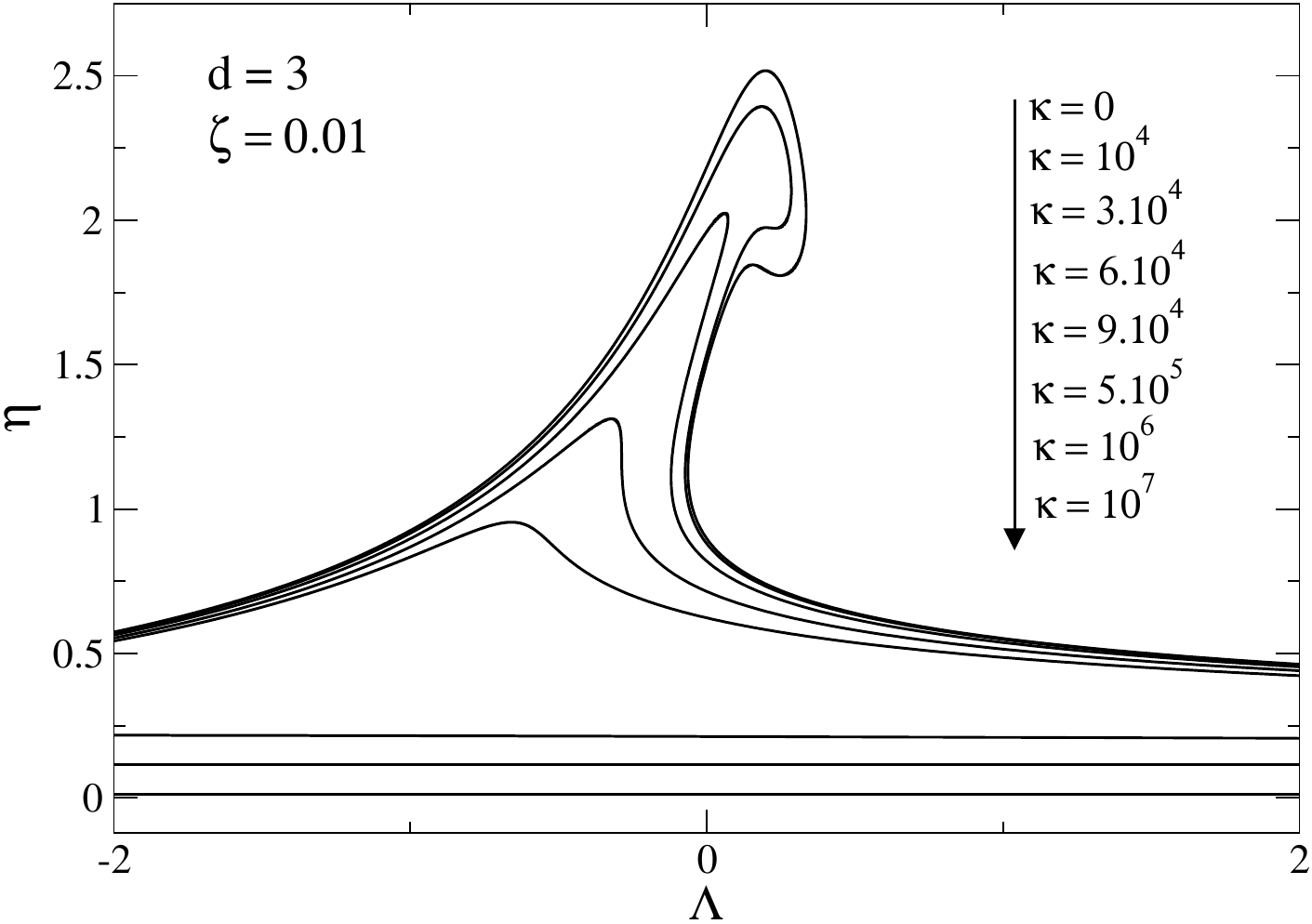}
\end{center}
\caption{\label{fig.condensecc}Deformation of the series of equilibria as we progressively increase the density $\kappa$ of the central
body for a fixed value of the core radius $\zeta = 0.01$.}
\end{figure}

\begin{figure}
\begin{center}
\includegraphics[width=0.9\linewidth,angle=0,clip]{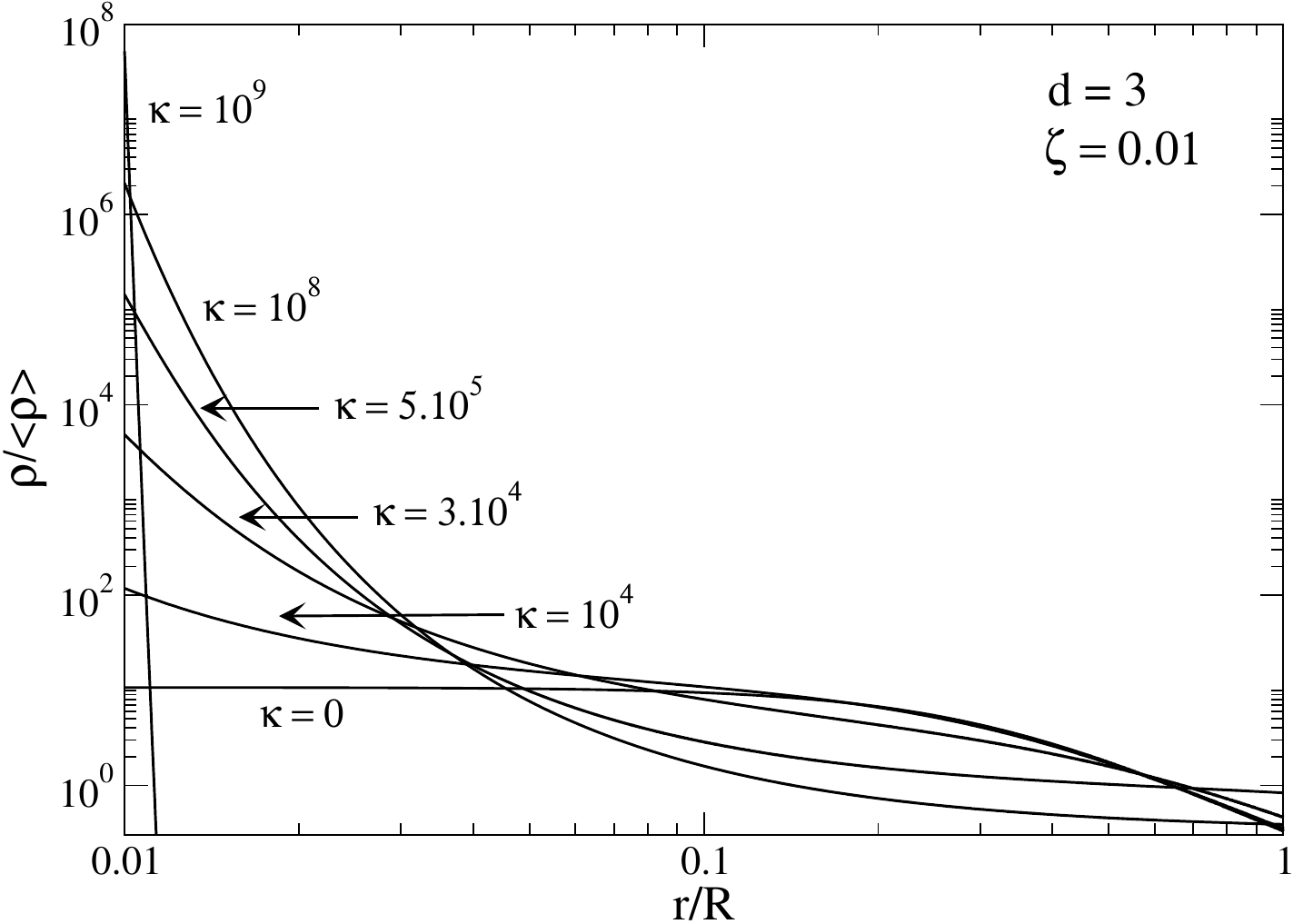}
\end{center}
\caption{\label{fig.prof} Density profiles corresponding to  the critical temperature $\eta = \eta_c$ as one increases the density of the central body. As $\kappa$
increases, the self-gravitating particles are more and more condensed
around the central body.}
\end{figure}

We consider how the series of equilibria changes when
the radius of the central body $R_{*}$ is fixed and its mass $M_*$ is
progressively increased so that the density
$\rho_*$ of the central body is larger and larger. 
The series of equilibria are represented in Fig.~\ref{fig.condensecc}. As the
density
$\rho_*$ increases, the collapse temperature $T_{c}$ and the collapse
energy $E_{c}$ increase, so that instability occurs
sooner. Typical density profiles at the verge of the canonical instability
($T=T_{c}$) are represented in Fig.~\ref{fig.prof}. The cusp is more and more
pronounced
as the density of the central body increases.

\begin{figure}
\begin{center}
\includegraphics[width=0.9\linewidth,angle=0,clip]{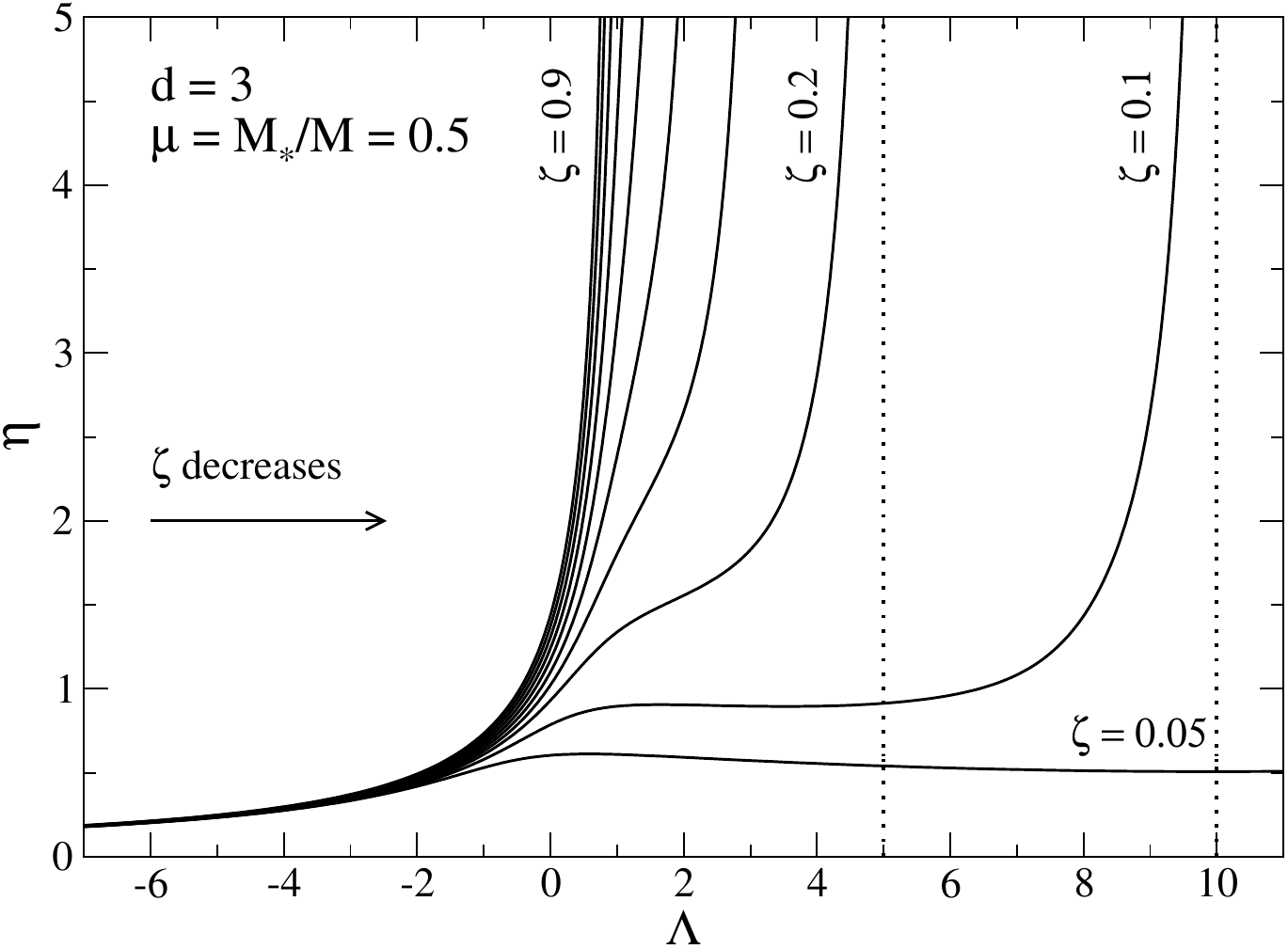}
\end{center}
\caption{\label{etalambda3dmu05} Caloric curves at fixed central mass
$\mu$ (we have taken
$\mu=0.5$ for illustration) for several values of the central body radius
$\zeta$. The case $\zeta\rightarrow 0$ corresponds to a central Dirac mass for
which there is no equilibrium state.}
\end{figure}

Finally, we consider the case where the mass $M_*$ of the central body is fixed
and its radius $R_*$ is progressively reduced. The corresponding caloric curves
are plotted in Fig.~\ref{etalambda3dmu05}.  For $R_*\rightarrow 0$, the central
body tends to a Dirac mass. However, we know from Sec.~\ref{sec_st} that there is no
equilibrium state in that case in $d=3$. This is why the caloric curves do not
tend to a well-defined limit for $R_*\rightarrow 0$.

\section{Caloric curves in the presence of a central body in $d=2$ and $d=1$
dimensions}
\label{sec_pt12}

In this section, we briefly describe the caloric curves of a classical
isothermal self-gravitating gas in the presence of a central body in $d=2$ and
$d=1$ dimensions. These caloric
curves and the corresponding density profiles can be obtained
analytically~\cite{companion}. In the figures, we fix the
mass $M_*$ of the central body and decrease its radius $R_*$, approaching
thereby a central Dirac mass when $R_*\rightarrow
0$.\footnote{In the biological problem, which is particularly justified in $d=1$
and
$d=2$ dimensions (see Appendix \ref{sec_analogy}), it is more relevant to
fix the mass of the chemoattractant rather than its density.}

We consider only
box-confined systems and refer to~\cite{companion} for complementary results.

\begin{figure}
\begin{center}
\includegraphics[width=0.9\linewidth,angle=0,clip]{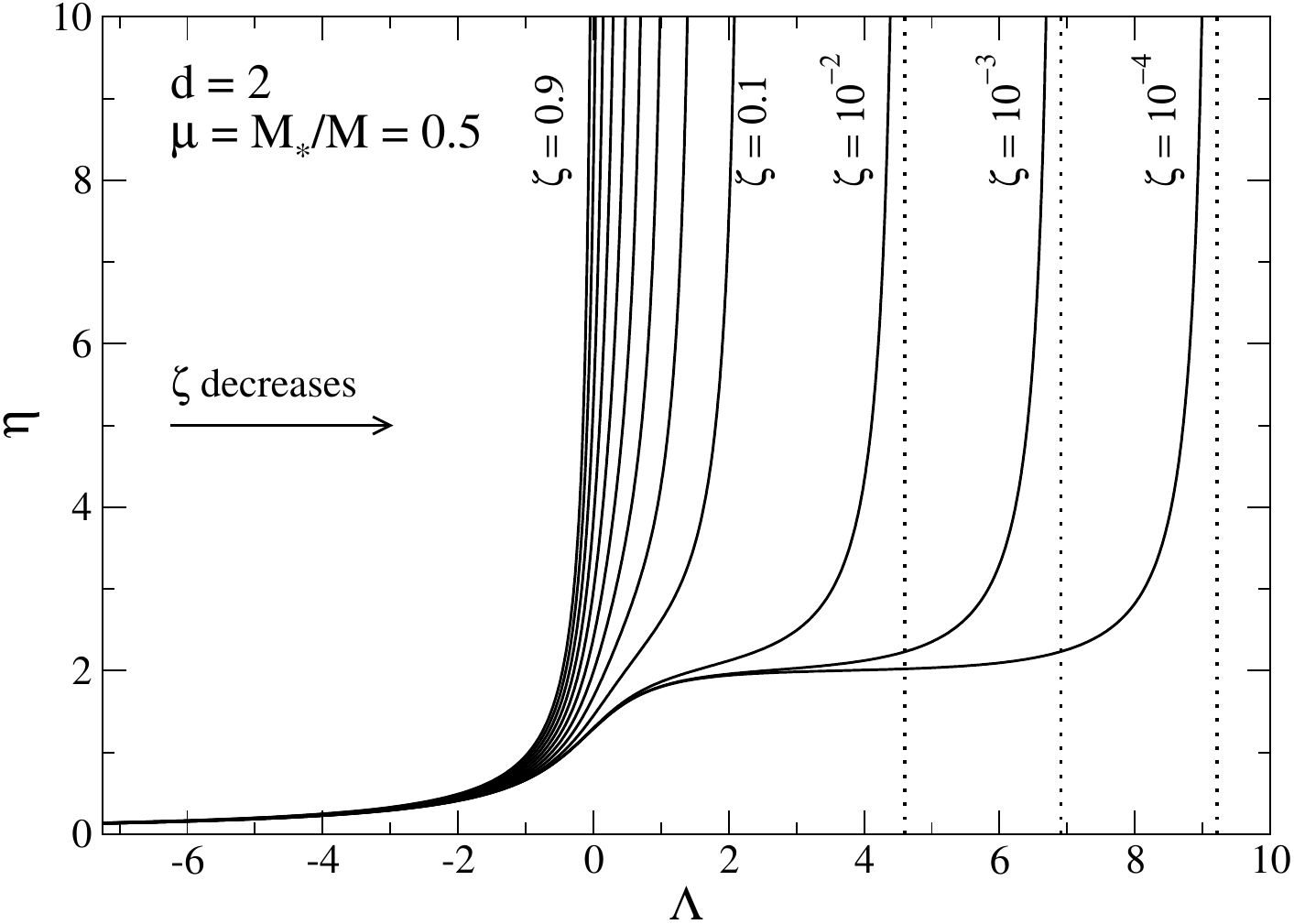}
\end{center}
\caption{\label{deux} Caloric curves at fixed central mass $\mu$ (we have taken
$\mu=0.5$ for illustration) for several values of the central body radius
$\zeta$. For small $\zeta$ (corresponding to a central Dirac mass), one observes the
appearance of the critical temperature $\eta_c=4/(1+2\mu)=2$.}
\end{figure}

The caloric curve of a classical self-gravitating gas without central body in
$d=2$ dimensions has been discussed by Katz and Lynden-Bell~\cite{klb} and Sire
and Chavanis~\cite{sc}. The caloric curve $\beta(E)$ is monotonic (which implies
thermodynamical stability) but the temperature tends to a constant $k_B T_c=
GMm/4$ when $E\rightarrow -\infty$. Stable equilibrium state exists for all
energies $E$ in the microcanonical ensemble, but only for temperatures
$T>T_c$ in the canonical ensemble. For $T\le T_c$ the system collapses and
ultimately forms a Dirac peak containing all the mass~\cite{sc}. In the presence
of a central body, the caloric curve is plotted in Fig.~\ref{deux}. There
is a minimum energy $E_{\rm min}$ at which the gas is concentrated on
the surface of the solid body (see Appendix \ref{sec_lmax}). A stable
equilibrium state exists for all energies $E\ge E_{\rm min}$ in the
microcanonical ensemble and for all temperatures $T\ge 0$ in the canonical
ensemble. As $R_*$ decreases, the minimum energy $E_{\rm min}$ is pushed towards
more and more negative values and a plateau forms at a critical
temperature $k_{B}T_{c}=(GMm/4)(1+2{M_{*}}/{M})$ modified by the presence of the
central Dirac mass $M_*$~\cite{companion}. These results are similar to
those obtained for a Fermi gas in $d=2$ dimensions
\cite{ptd,exclusion,kmcfermions}.

\begin{figure}
\begin{center}
\includegraphics[width=0.9\linewidth,angle=0,clip]{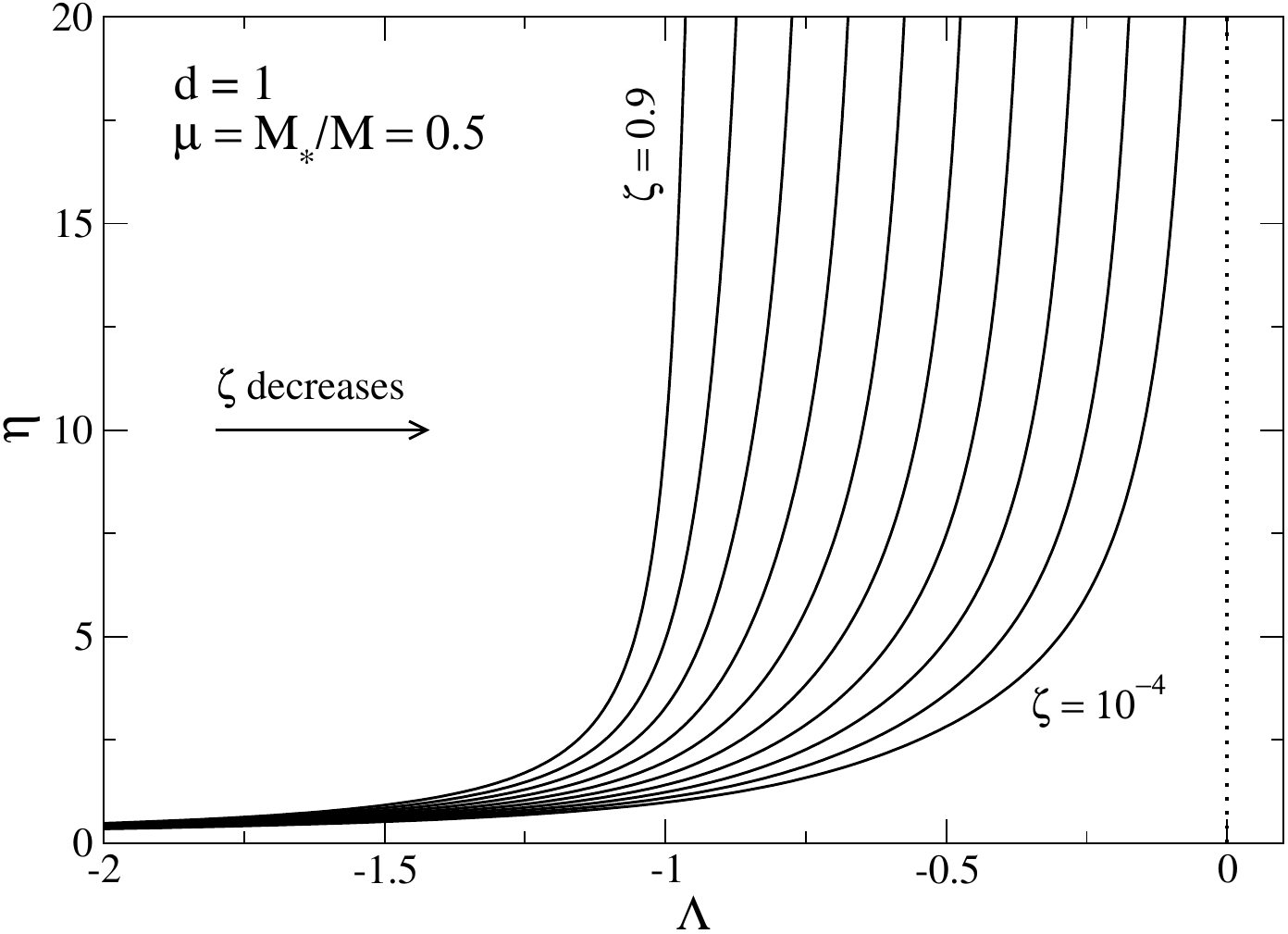}
\end{center}
\caption{\label{un} Caloric curves at fixed central mass
$\mu$ (we have taken
$\mu=0.5$ for illustration) for several values of the central body radius
$\zeta$. For small $\zeta$ (corresponding to a central Dirac mass), the minimum
energy approaches the absolute minimum
value $\Lambda=0$.}
\end{figure}

The caloric curve of a classical self-gravitating gas without central body  in
$d=1$ dimension has been discussed by Katz and Lecar~\cite{kl} and Sire and
Chavanis~\cite{sc}. The caloric curve $\beta(E)$ is monotonic (which implies
thermodynamical stability). Stable equilibrium state exists for all
accessible energies $E\ge 0$ in the microcanonical ensemble, and for all
temperatures
$T\ge 0$ in the canonical ensemble. There are no mechanical instabilities. 
In the presence of a central body, the caloric curve is plotted in Fig.
\ref{un}. There
is a minimum energy $E_{\rm min}>0$ at which the gas is concentrated on
the surface of the solid body (see Appendix \ref{sec_lmax}). A stable
equilibrium
state exists for all energies $E\ge E_{\rm min}$ in the microcanonical ensemble,
and for all temperatures $T\ge 0$ in the canonical ensemble. As $R_*$ decreases, the
minimum energy is pushed towards $E=0$. These results
are similar to those obtained for a Fermi gas in $d=1$
\cite{ptd,exclusion,kmcfermions}.

\section{Conclusion}
\label{sec.conclusion}

In this work, we have studied the statistical equilibrium states of a
classical self-gravitating gas around a central body. The central body can
represent a planetary core or mimic a black hole at the center of a galaxy or at
the center of a globular cluster. The gas is described by the Boltzmann
distribution which is the fundamental distribution function predicted by
statistical
mechanics. Like in previous studies, we must
enclose the system within a box in order to have a well-defined equilibrium
(maximum entropy state). Otherwise the gas evaporates and there is no
statistical equilibrium state in a strict sense.

We have studied the phase transitions of this system in both
microcanonical (fixed energy) and canonical (fixed temperature) ensembles in
different dimensions of space. In $d=3$ dimensions, and for sufficiently large
systems, we have evidenced both microcanonical and canonical phase
transitions.
Below a critical energy in the microcanonical ensemble the atmosphere
experiences a gravothermal
catastrophe and below a critical temperature in the canonical ensemble it
experiences an isothermal
collapse. In the presence of a central body, the collapse stops when the gas
comes into contact with the central body and forms a thin layer (spike) around
it. This leads to a  zeroth order
phase transition between a dilute (gaseous) phase and a condensed phase
with a ``cusp-halo'' structure. For intermediate-size systems, there are only
canonical phase transitions (of zeroth order) and for small systems there is no
phase transition at all. In $d=2$ dimensions, there is no phase
transition in a strict sense (no discontinuity of any thermodynamical parameter)
but, for sufficiently large systems, a plateau temperature indicates a
rapid change between a dilute phase and a condensed phase. In $d=1$ dimension,
there is no phase transition. In $d=3$ dimensions there are regions of ensemble
inequivalence (except for small systems) associated with negative specific
heats while in $d=2$ and $d=1$ dimensions the specific heat is positive and the
ensembles
are equivalent. These results are similar to those obtained previously for
self-gravitating fermions~\cite{ijmpb}. There is, however, a structural
difference.
In the case of fermions, the small-scale regularization is intrinsic to the
quantum system (it arises from the Pauli exclusion principle contained in the
Fermi-Dirac distribution) while, in the present case, the distribution function
is classical and
the small-scale regularization is external to the system (it is due to the
finite radius of the central body).

Of course, the box is artificial and we can have two points of view on this
problem. If we are mainly interested in statistical mechanics, it is relevant to
use the Boltzmann distribution (which is the fundamental distribution function 
of
statistical mechanics) and introduce a box to make the
problem
well-defined mathematically. In this sense, our thermodynamic
approach is rigorously justified. If we are rather interested in astrophysical
applications, we can use a truncated Boltzmann
distribution, e.g., the King~\cite{king} distribution, to take into account the
evaporation
of the particles and avoid the artificial box. Although we gain in realism, the
problem becomes less well-justified from a statistical mechanics point of view
because we now have to deal with an out-of-equilibrium problem. We can
nevertheless try to apply a thermodynamical
approach as discussed by Katz~\cite{katzking} and Chavanis {\it
et al.}~\cite{clm1} for the classical King model. The study of phase transitions
in the
fermionic King model~\cite{stella,chavmnras} has been performed in
\cite{clm2,adky} both for
nonrelativistic and relativistic systems, giving results qualitatively similar
to those obtained in a box~\cite{ijmpb,calettre,acepjb}. Recently, Bonsor {\it
et al.}~\cite{bonsor} have considered the equilibrium states of a loaded
King distribution around a central body. In this study, the central body models
a black hole at the center of a globular cluster. They report a transition
between equilibria which are dominated by the mass of the host stellar
system or by the mass of the central black hole. Their results are complementary
to those obtained for the classical and fermionic King models studied in~\cite{clm1,clm2} and for the box-confined classical isothermal gas surrounding a
central body considered in the present article. These works provide a
rather complete
picture of gravitational phase transitions in these different contexts.


\begin{appendix}

\section{Analogy between self-gravitating systems and bacterial populations}
\label{sec_analogy}

In this Appendix, we point out some analogies between a
gas of self-gravitating Brownian particles~\cite{crs,sc,post} and the chemotaxis
of bacterial
populations~\cite{murray} (see, e.g.,~\cite{crrs,cskin,chemocoll,kss,cd} for
more details on this analogy). The name chemotaxis refers to the motion
of organisms induced by chemical signals. In some cases, the
biological organisms
secrete a substance (chemoattractant, pheromone) that has an attractive effect on the organisms
themselves.
Therefore, in addition to their diffusive motion, they move systematically along
the gradient of concentration of the chemical they secrete (chemotactic flux).
When attraction prevails over diffusion, the chemotaxis can trigger a
self-accelerating process (chemotactic collapse) until a point at which
aggregation takes place. This
is the case for the slime mold {\it Dictyostelium discoideum} and for the
bacteria {\it Escherichia coli}. The Keller-Segel~\cite{ks} model describing the
chemotaxis of biological populations can be written as
\be
\label{ana1}
\frac{\partial\rho}{\partial t}=\nabla\cdot(D\nabla\rho-\chi\rho\nabla c),
\ee
\be
\label{ana2}
\frac{1}{D'}\frac{\partial c}{\partial
t}=\Delta c-k^2 c+\lambda\rho,
\ee
where $\rho({\bf r},t)$ is the concentration of the
biological species (e.g., bacteria) and $c({\bf r},t)$ is the concentration of
the secreted chemical. The bacteria diffuse with a diffusion coefficient $D$ and
undergo a chemotactic drift with strength $\chi$ along the gradient
of chemical. The chemical is produced by the bacteria at a
rate $D'\lambda$, is degraded at a rate $D'k^2$, and diffuses with a
diffusion coefficient $D'$. In the limit of large diffusivity of the chemical
$D'\rightarrow +\infty$ and a vanishing degradation rate $k^2=0$, the
reaction-diffusion equation (\ref{ana2}) reduces to the Poisson
equation $\Delta c=-\lambda\rho$ (see Appendix C of~\cite{cskin} for more
details). In that case, the Keller-Segel
model becomes isomorphic to the Smoluchowski-Poisson equations describing
self-gravitating Brownian particles in a high
friction limit $\xi\rightarrow
+\infty$~\cite{crs,sc,post}: 
\be
\label{ana3}
\xi\frac{\partial\rho}{\partial t}=\nabla\cdot \left
(\frac{k_B T}{m}\nabla\rho+\rho\nabla \Phi\right ),
\ee
\be
\label{ana4}
\Delta\Phi=S_d G\rho.
\ee
In this analogy, the concentration of chemical $c({\bf
r},t)$, plays the role of minus the gravitational potential $-\Phi({\bf r},t)$.
The effective statistical ensemble associated
with the Keller-Segel model is the canonical ensemble.\footnote{The
drift-diffusion equation (\ref{ana1}) and the Smoluchowski equation (\ref{ana3})
can be interpreted as Fokker-Planck equations in position space describing an
overdamped dynamics. In the Brownian picture, the particles are in contact with
a heat bath fixing the temperature.} The
steady states of the Keller-Segel model are of the form $\rho=Ae^{(\chi/D)c}$
which is similar to the Boltzmann distribution $\rho=Ae^{-\beta m\Phi}$ with an
effective temperature $T_{\rm eff}=D/\chi$ given by a form of Einstein relation.
The Lyapunov functional
associated with the Keller-Segel model is $F=-\frac{1}{2}\int\rho c\, d{\bf
r}+\frac{D}{\chi}\int\rho\ln\rho\, d{\bf r}$. It is similar to a free energy
$F=E-TS$ in
thermodynamics, where $E$ is the energy and $S$ is the Boltzmann entropy. The
Keller-Segel model conserves the mass and satisfies an $H$-theorem for the
free energy, i.e., $\dot F\le 0$ with an equality if, and only if, $\rho$
is the Boltzmann distribution discussed above. Furthermore, the Boltzmann
distribution is dynamically stable if, and only if, it is a
local minimum of free energy at fixed mass. In that context,
the minimization problem $\min_{\rho}\lbrace F[\rho] |
M[\rho]=M\rbrace$
determines a steady state of the Keller-Segel model that is dynamically
stable. This is similar to a condition of thermodynamical
stability in the canonical ensemble.

\section{The virial theorem}
\label{sec_vt}

In this Appendix we derive the equilibrium scalar virial theorem for a
self-gravitating system in the presence of a central body in $d$ dimensions.
We refer to Appendix G of~\cite{ggpp} and Appendix B of~\cite{ggppbh} for more
details and
generalizations.

\subsection{The virial of the gravitational
force produced by the central body}

Assuming that the central body is spherically symmetric, and using 
Newton's law (see
Appendix \ref{sec_pfd}), the gravitational field
that it creates in $r\ge R_*$ is 
\be
\label{dev}
-\nabla\Phi_{\rm ext}=-\frac{GM_*}{r^{d-1}}{\bf e}_r\quad {\rm
or}\quad\frac{d \Phi_{\rm ext}}{dr}= \frac{GM_{*}}{r^{d-1}}.
\ee
In particular, we have
\be
\label{lg}
\frac{d \Phi_{\rm ext}}{dr}(R_*)= \frac{GM_{*}}{R_*^{d-1}}.
\ee
The corresponding gravitational potential in $r\ge R_*$ is
\be
\label{pcore1}
\Phi_{\rm ext}=-\frac{1}{d-2}\frac{GM_*}{r^{d-2}}\quad (d\neq 2),
\ee
\be
\label{pcore2}
\Phi_{\rm ext}=GM_*\ln \left (\frac{r}{R}\right )\quad (d= 2).
\ee
This is the solution of the Laplace equation
\be
\Delta\Phi_{\rm ext}=0.
\ee
The gravitational energy of the gas in the potential created by the central
body is
\be
W_{\rm ext}=\int \rho\Phi_{\rm ext}\, d{\bf r}.
\ee
Using Eqs.~(\ref{pcore1}) and (\ref{pcore2}) we get
\be
\label{dev3}
W_{\rm ext}=-\frac{1}{d-2}\int \rho \frac{GM_*}{r^{d-2}}\, d{\bf
r}\quad (d\neq 2),
\ee
\be
W_{\rm ext}= GM_*\int \rho \ln \left (\frac{r}{R}\right )\, d{\bf
r}\quad (d\neq 2).
\ee

The virial of the external force is defined by
\be
\label{vir1}
W^{\rm ext}_{ii}=-\int \rho {\bf r}\cdot\nabla\Phi_{\rm ext}\, d{\bf
r}.
\ee
Using Eq.~(\ref{dev}), we get
\be
\label{ad}
W^{\rm ext}_{ii}=-\int \rho \frac{GM_*}{r^{d-2}}\, d{\bf
r}.
\ee
For $d\neq 2$, comparing Eqs.~(\ref{dev3}) and (\ref{ad}) we find that
\be
\label{all2}
W^{\rm ext}_{ii}=(d-2)W_{\rm ext}.
\ee
For $d=2$, Eq.~(\ref{ad}) yields
\be
\label{deu2}
W^{\rm ext}_{ii}=-GM_*M.
\ee

\subsection{The virial of the gravitational force produced by the gas}

The gravitational field produced by the gas is
\be
\label{epo}
-\nabla\Phi({\bf r})=-G\int \rho({\bf r}')\frac{{\bf r}-{\bf
r}'}{|{\bf r}-{\bf
r}'|^{d}}\, d{\bf r}'.
\ee
The corresponding gravitational potential is
\be
\label{vi1}
\Phi({\bf r})=-\frac{G}{d-2}\int \frac{\rho({\bf r}')}{|{\bf r}-{\bf
r}'|^{d-2}}\, d{\bf r}' \qquad (d\neq 2),
\ee
\be
\label{vi2}
\Phi({\bf r})=G\int \rho({\bf r}')\ln \frac{|{\bf r}-{\bf r}'|}{R}\, d{\bf
r}' \qquad (d=2).
\ee 
This is the solution of the Poisson equation
\be
\Delta\Phi=S_d G\rho.
\ee
The gravitational energy  of the gas  due to the interaction of the
particles
between themselves is
\be
W=\frac{1}{2}\int \rho\Phi\, d{\bf r}.
\ee
Using  Eqs.~(\ref{vi1}) and (\ref{vi2}) we get
\be
\label{tim}
W=-\frac{1}{2}\frac{G}{d-2}\int \frac{\rho({\bf r})\rho({\bf
r}')}{|{\bf r}-{\bf
r}'|^{d-2}}\, d{\bf r}d{\bf r}'\qquad (d\neq 2),
\ee
\be
W=\frac{1}{2}G\int {\rho({\bf r})\rho({\bf r}')}\ln \frac{|{\bf r}-{\bf
r}'|}{R}\, d{\bf
r}d{\bf r}'\qquad (d=2).
\ee

The virial of the gravitational force produced by the gas is
\be
\label{mic}
W_{ii}=-\int \rho {\bf r}\cdot\nabla\Phi\, d{\bf r}.
\ee
Substituting Eq.~(\ref{epo}) into Eq.~(\ref{mic}), we get
\be
\label{ca}
W_{ii}=-G\int \rho({\bf r})\rho({\bf r}')\frac{{\bf r}\cdot ({\bf r}-{\bf
r}')}{|{\bf r}-{\bf r}'|^{d}}\, d{\bf r}d{\bf r}'.
\ee
Interchanging the dummy variables ${\bf r}$ and ${\bf r}'$
and adding the
resulting expression to Eq.~(\ref{ca}), we obtain 
\be
\label{pra}
W_{ii}=-\frac{1}{2}G\int \frac{\rho({\bf r})\rho({\bf r}')}{|{\bf r}-{\bf
r}'|^{d-2}}\, d{\bf r}d{\bf r}'.
\ee
For $d\neq 2$, comparing Eqs.~(\ref{tim}) and (\ref{pra}) we find that
\be
\label{all}
W_{ii}=(d-2)W.
\ee
For $d=2$, Eq.~(\ref{pra}) yields
\be
\label{deu}
W_{ii}=-\frac{GM^2}{2}.
\ee

These results are valid for an arbitrary distribution of the particles of the
gas. In particular, for $d\neq 2$, we have
\be
\label{nuit}
W=-\frac{1}{d-2}\int \rho {\bf r}\cdot\nabla\Phi\, d{\bf r}.
\ee
Now, for a spherically symmetric system, according to Newton's law (see Appendix
\ref{sec_pfd}),
we have
\begin{eqnarray}
\label{virial25q}
-\nabla\Phi=-\frac{GM(r)}{r^{d-1}}{\bf e}_r\quad {\rm
or}\quad \frac{d\Phi}{dr}=\frac{GM(r)}{r^{d-1}},
\end{eqnarray}
where
\begin{eqnarray}
\label{virial25b}
M(r)=\int_{R_*}^r \rho(r') S_d {r'}^{d-1}\, dr'
\end{eqnarray}
is the mass of the gas contained within the sphere of radius $r$. Using
Eq.~(\ref{virial25q}) or Eq.~(\ref{nuit}), the virial of the gravitational
force (\ref{mic}) can be written as
\begin{equation}
\label{virial26}
W_{ii}=-S_dG\int_{R_*}^R \rho(r)M(r)r\, dr=-\int_{R_*}^R\frac{GM(r)}{r^{d-2}}\,
dM(r).
\end{equation}
For $d=2$, we immediately recover Eq.~(\ref{deu}). For $d\neq 2$, using Eq.
(\ref{all}), we obtain the formula
\begin{equation}
\label{virial27b}
W=-\frac{1}{d-2}\int_{R_*}^R \rho(r)\frac{GM(r)}{r^{d-2}}S_d r^{d-1}\, dr,
\end{equation}
which is useful to calculate the gravitational potential energy of a spherically
symmetric distribution of matter.
This expression can be directly obtained by approaching from
infinity a succession of spherical shells of
mass $dM(r)=\rho(r) S_dr^{d-1}dr$ with potential energy
$-GM(r)dM(r)/[(d-2)r^{d-2}]$ in the field of the mass $M(r)$ already
present, and integrating over $r$.

{\it Remark:} Combining Eqs.~(\ref{nrj}), (\ref{dev3}) and (\ref{virial27b}) we
get
\begin{equation}
W_{\rm tot}=-\frac{1}{d-2}\int_{R_*}^R \rho(r)\frac{G(M_*+M(r))}{r^{d-2}}S_d
r^{d-1}\, dr
\end{equation}
or, equivalently,
\be
W=-\frac{S_d G}{d-2}\int_{R_*}^R \rho(r) (M_*+M(r))r\, dr.
\ee

\subsection{The equilibrium virial theorem}

The condition of hydrostatic equilibrium from Eq. (\ref{eqhydro}) can be
rewritten as
\be
\nabla P+\rho\nabla\Phi+\rho\nabla\Phi_{\rm ext}={\bf 0}.
\ee
Taking the scalar product of this relation with ${\bf r}$ and integrating over
the region containing the gas we get
\be
\int {\bf r}\cdot \nabla P\, d{\bf r}-W_{ii}-W^{\rm
ext}_{ii}=0,
\ee
where  $W^{\rm ext}_{ii}$  and $W_{ii}$ are defined by Eqs.~(\ref{vir1}) and
(\ref{mic}). Integrating by parts, we obtain the scalar virial theorem
\be
d\int P\, d{\bf r}+W_{ii}+W^{\rm
ext}_{ii}=\oint P {\bf r}\cdot  d{\bf S}.
\ee
For a nonrelativistic gas, the local pressure is given by
\begin{eqnarray}
P=\frac{1}{d}\int f v^2\, d{\bf v}.
\end{eqnarray}
This relation is valid for an arbitrary distribution function. The kinetic
energy is
\begin{eqnarray}
E_{\rm kin}=\int f \frac{v^2}{2}\, d{\bf r}d{\bf v}.
\end{eqnarray}
Therefore, we have the relation
\begin{eqnarray}
\label{ju}
E_{\rm kin}=\frac{d}{2}\int P \, d{\bf r}.
\end{eqnarray}
The scalar virial theorem can then be rewritten as
\begin{eqnarray}
2E_{\rm kin}+W_{ii}+W^{\rm ext}_{ii}=\oint P {\bf r}\cdot d{\bf S}.
\end{eqnarray}
For $d\neq 2$, using Eqs.~(\ref{all2}) and (\ref{all})
we obtain
\begin{eqnarray}
\label{ju1}
2E_{\rm kin}+(d-2)W_{\rm tot}=\oint P {\bf r}\cdot d{\bf S},
\end{eqnarray}
where we have introduced $W_{\rm tot}=W+W_{\rm ext}$.
For $d=2$, using Eqs.~(\ref{deu2}) and (\ref{deu}) we get
\begin{eqnarray}
\label{ju2}
2E_{\rm kin}-\frac{GM^2}{2}\left (1+2\frac{M_*}{M}\right )=\oint P
{\bf r}\cdot
d{\bf S}.
\end{eqnarray}
To obtain these identities, we have not assumed that the gas is spherically
symmetric (only that the central body is spherically symmetric). Now, for a
spherically symmetric system enclosed between
the spheres of radius $R_*$ and $R$, owing to the fact that the pressure
$P$ is uniform on these spheres, we have 
\begin{eqnarray}
\oint P {\bf r}\cdot d{\bf S}=d(P(R)V-P(R_*)V_*),
\end{eqnarray}
where $V$ and $V_*$ are the volumes of the spheres of radius $R$ and $R_*$
respectively. Since $V=\frac{1}{d}S_dR^d$ this relation can be rewritten as
\begin{eqnarray}
\label{ju3}
\oint P {\bf r}\cdot d{\bf S}=P(R)S_d R^d-P(R_*)S_d R_*^d.
\end{eqnarray}
The scalar virial theorem is then given by Eqs.~(\ref{ju1}) and (\ref{ju2})
with Eq.~(\ref{ju3}).

\section{Condition of hydrostatic equilibrium for a spherical and
isotropic stellar system}
\label{sec_che}

Let us consider a stellar system described by a
distribution function $f({\bf r},{\bf v})$. For a spherical and isotropic system
we have $f=f(\epsilon)$ where $\epsilon=v^2/2+\Phi({\bf r})$ is the individual
energy of the stars by unit of mass. We define the density
\begin{eqnarray}
\rho({\bf r})=\int f\, d{\bf v}=\int f\lbrack
v^2/2+\Phi({\bf r})\rbrack\, d{\bf v}
\end{eqnarray}
and the pressure
\begin{eqnarray}
\label{pr}
P({\bf r})=\frac{1}{d}\int f v^2\, d{\bf
v}=\frac{1}{d}\int f\lbrack v^2/2+\Phi({\bf r})\rbrack v^2\, d{\bf v}.
\end{eqnarray}
We note that $\rho({\bf r})=\rho\lbrack \Phi({\bf r})\rbrack$ and $P({\bf
r})=P\lbrack \Phi({\bf r})\rbrack$ depend only on the gravitational potential.
Eliminating formally $\Phi({\bf r})$ between
these two expressions, we find that the equation of state is barotropic in the
sense that the pressure depends only on the density: $P({\bf
r})=P\lbrack\rho({\bf r})\rbrack$. We can therefore associate to the stellar
system with distribution function $f(\epsilon)$ a corresponding barotropic gas
with an equation of state $P=P(\rho)$. Taking the gradient of the
pressure in
Eq.~(\ref{pr}) and making straightforward manipulations, we get
\begin{eqnarray}
\nabla P&=&\frac{1}{d}\nabla\Phi \int f'(\epsilon) v^2\,
d{\bf v}\nonumber\\
&=&\frac{1}{d}\nabla\Phi \int\frac{\partial f}{\partial {\bf v}}\cdot
{\bf v}\,
d{\bf v}\nonumber\\
&=&-\frac{1}{d}\nabla\Phi \int f \nabla\cdot {\bf v}\,
d{\bf v}\nonumber\\
&=&-\nabla\Phi \int f\,
d{\bf v}.
\end{eqnarray}
We therefore recover the condition of hydrostatic equilibrium for a gas
\begin{eqnarray}
\nabla P+\rho\nabla\Phi={\bf 0}.
\end{eqnarray}
In the presence of an external potential $\Phi_{\rm
ext}$, we just have to make the substitution $\Phi\rightarrow \Phi+\Phi_{\rm
ext}$.

\section{Newton's law in $d$ dimensions}
\label{sec_pfd}

If we consider  a spherically symmetric distribution of matter with density
$\rho(r)$, the Poisson equation
\be
\label{poi}
\Delta\Phi=S_{d}G\rho
\ee
becomes
\be
\frac{1}{r^{d-1}}\frac{d}{dr}\left (r^{d-1}\frac{d\Phi}{dr}\right
)=S_{d}G\rho.
\ee
Integrating this equation between $0$ and $r$, we
obtain
\be
\label{amo}
\frac{d\Phi}{dr}=\frac{G M(r)}{r^{d-1}},
\ee
where
\be
M(r)=\int_{0}^{r}\rho(r')S_d {r'}^{d-1}\, dr'
\ee
denotes the mass contained within the sphere of radius~$r$. We have
\be
\rho(r)=\frac{M'(r)}{S_d r^{d-1}}.
\ee
Eq.~(\ref{amo}) expresses Newton's second theorem (in $d$ dimensions)
saying that the gravitational force produced in $r$ by a spherically
symmetric distribution of matter is the same as it would be if all the mass
$M(r)$ were concentrated into a point at its center. In vectorial form, 
\be
-\nabla\Phi=-\frac{GM(r)}{r^{d-1}}{\bf e}_r.
\ee
This equation can also be obtained by integrating Eq.~(\ref{poi}) over a sphere
of radius $r$ and using the Gauss (or Ostrogradsky) theorem.

If the density $\rho(r)$ vanishes above a certain radius $R$, then for $r\ge R$,
we have
\be
\label{monde}
\frac{d\Phi}{dr}=\frac{G M}{r^{d-1}},
\ee
where $M=M(R)$ is the mass enclosed within the sphere of radius $R$.
In particular, we obtain
\be
\frac{d\Phi}{dr}(R)=\frac{G M}{R^{d-1}}.
\ee
For $r\ge R$, the gravitational potential is given by
\be
\label{nu1}
\Phi(r)=-\frac{1}{d-2}\frac{GM}{r^{d-2}}\quad (d\neq 2),
\ee
\be
\label{nu2}
\Phi(r)=GM\ln \left (\frac{r}{R}\right )\quad (d= 2).
\ee
In particular,
\be
\label{nu1p}
\Phi(R)=-\frac{1}{d-2}\frac{GM}{R^{d-2}}\quad (d\neq 2),
\ee
\be
\label{nu2p}
\Phi(R)=0\quad (d= 2).
\ee

\section{The potential in $d=3$ dimensions for a spherically symmetric system}
\label{app.varenergiepot}

For a spherically symmetric system, the gravitational potential in $d=3$ can be
written as
\be
\label{eq.phispherique}
\Phi({r})=-G\int_{R_*}^{R}\int_0^{2\pi}\int_0^{\pi} \frac{\rho({r}_1)}{|{\bf
r}-{\bf r}_1|}\sin\theta \, d\theta d\phi  r_1^{2}d{r}_1.
\ee
Introducing the expansion
\be
\frac{1}{|{\bf r}-{\bf
r}_1|}=\sum_{l=0}^{+\infty}\frac{r_{<}^{l}}{r_{>}^{l+1}}P_{l}(\cos\theta),
\ee
where $r_<={\rm min}(r,r_1)$ and $r_>={\rm max}(r,r_1)$ in Eq.
(\ref{eq.phispherique}), making the change of variable
$x=\cos\theta$ and using the identity $\int_{-1}^{+1}P_{l}(x)dx=2\delta_{l0}$,
we
find that
\bea
\Phi({r})=-4\pi G\left\lbrack
\frac{1}{r}\int_{R_*}^{r}\rho(r_1)r_1^{2}dr_1+\int_{r}^{R}\rho(r_1)r_1
dr_1\right\rbrack.\nonumber\\
\eea
From this expression, we get
\be
\Phi(R_{*})=-4\pi G\int_{R_{*}}^{R}\rho(r) r dr.
\ee

\section{The energy of the ground state}
\label{sec_lmax}

At $T=0$, all the gas is uniformly concentrated on the surface of the central
body.  Therefore, the energy of the ground state
$E_{\rm min}=W_{\rm tot}^{\rm min}$
corresponds to the potential energy associated to this configuration.
To compute this minimum energy, we start from Eq.~(\ref{nrj}).
Accounting for the fact that the total mass of the gas $M$ is located at
$r=R_{*}$, we have
\be
E_{\rm min}=\frac{1}{2}M\Phi(R_{*})+\Phi_{\rm ext}(R_{*})M.
\ee
For $d\neq 2$, using Eqs.~(\ref{pcore1}) and (\ref{nu1}),
we obtain
\be
E_{\rm min}=-\frac{1}{d-2}(1+2\mu)\frac{GM^{2}}{R_{*}^{d-2}}.
\ee
The normalized minimum energy is therefore
\be
\Lambda_{\rm max}  = \frac{1+2\mu}{2(d-2)\zeta^{d-2}}.
\ee
For $d=2$, using Eqs.~(\ref{pcore2}) and (\ref{nu2}),
we obtain
\be
E_{\rm min} = \frac{GM^2}{2}(1+2\mu)\ln\left(\frac{R_*}{R}\right).
\ee
The normalized minimum energy is therefore
\be
\Lambda_{\rm max} =  -\frac{1}{2}(1+2\mu)\ln\zeta.
\ee

\end{appendix}



\end{document}